\documentclass[fleqn,usenatbib]{mnras}

\usepackage[T1]{fontenc}
\usepackage{ae,aecompl}

\usepackage{graphicx}	% Including figure files
\usepackage{amsmath}	% Advanced maths commands
\usepackage{amssymb}	% Extra maths symbols

\title[Three Galactic star clusters discovered with Gaia]{Three new Galactic star clusters discovered in the field of the open cluster NGC\,5999 with  Gaia DR2}

\author[F. A. Ferreira et al.]{
Filipe A. Ferreira$^{1}$\thanks{E-mail: filipe1906@ufmg.br},
J. F. C. Santos Jr.$^{1}$,
W. J. B. Corradi$^{1}$,
F. F. S. Maia$^{1,2}$ \newauthor
\ and M. S. Angelo$^{3}$
\\
$^{1}$Universidade Federal de Minas Gerais, Departamento de F\'isica, Av. Ant\^onio Carlos 6627, 31270-901, Brazil \\
$^{2}$Universidade de S\~ao Paulo, Instituto de Astronomia, Geof\'isica e Ci\^encias Atmosf\'ericas, Rua do Mat\~ao 1226, 05508-090, Brazil 
\\
$^{3}$Centro Federal de Educa\c c\~ao Tecnol\'ogica de Minas Gerais, Av. Monsenhor Luiz de Gonzaga, 103, 37250-000, Brazil
}

\date{Accepted XXX. Received YYY; in original form ZZZ}

\pubyear{2018}

\begin{document}
\label{firstpage}
\pagerange{\pageref{firstpage}--\pageref{lastpage}}
\maketitle

\begin{abstract}
We report the serendipitous discovery of three new open clusters, named UFMG\,1, UFMG\,2 and UFMG\,3 in the field of the intermediate-age cluster NGC 5999, by using Gaia DR2 data. A colour-magnitude filter tailored for a proper selection of  main-sequence stars and red clump giants turned evident the presence of NGC\,5999 and these three new stellar groups in proper motion space. Their structural parameters were derived from King-profile fittings over their projected stellar distributions and isochrone fits were performed on the clusters cleaned colour-magnitude diagrams built with Gaia bands to derive their astrophysical parameters. The clusters projected sky motion were calculated for each target using our members selection. Distances to the clusters were inferred from stellar parallaxes through a bayesian model, showing that they are marginally consistent with their isochronal distances, considering the random and systematic uncertainties involved. The new clusters are located in the nearby Sagittarius arm ($d\sim 1.5$\,kpc) with NGC\,5999 at the background ($d\sim 1.8$\,kpc). They contain at least a few hundred stars of nearly solar metallicity and have ages between 100 and 1400\,Myr.

\end{abstract}


%%%%%%%%%%%%%%%%%%%%%%%%%%%%%%%%%%%%%%%%%%%%%%%%%%

%%%%%%%%%%%%%%%%% BODY OF PAPER %%%%%%%%%%%%%%%%%%

\section{Introduction}

The spatial density fluctuations of low Galactic latitude stellar populations make difficult the identification and cha\-racterization of open clusters for studies of the history and structure of the Galactic disk. The recent Gaia DR2 \citep{Prusti:2016,Brown:2018,Evans:2018} provides precise astrometric and photometric data for an unprecedented number of stars, allowing us to better investigate and/or find many of these objects, normally suppressed by both high density stellar fields and extinction.

Nowadays there are more than  2500 known open clusters \citep{Dias:2002,Kharchenko:2013}. With the advent of near-infrared surveys such as 2MASS \citep{Skrutskie:2006} and VVV \citep{Minniti:2010}, new objects have been discovered, notably young clusters embedded in molecular clouds that were invisible in the optical due to high absorption \citep[e.g.][]{Barba:2015,Borissova:2014,Bica:2003}. In the optical, the Gaia mission has acquired whole sky high precision proper motions and paralaxes which are suited to an accurate distinction between cluster and field stars due to the expected confined loci of cluster stars in the astrometric space. This has led to an increase of the  number of open clusters discovered recently \citep{Ryu:2018,Cantat:2018,Castro:2018,Torrealba:2018}. 

We have been carrying out a study of open clusters in dense stellar fields using VVV, 2MASS and more recently Gaia to fully characterize such objects. One of these objects is NGC\,5999, located in the direction of the Galactic disk, with Galactic coordinates $\ell = 326^\circ$ and $b = -1.93^\circ$ \citep{Dias:2002}. NGC\,5999 is approximately 400 Myr old, with distance determinations ranging from 1.6 to 2.5\,kpc and reddening $E(B-V) = 0.45\pm 0.05$ \citep{Dias:2002,Piatti:1999,Kharchenko:2013,Netopil:2007,Santos:1993,Moni:2014}. 
The Milky Way Star Clusters project \citep[MWSC;][]{Kharchenko:2013}, which is based on 2MASS photometry and PPMXL \citep{Roeser:2010} astrometry, gives for NGC\,5999: $\log(t({\rm yr}))=8.600\pm0.095$ (with 2 stars used to calculate the age), $d=1629$\,pc, $E(B-V)=0.437$, core radius $r_c=0.71\pm0.09$\,pc, and tidal radius $r_t=6.05\pm0.83$\,pc. \cite{Piatti:1999} employed $BVI$ observations aided by integrated spectroscopy from \cite{Santos:1993} to derive for NGC\,5999 the following parameters: $t=400\pm100$\,Myr, $d=2.2\pm0.4$\,kpc, and $E(B-V)=0.45\pm0.05$. \cite{Moni:2014} investigated a possible connection between NGC\,5999 and the planetary nebulae VBe\,3, located at 5\,arcmin from the cluster centre. To do this, they observed spectroscopically 4 members of the cluster and the planetary nebulae, caming to the conclusion that there is no association between them as indicated by their large radial velocity difference. An average of $v_r=-39\pm3$\,kms$^{-1}$ for the cluster radial velocity was derived from the velocities of the 4 member stars. 
\cite{Netopil:2007} searched for chemically peculiar stars in NGC\,5999 employing observations in the $\Delta a$ photometric system, specifically designed to detect such objects. They obtained $\log(t({\rm yr}))=8.50\pm0.01$, $E(B-V)=0.48\pm0.05$ and $d=2.20\pm0.36$\,kpc by fitting isochrones to a colour-magnitude diagram (CMD) in that photometric system. In this paper we propose to improve NGC\,5999 parameters taking advantage of the Gaia DR2 data precision, both astrometric and photometric, leading to the possibility of a clear distinction of members and field stars. 

A method often employed to detect a star cluster is the search for spatial stellar overdensities with respect to the background \cite[e.g.][]{Froebrich:2007,Kronberger:2006}.
However, crowded fields make difficult the analysis and overdensities are not always perceptible. The Gaia DR2 allowed automatic overdensity searches based on 5-parameters, efficiently exploring the mission capabilities \citep{Cantat:2018,Castro:2018,Babusiaux:2018}. Although operating on the full Gaia DR2 database needs data mining techniques, serendipitous discoveries based on local strategies to enhance the contrast between field and a hidden cluster are not improbable as the ones reported in this study.

This paper is structured as follows. In Sect. 2 the data 
is presented. In Sect. 3  the method used in the detection of the new clusters is described. In Sect. 4 the analysis procedures are developed, including membership assessment, determination of astrophysical 
and structural parameters, and distance inference via a bayesian method. The sky region investigated is discussed in the context of the newly found clusters in Sect. 5 and 
the concluding remarks  are given in Sect. 6.

\begin{figure*}
\includegraphics[width=0.325\linewidth]{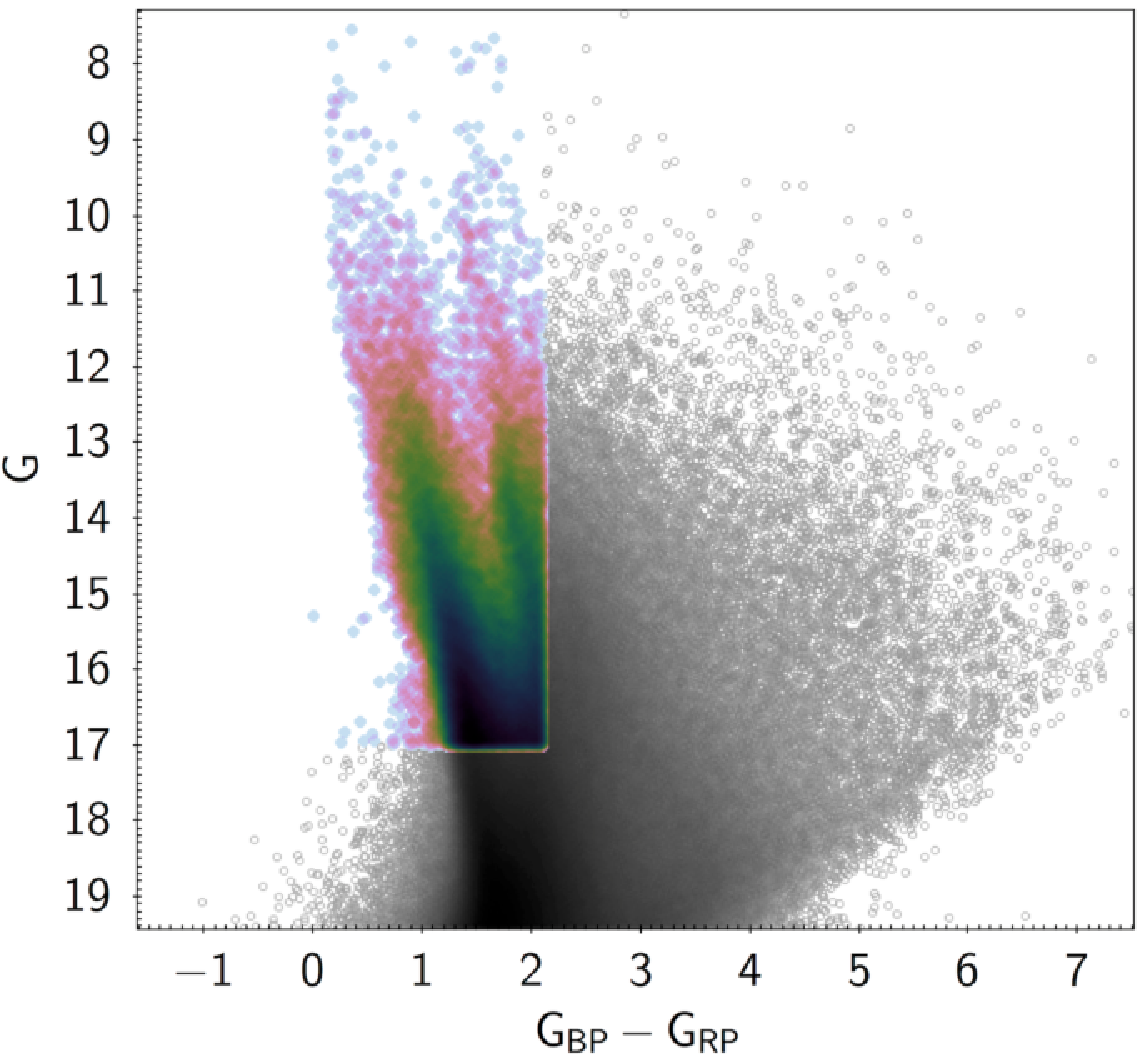}
\includegraphics[width=0.32\linewidth]{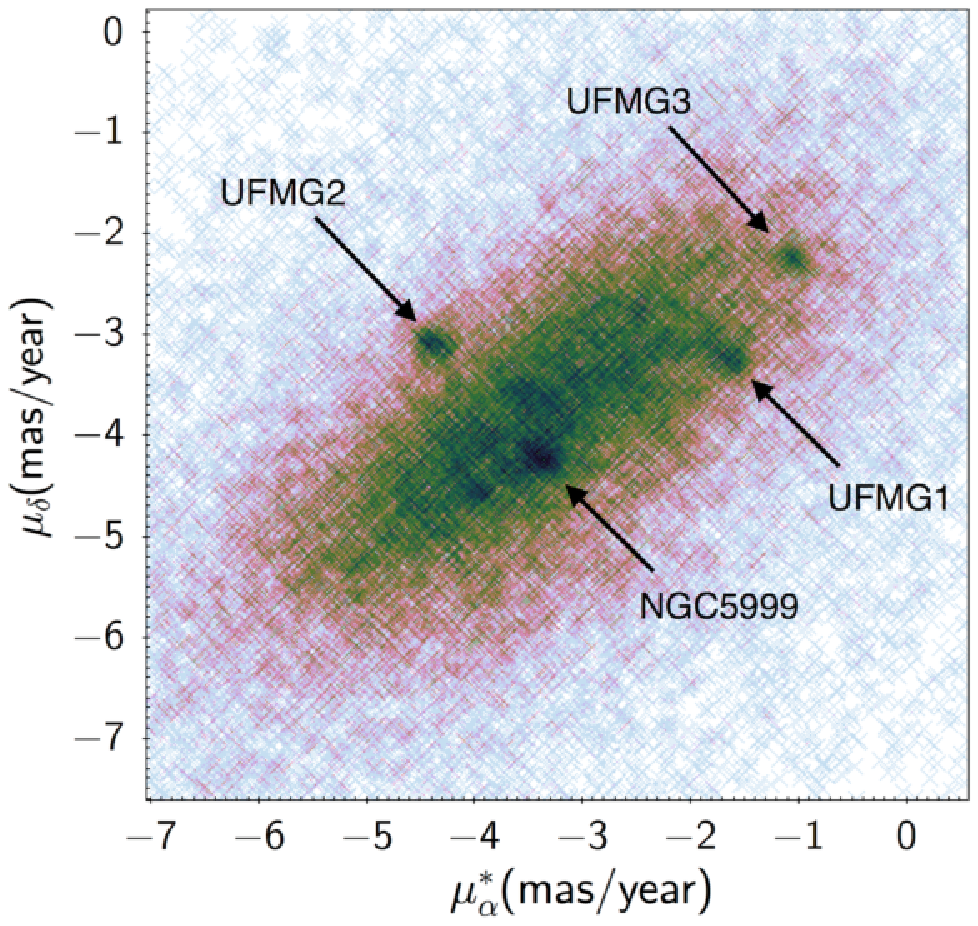}
\includegraphics[width=0.33\linewidth]{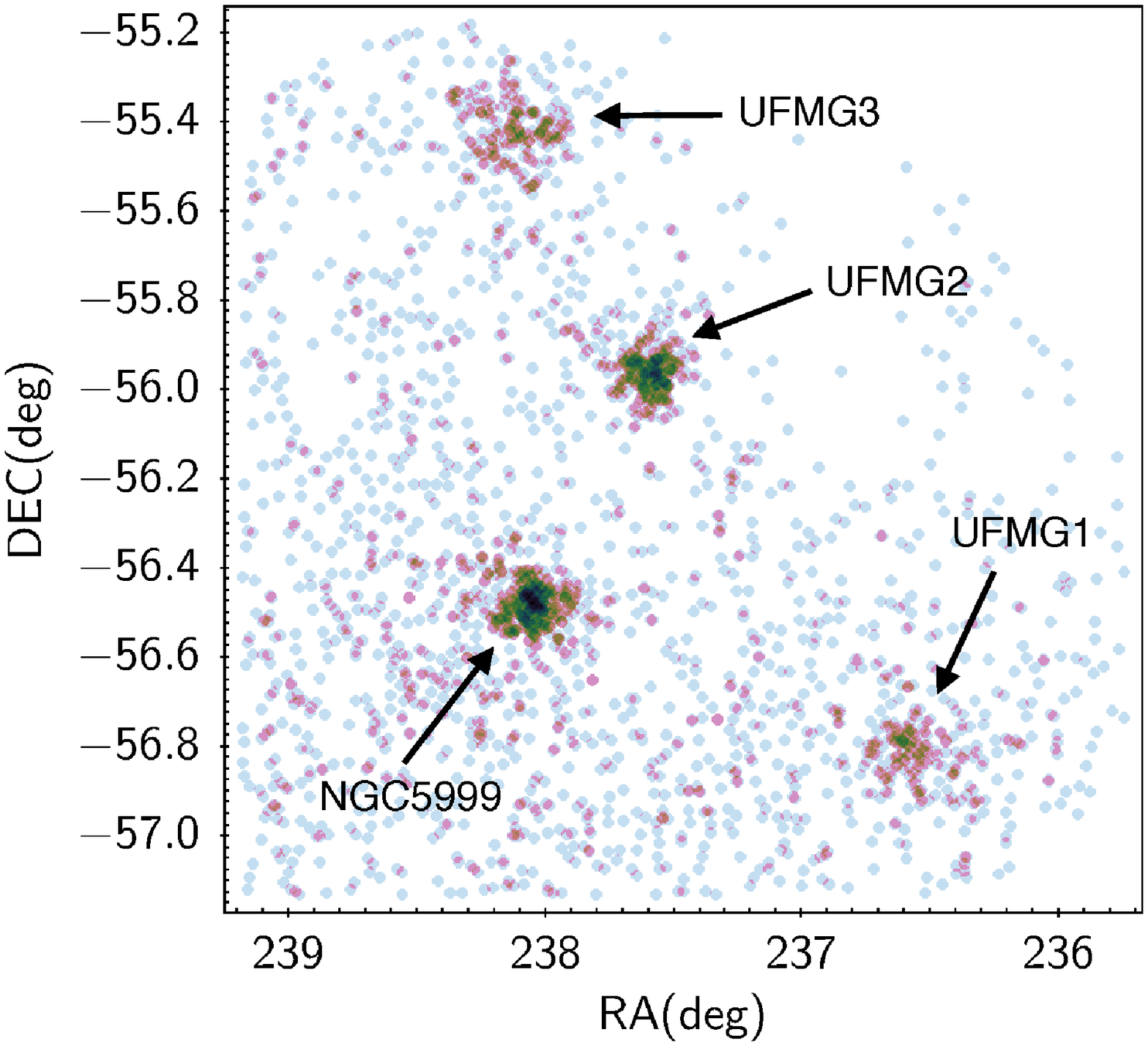}
\caption{Left: Colour-magnitude diagram of the  NGC\,5999 region.  Colored/dark symbols highlight the stars filtered by colour and magnitude. Middle: Corresponding vector point diagram of the  subsample. Right: The spatial distribution of NGC\,5999 and the three discovered clumps. Without enhancing the contrast between the objects and the field such discovery would have not been possible.}
\label{fig:color_cut}
\end{figure*}

\section{Data}

The Gaia DR2 catalogue provides positions, proper motions in right ascension and declination, parallaxes and magnitudes in three bands ($G$, $G_{BP}$ and $G_{RP}$) for more than 1.3 billion sources \citep{Brown:2018,Evans:2018}. Parallax uncertainty goes from 0.04\,mas for sources at $G < 15$\,mag, to about 0.1\,mas at $G \approx 17$\,mag and up to 0.7\,mas at $G = 20$\,mag. The corresponding uncertainty in the respective proper motion components goes from 0.06\,mas\,yr$^{-1}$ (for $G < 15$\,mag) to about 0.20\,mas\,yr$^{-1}$ (for $G \approx 17$\,mag) and up to 1.2\,mas\,yr$^{-1}$ (for $G = 20$\,mag).
 Both parallax and proper motion are also affected by systematic errors, on the order of 0.1\,mas and 0.1\,mas yr$^{-1}$, respectively \citep{Lindegren:2018,Luri:2018}.
 
 We have extracted Gaia DR2 data using the Vizier service.  In order to clean our work sample from contamination due to double stars, astrometric effects from binary stars and calibration problems, we used equations (1), (2) and (3) from \cite{Arenou:2018}. This is a recommended basic filter aimed to assure the best quality of the data for analysis, after which around 20\% of the total sample was kept.

\section{Discovery of the new clusters}

Analyzing the region adjacent to NGC\,5999, we have noticed the existence of other clusters not yet reported in the literature as to our best knowledge.  To arrive at this conclusion, we took the following steps. After the basic filter was applied, we selected data from a region of 1.3 degrees radius centred in NGC\,5999. Since the cluster is immersed in a dense star field, it was impossible to contrast it over the background. 
In order to highlight the cluster members in the proper motion diagram (VPD), we performed cuts in colour and magnitude (left panel of Fig.~\ref{fig:color_cut}), limiting our sample to  magnitude $G<17$ and colour ($G_{BP}-G_{RP}$)$<2.1$, thus excluding faint and highly reddened stars. This subsample clearly revealed NGC\,5999 and three unexpected overdensities in the VPD (middle panel of Fig.~\ref{fig:color_cut}). 

 Exploration of the spatial region associated with each of these clumps in the VPD, showed that their stars, besides having similar movements, also own a spatial bond. Specifically, VPD boxes of 2\,mas\,yr$^{-1}$ were defined to contain each of the four comoving structures as a constraint to look for members in right ascension and declination. The boxes size was big enough to completely select the  proper motion clumps and small enough to mitigate the field background, therefore optimizing the process of obtaining the clusters' centres and sizes. From the stellar positions in this subsample, a preliminary centre was determined, which was then used to analyse the stellar distribution around 30\,arcmin from each clump centre. We advance that the stellar distribution in the colour-magnitude diagram (see Sect. \ref{sect:cmd}) also made evident that the overdensities are indeed star clusters. 
The spatial distribution of NGC\,5999 stars and those of the three discovered clusters, named UFMG\,1, UFMG\,2 and UFMG\,3, is shown in the right panel of Fig.~\ref{fig:color_cut}.  The abbreviation UFMG accounts for Universidade Federal de Minas Gerais.

\begin{figure}
\centering
\includegraphics[width=0.45\linewidth]{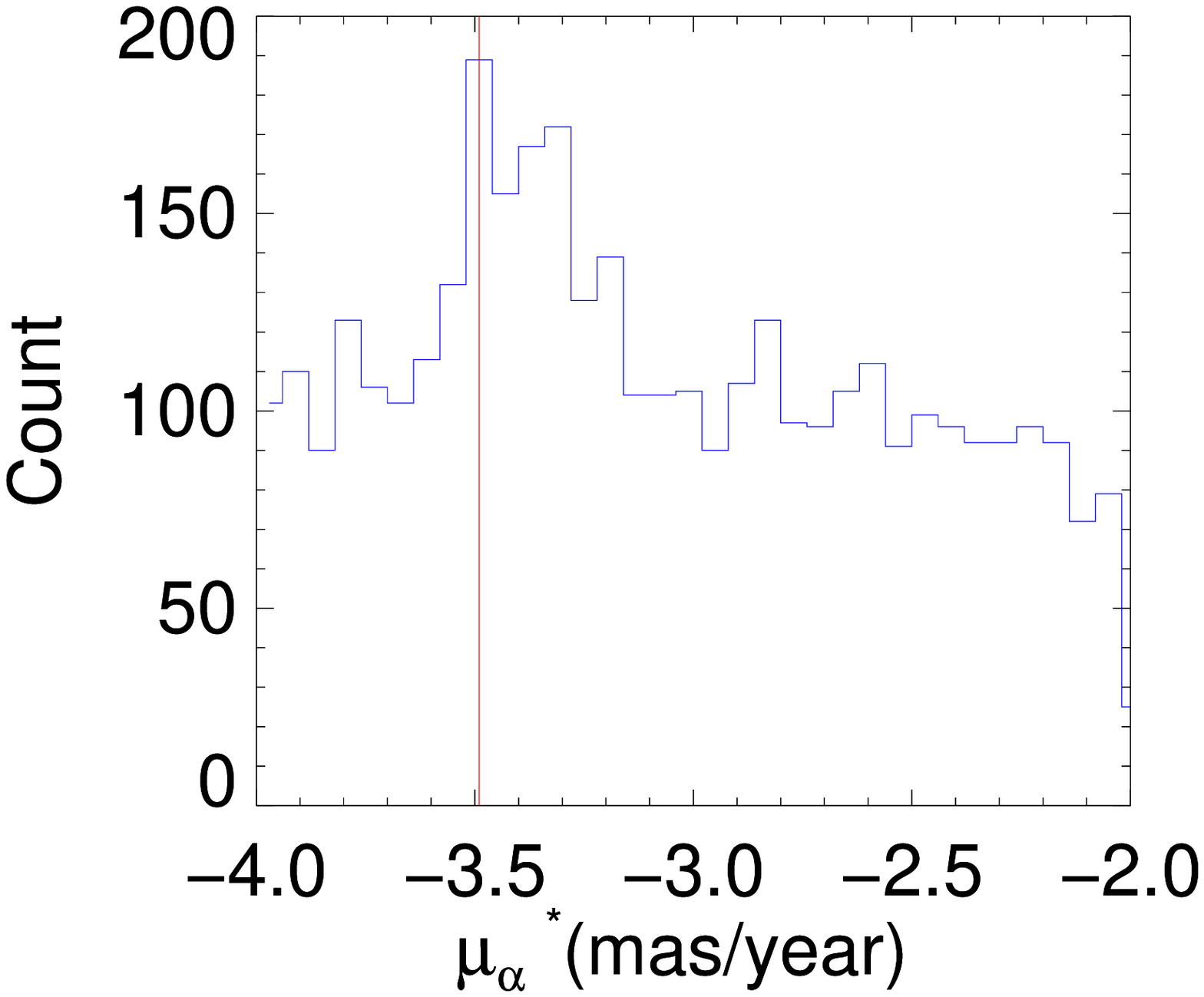}
\includegraphics[width=0.44\linewidth]{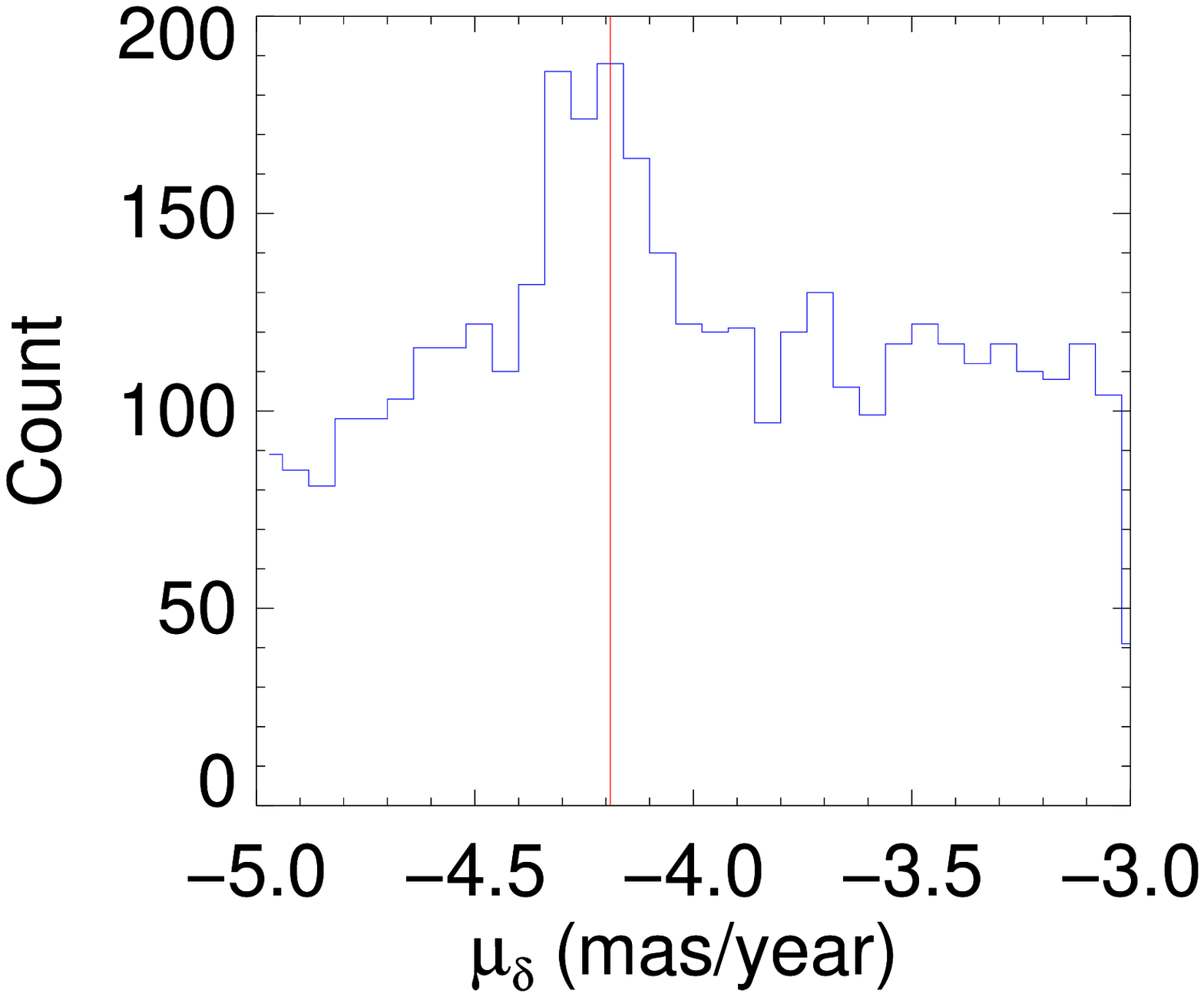} \\
\caption{Histogram of the proper motion components of the cluster NGC 5999. The mode of the distributions is indicated.}
\label{fig:pm_peak}
\end{figure}

\begin{figure}
\centering
\includegraphics[width=0.45\linewidth]{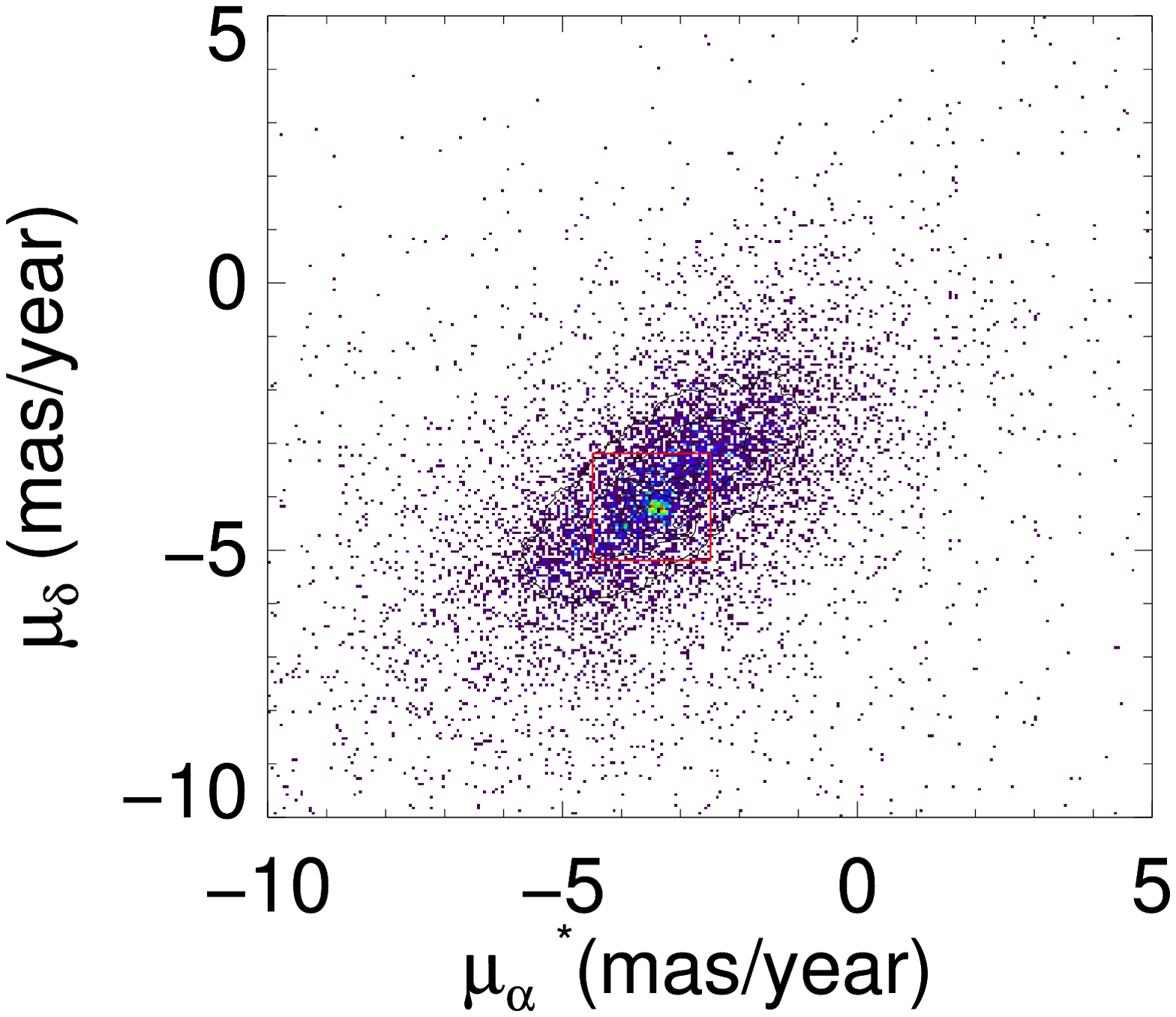}
\includegraphics[width=0.44\linewidth]{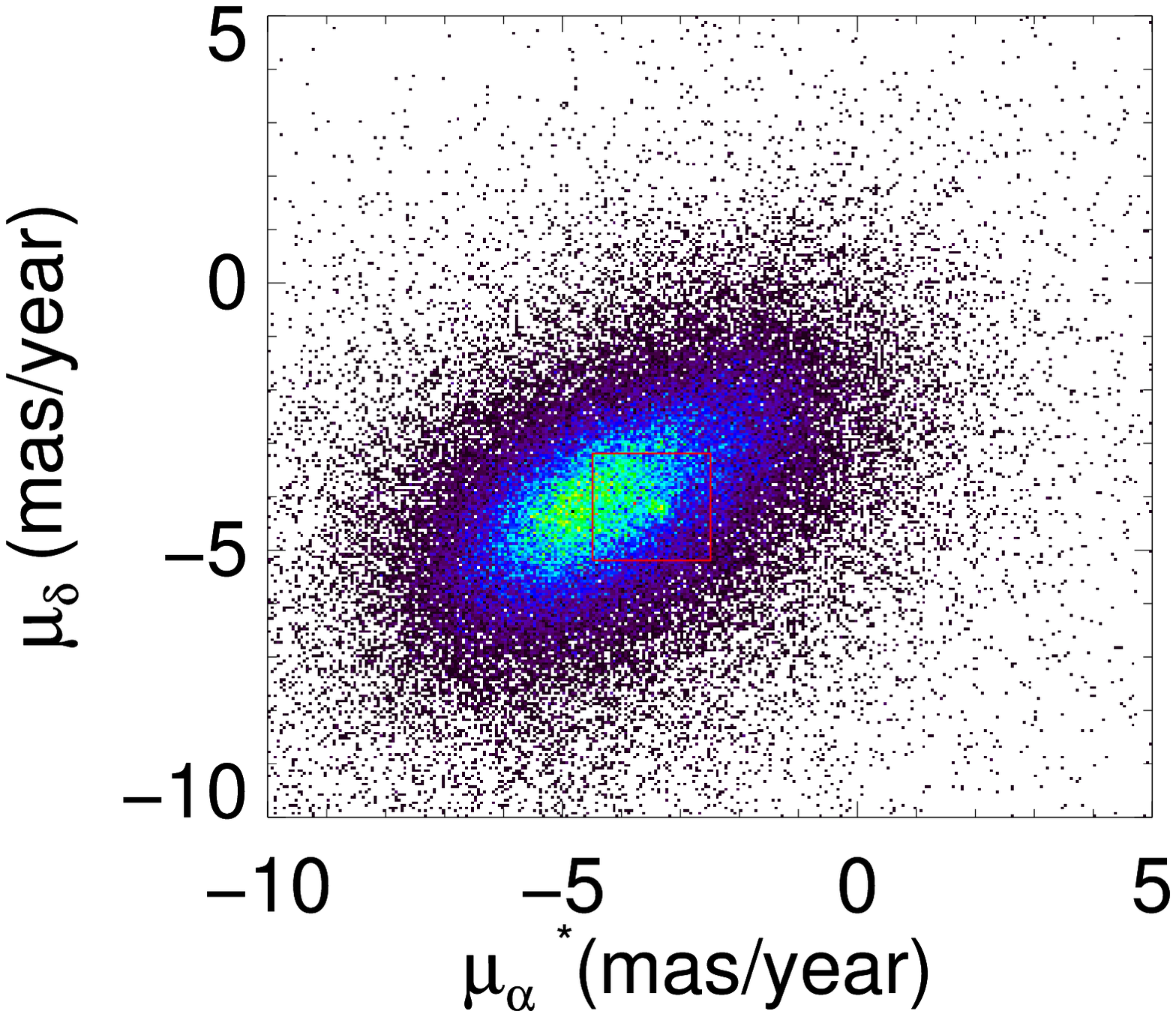} \\
\caption{Proper motion data within 30 arcmin of NGC 5999. Left: the data filtered by colour and magnitude and the recommended basic filter, highlighting the cluster overdensity. Right: the data selected only by the basic filter, employed in the subsequent analysis. The red square indicate the box of 2\,mas\,yr$^{-1}$ in $\mu_{\alpha}^\ast$ and $\mu_{\delta}$.}
\label{fig:pm_mask}
\end{figure}

 We emphasize that the constraints defined by the colour and magnitude cuts in the sample were only performed to enhance the overdensities in the VPD, as exemplified in Fig. \ref{fig:pm_peak}, allowing us to locate them and compute the mode of the proper motion components distribution by making histograms in $\mu_{\alpha}^\ast$ and $\mu_{\delta}$ for the data in the 2\,mas\,yr$^{-1}$ box defined for each clump (Fig. \ref{fig:pm_mask}). Thereafter the colour-magnitude filter was dismissed and the sample only subjected to the basic filter was employed in the subsequent analysis.

\section{Analysis}

\subsection{Centre and size determination}
\label{sect:4.2}

 Having defined our working sample, i.e., best quality Gaia data within the 2\,mas\,yr$^{-1}$ proper motion box and 30\,arcmin from the objects centres, we built radial density profiles (RDP) to determine the size of the objects, the local background and fine tune their centres. The stellar density was calculated by measuring the number of stars in concentric rings around the initial visual centre and dividing it by the rings area. Independent RDPs were obtained by using four different ring widths and the results merged in a single profile. 

\begin{figure*}
\centering
\includegraphics[width=0.45\linewidth]{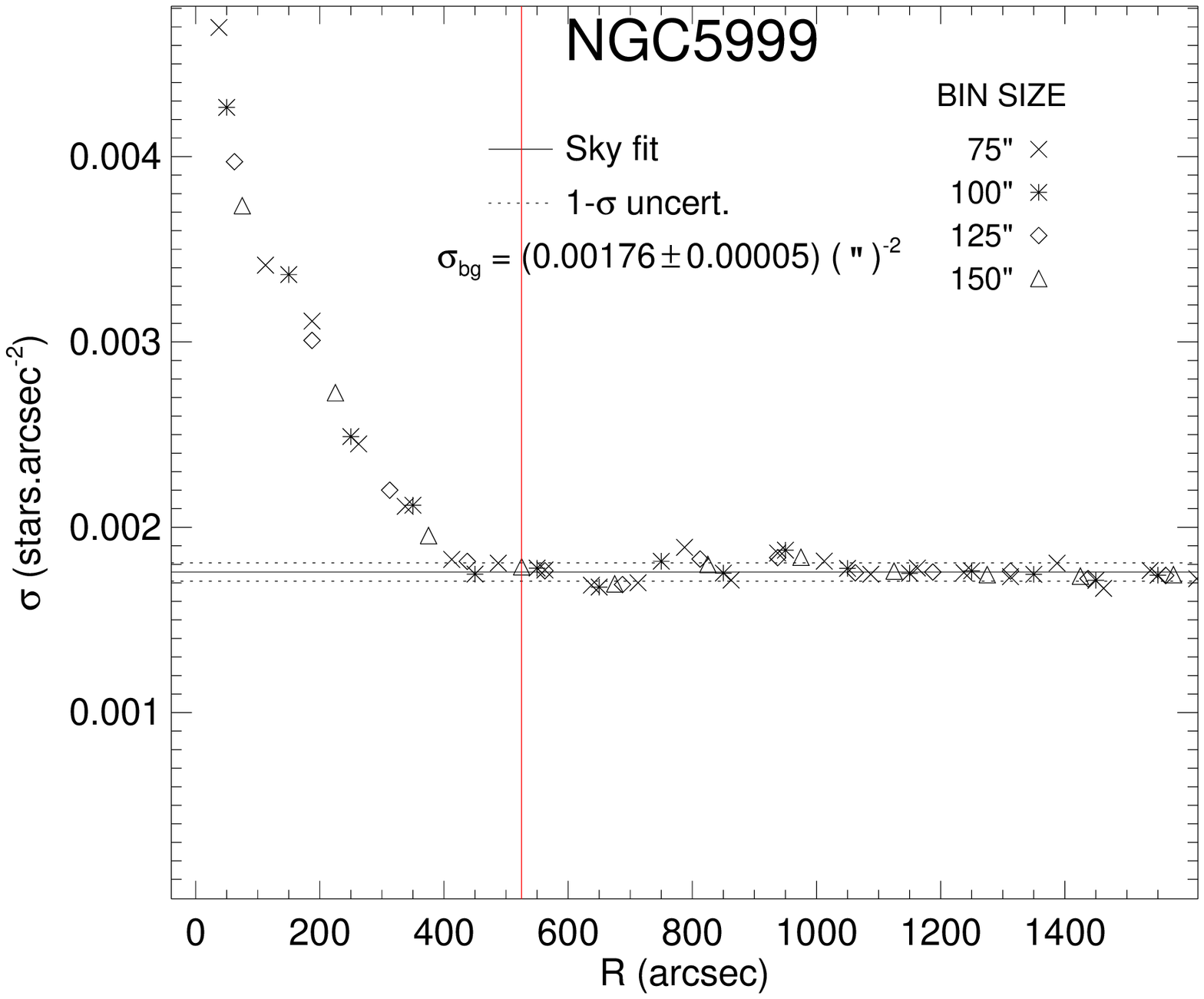} \hspace{0.25cm}
\includegraphics[width=0.46\linewidth]{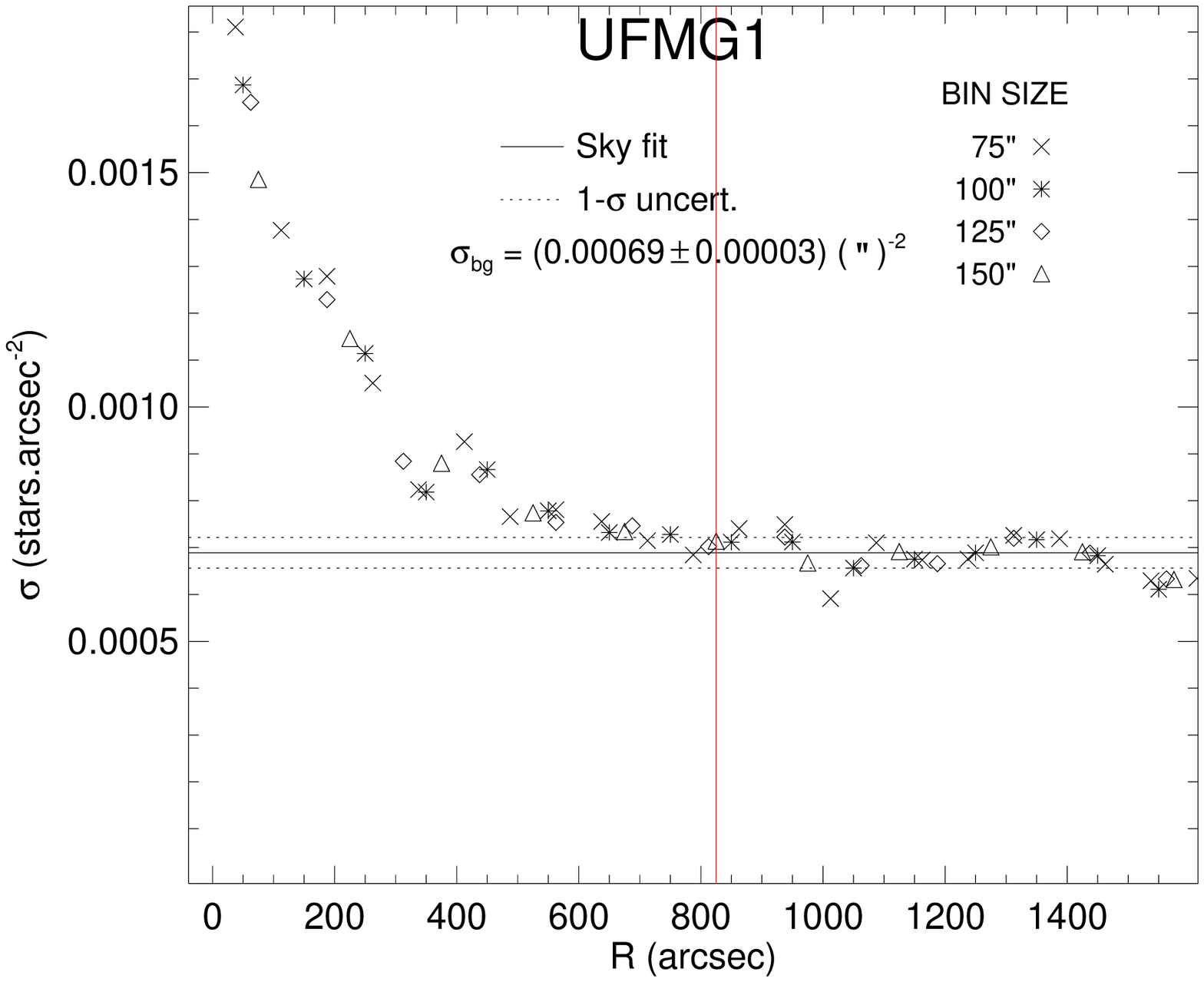} \\ \vspace{0.25cm}
\includegraphics[width=0.45\linewidth]{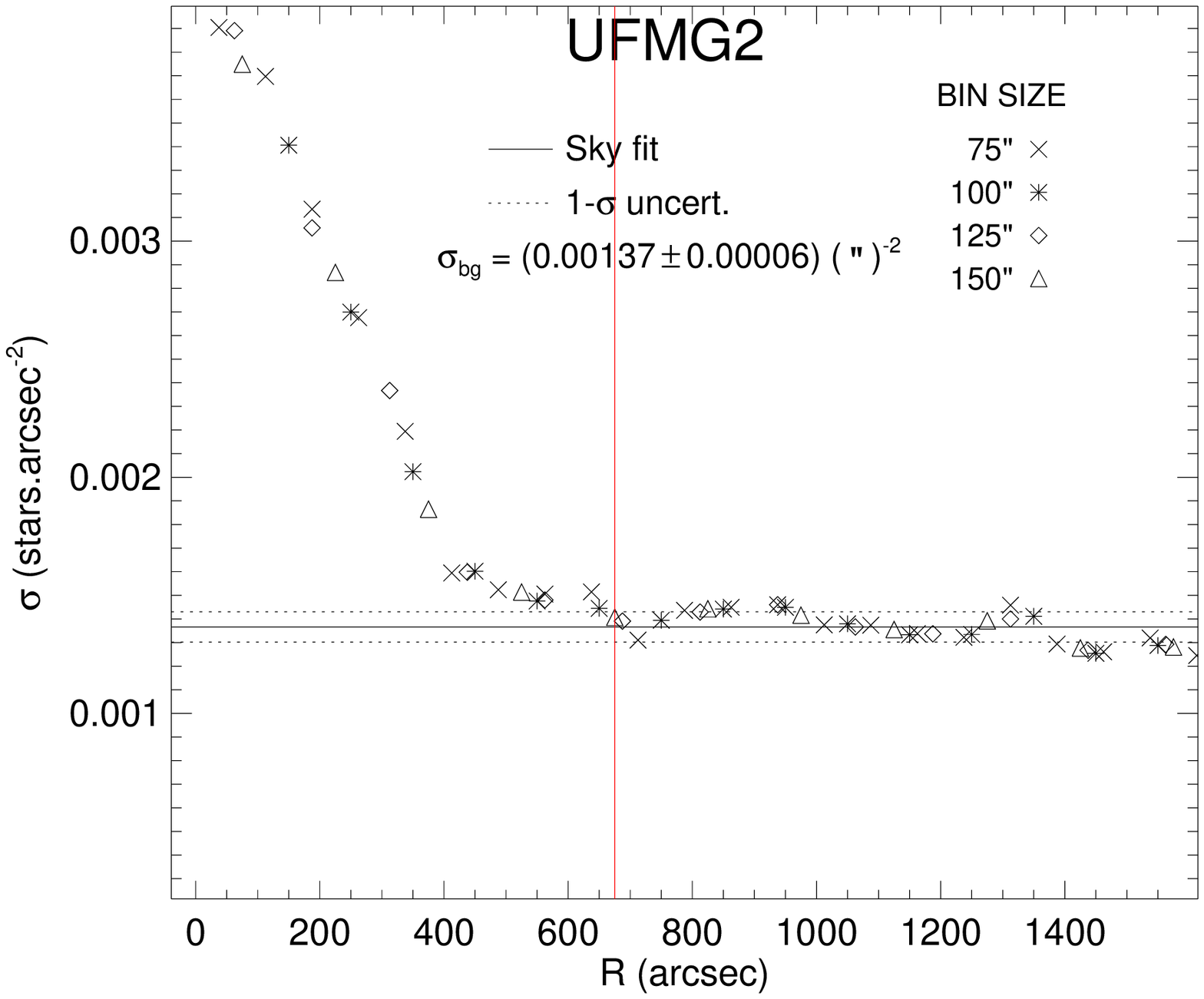} \hspace{0.25cm}
\includegraphics[width=0.46\linewidth]{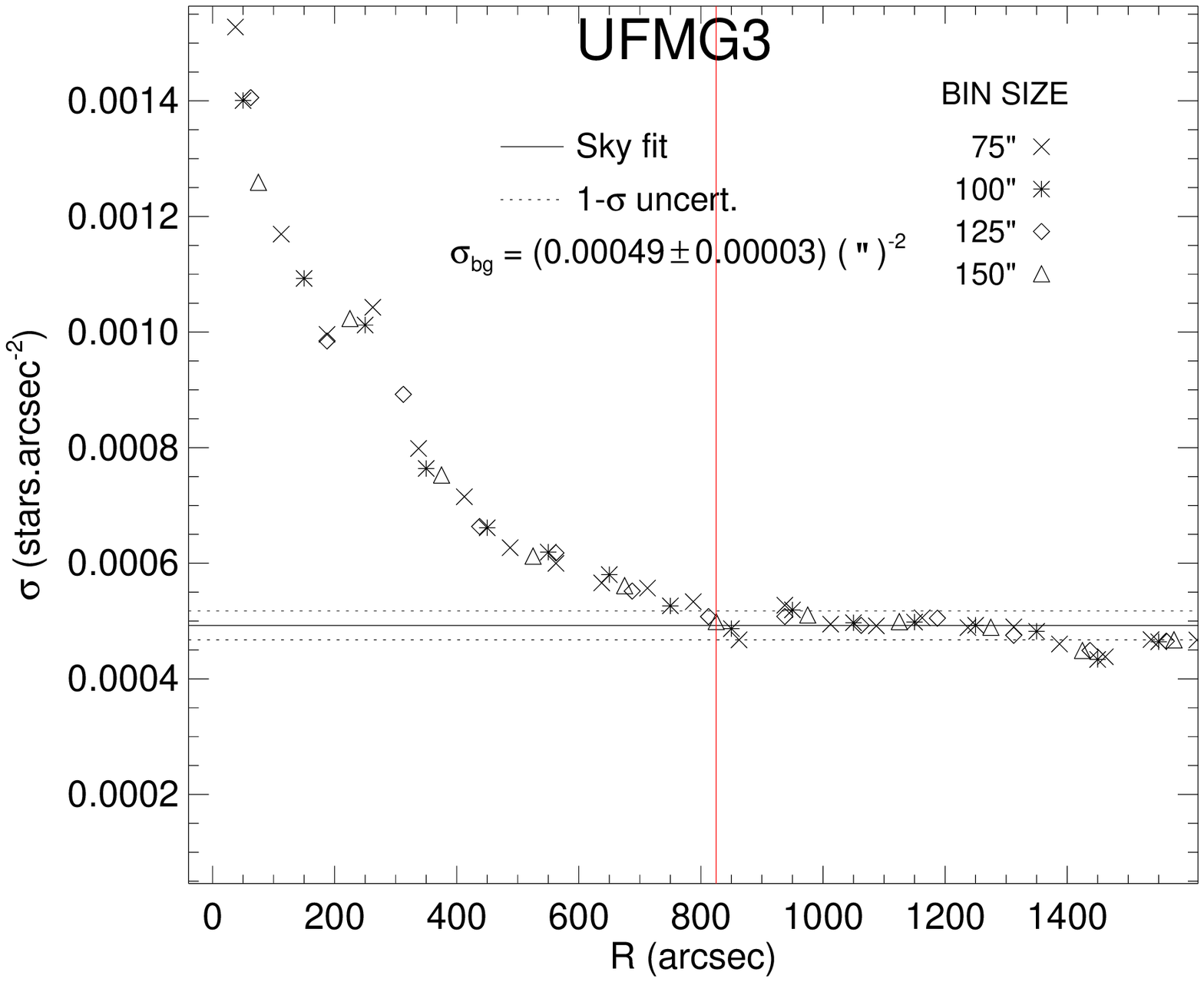}
\caption{Radial density profiles of the studied clusters used to determine their limiting radius (vertical line). The horizontal lines represent the mean background density level (solid line) and its
$1-\sigma$ fluctuations (dashed lines).} 
%It can be seen that UFMG\,1 UFMG\,3 are both sparser than UFMG\,2.}
\label{fig:rdp1}
\end{figure*}

 We refined the visual center by making small adjustments to its coordinates, searching for the maximum central density of the profile. The sky background level ($\sigma_{bkg}$) and its uncertainty was determined by fitting a straight line to the stellar density for rings well beyond the objects core, between 15 and 25\,arcmin. The limiting radius ($r_{lim}$), defined as the radius where the stellar density reaches the sky level, resulted between 7 and 15\,arcmin for all clusters. Fig.~\ref{fig:rdp1} shows the radial density profiles of each cluster and the derived sky density level used to determine the limiting radius, as given in Table~\ref{tab:parameters}.

%\subsection{Detection and proper motion mask}

%The proper motion mask consist in a 2x2 box centered on the proper motion peak values, the size value was big enough to contain the entery proper motion clumps and small enough to be possible to hightlight the clusters over the background and to determine the clusters centres and sizes.

% In the left panel of Fig. \ref{fig:mask} we can see how the color and magnitude cut reveal the clusters proper motion and, in the left panel,  the proper motion mask being aplied on the data.  We filtered the data for all the 4 targets  by keeping  their proper motion components inside the mask limit.

%The  proper motion mask was used as a first filter to restrict our sample, this one was important  to highlight the cluster over foreground/background stars. With the filtered sample, to determine the centre coordinate of the clusters and their sizes we have built radial density profiles to determine their limiting radii. It was done by measuring 

\begin{table}
\small
\caption{Astrophysical parameters for the new clusters and NGC\,5999}
\begin{tabular}{lcccc} \hline
                & NGC\,5999     & UFMG\,1        & UFMG\,2        & UFMG\,3        \\ \hline
$\alpha$(J2000) & 15 52 11.3    & 15 46 24.5     & 15 50 23.3     & 15 52 26.2     \\
$\delta$(J2000) & -56 29 17     & -56 48 29      & -55 57 32      & -55 25 19      \\
$\ell$ ($^\circ$)    & 326.00        & 325.19         & 326.14         & 326.70         \\
$b$ ($^\circ$)            & -1.94         & -1.70          & -1.37   & -1.13          \\
$\mu_\alpha^\ast$ ($\frac{\mathrm{mas}}{\mathrm{yr}}$)    &  -3.374(3)$^a$       &  -1.634(4)$^a$       &  -4.420(3)$^a$   &  -1.040(4)$^a$   \\
$\mu_\delta$ ($\frac{\mathrm{mas}}{\mathrm{yr}}$)  & -4.216(3)$^a$     &  -3.220(4)$^a$   & -3.068(3)$^a$   & -2.238(4)$^a$               \\
$r_\mathrm{lim}$ (pc)& 4.6$\pm$0.6   & 6.3$\pm$0.7    & 4.8$\pm$0.6    & 6.0$\pm$0.7    \\
$r_c$ (pc)& 1.5$\pm$0.2   & 1.9$\pm$0.2    & 1.6$\pm$0.2    & 2.5$\pm$0.4    \\
$r_t$ (pc)& 6.0$\pm$0.7   & 10.4$\pm$1.5    & 6.4$\pm$0.7    & 7.3$\pm$1.0    \\
$\log{t(\mathrm{yr})}$     & 8.50$\pm$0.10   & 8.90$\pm$0.05   & 9.15$\pm$0.05   & 8.0$\pm$0.1    \\
$(m-M)$     & 11.3$\pm$0.2   & 11.0$\pm$0.2   & 10.85$\pm$0.2   & 10.9$\pm$0.2    \\
$c$ & 0.60 & 0.74 & 0.60 & 0.46 \\
$[\mathrm{Fe/H}]$     & -0.2$\pm$0.2   & $-0.2\pm$0.2    & 0.0$\pm$0.2   & 0.0$\pm$0.2    \\
d (kpc)         & 1.82$\pm$0.19 & 1.58$\pm$0.16  & 1.48$\pm$0.15  & 1.51$\pm$0.14   \\
$E(B-V)$ & 0.64$\pm$0.04 & 0.75$\pm$0.03   &  1.06$\pm$0.04   & 0.99$\pm$0.02 \\
$N$             & 405           &  191    & 592     &  261              \\
\hline
\end{tabular}

$^a$ uncertainties on $\mu_\alpha^\ast$ and $\mu_\delta$ are given within parenthesis in $\mu$as yr$^{-1}$;  $\mu_\alpha^\ast$ and $\mu_\delta$ are also affected by systematic uncertainties of 0.1\,mas yr$^{-1}$
\label{tab:parameters}
\end{table}

%\begin{figure}
%\centering
%\includegraphics[width=0.9\linewidth]{figs_novas_eps/rprof_ufmg2.eps}\\
%\includegraphics[width=0.9\linewidth]{figs_novas_eps/rprof_ufmg3.eps}
%\caption{Radial density profiles of UFMG\,2 (top) and UFMG\,3 (botton) used to determine their limiting radius. The dashed lines represent the calculated sky density level.}
%It can be seen that UFMG\,1 and UFMG\,3 are both sparser than UFMG\,2.}
%\label{fig:rdp2}
%\end{figure}

\subsection{Assessing membership}
\label{sect:membership}

\subsubsection{Proper motion selection}
\label{sect:pms}

%A rectangular mask over the proper motion overdensity was a procedure enough to highlight the cluster over the background to estimate its centre and limit radii, on the other hand it does not infer any  feature of the proper motion distribution and  astrometric errors limits and this procedure introduce contamination from background stars with proper motions components out of the distributions limits and with high errors.  

 Histograms of the proper motion components $\mu_{\alpha}^{*}$ and $\mu_{\delta}$ were built for stars inside the limiting radii of each cluster and for an adjacent annular field with the same respective area. 
The proper motion distributions in the clusters' regions were then subtracted from the field contribution to determine a clean sample.

\begin{figure*}
\centering
\includegraphics[width=0.22\linewidth]{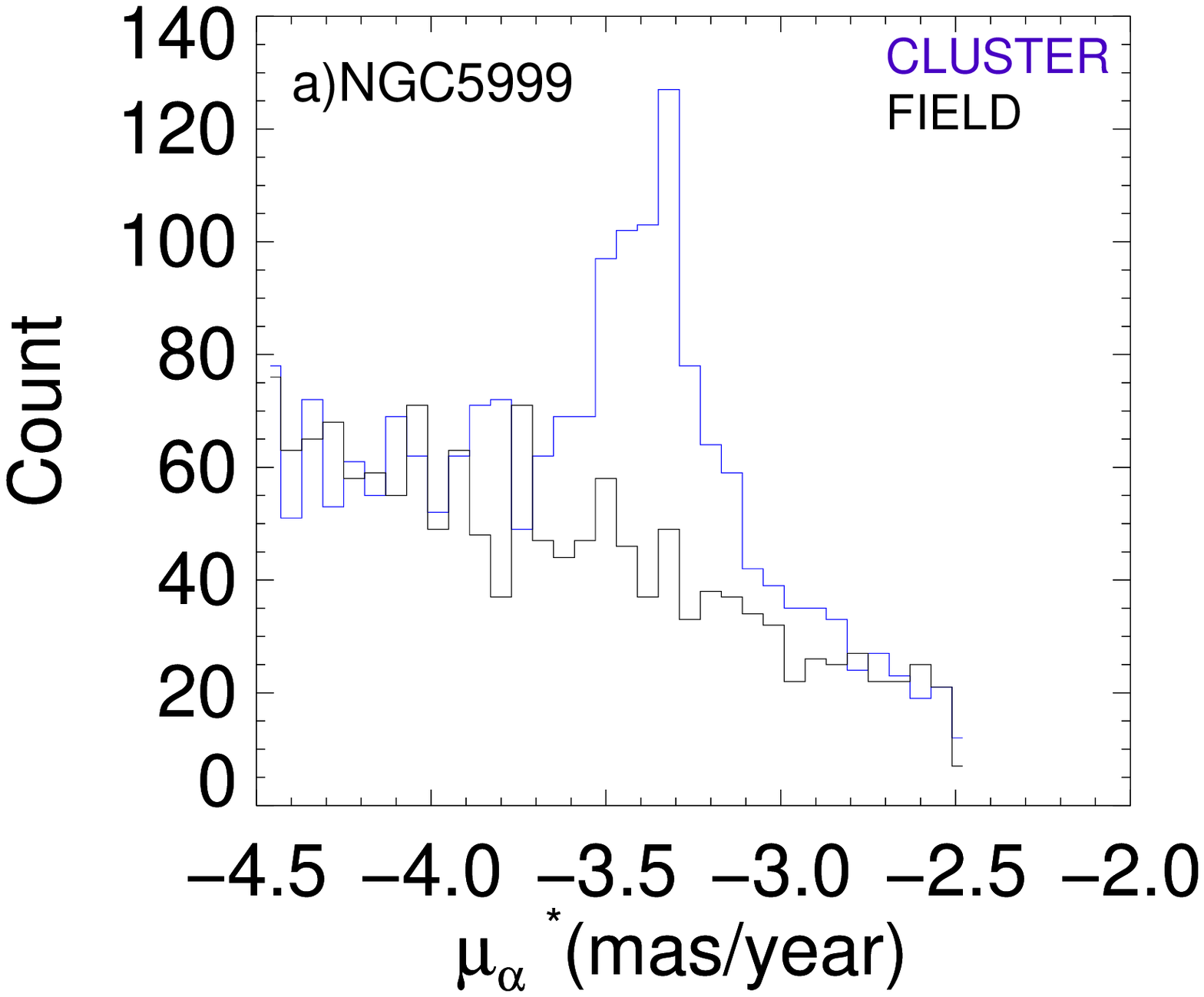} 
\includegraphics[width=0.22\linewidth]{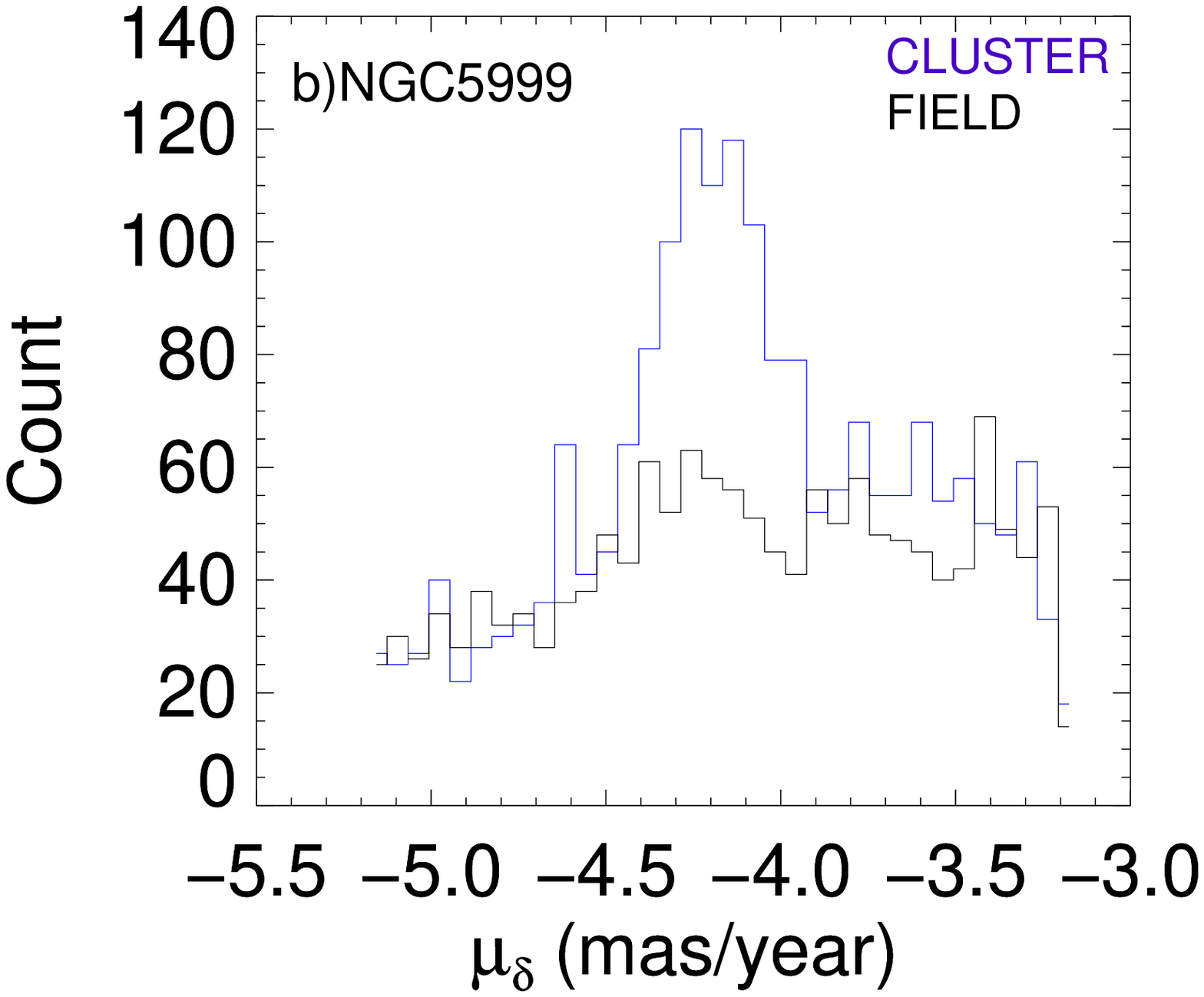} \hspace{0.75cm}
\includegraphics[width=0.22\linewidth]{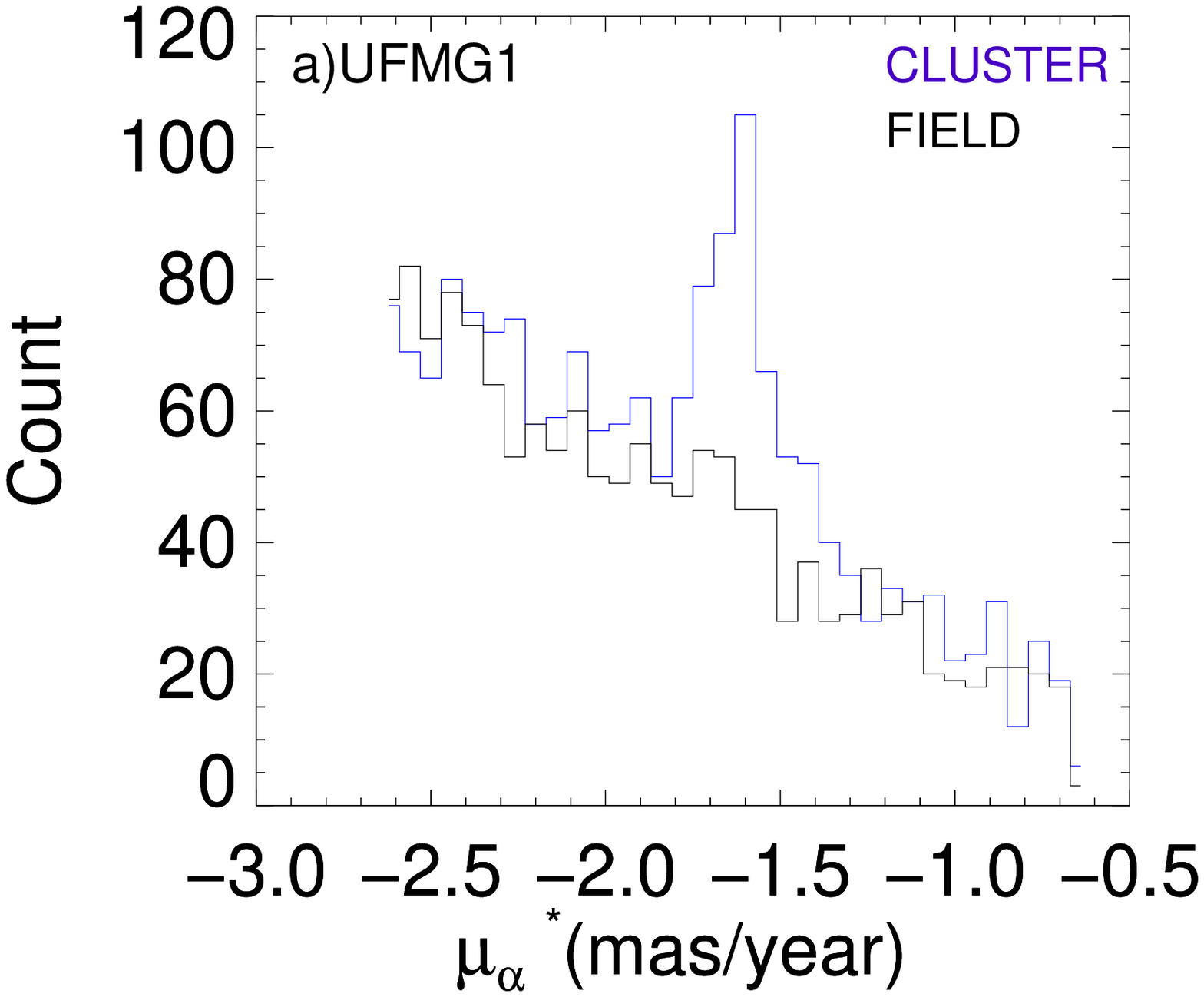} \includegraphics[width=0.22\linewidth]{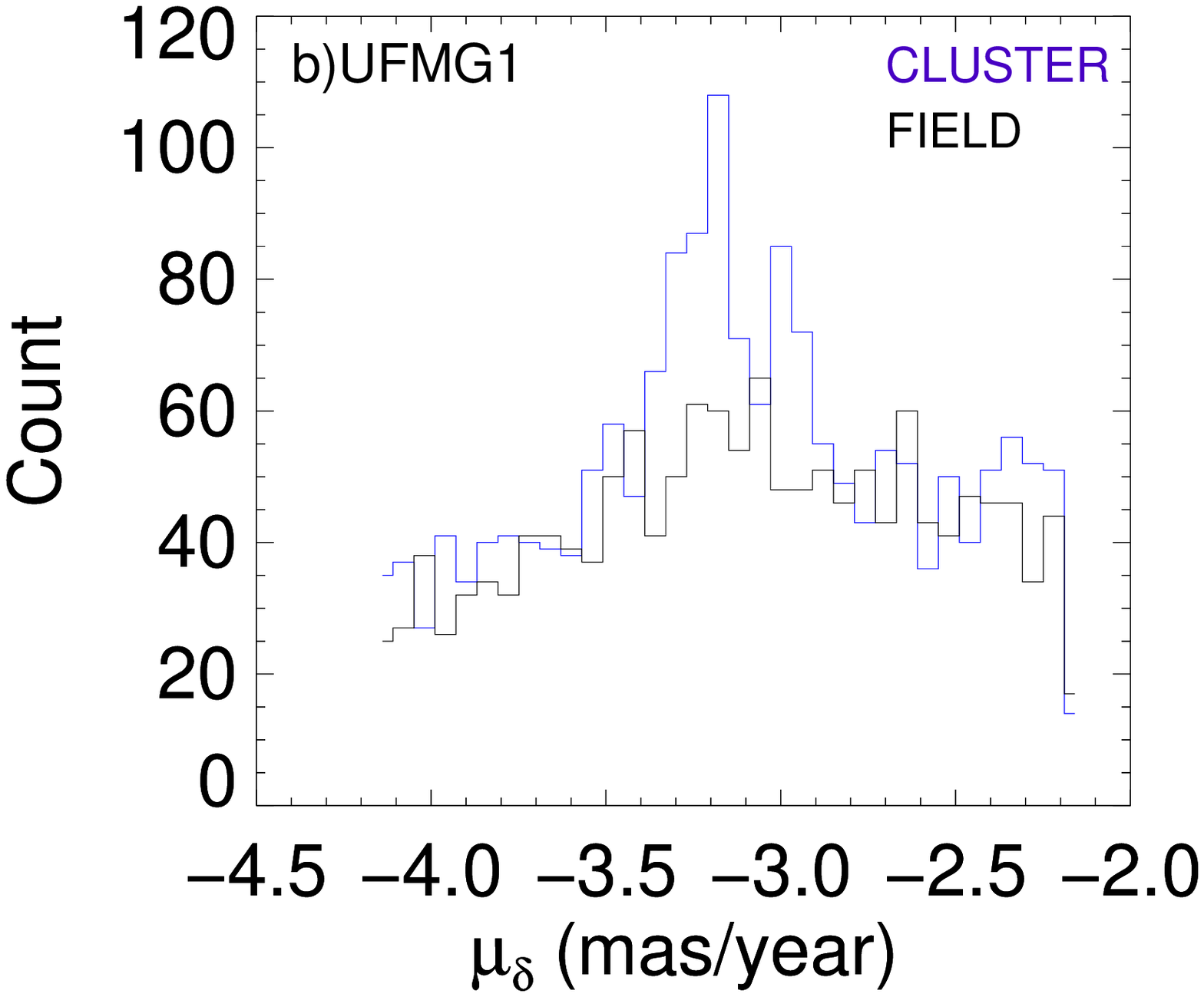} \\

\includegraphics[width=0.22\linewidth]{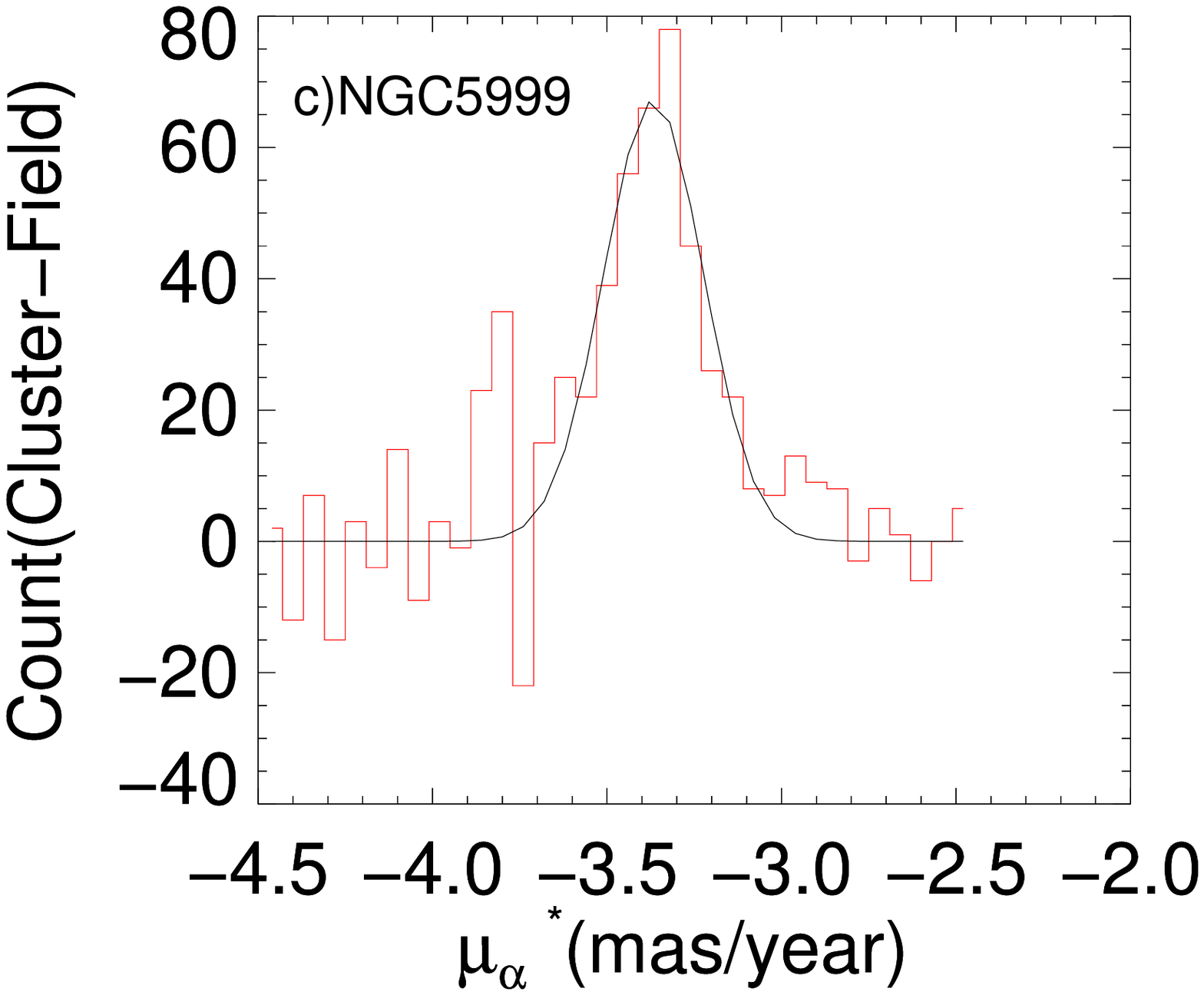} 
\includegraphics[width=0.22\linewidth]{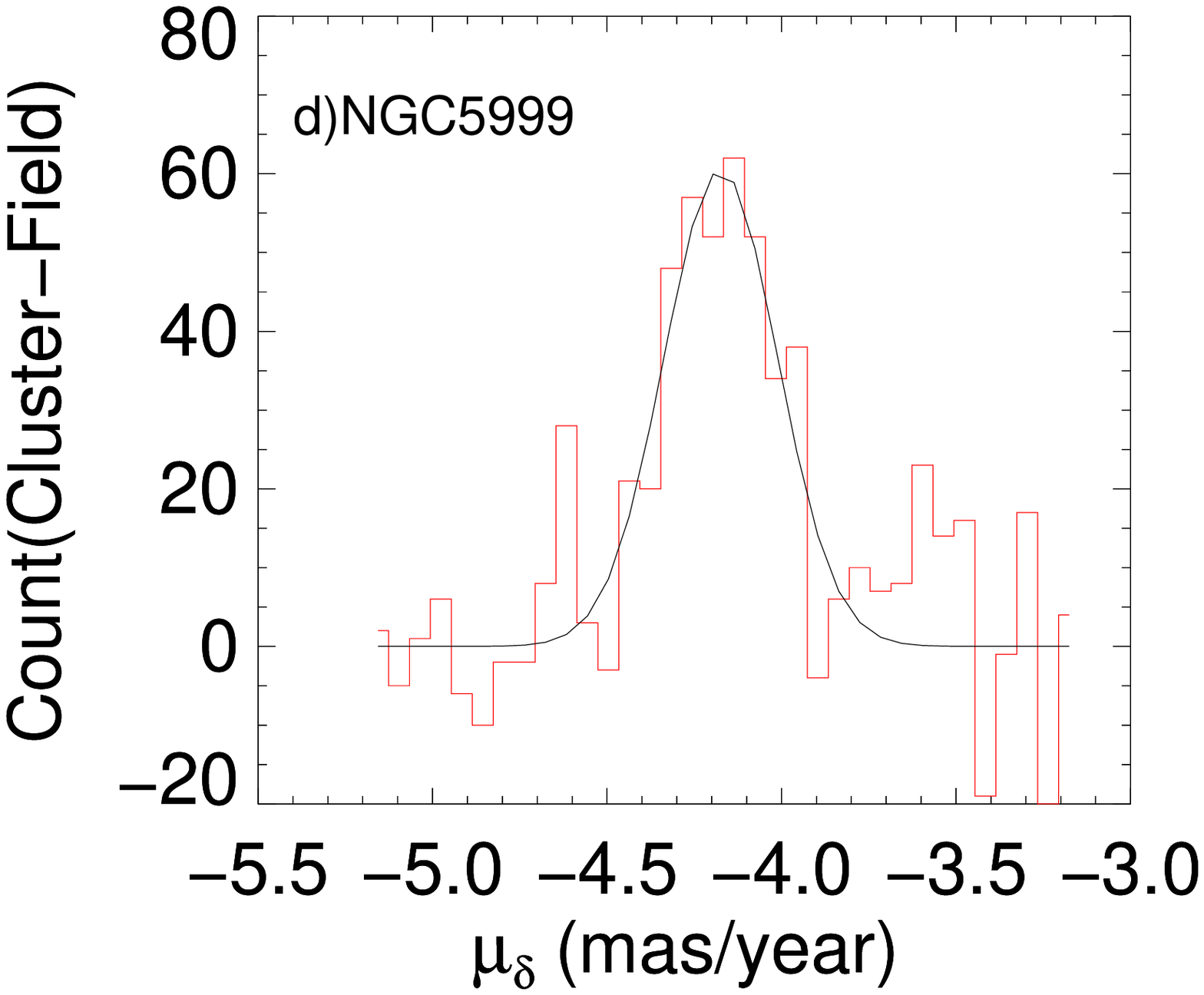} \hspace{0.75cm}
\includegraphics[width=0.22\linewidth]{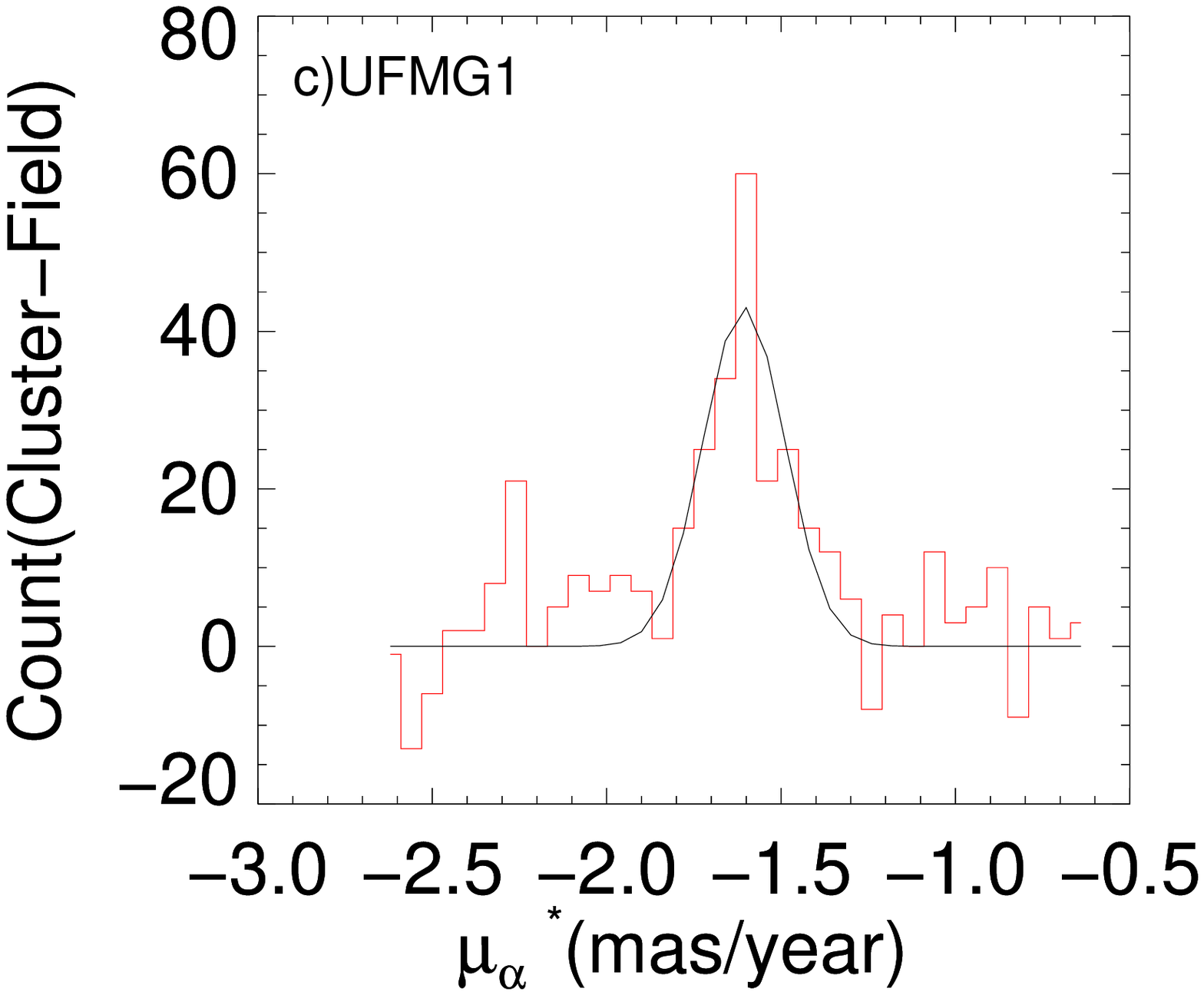}
\includegraphics[width=0.22\linewidth]{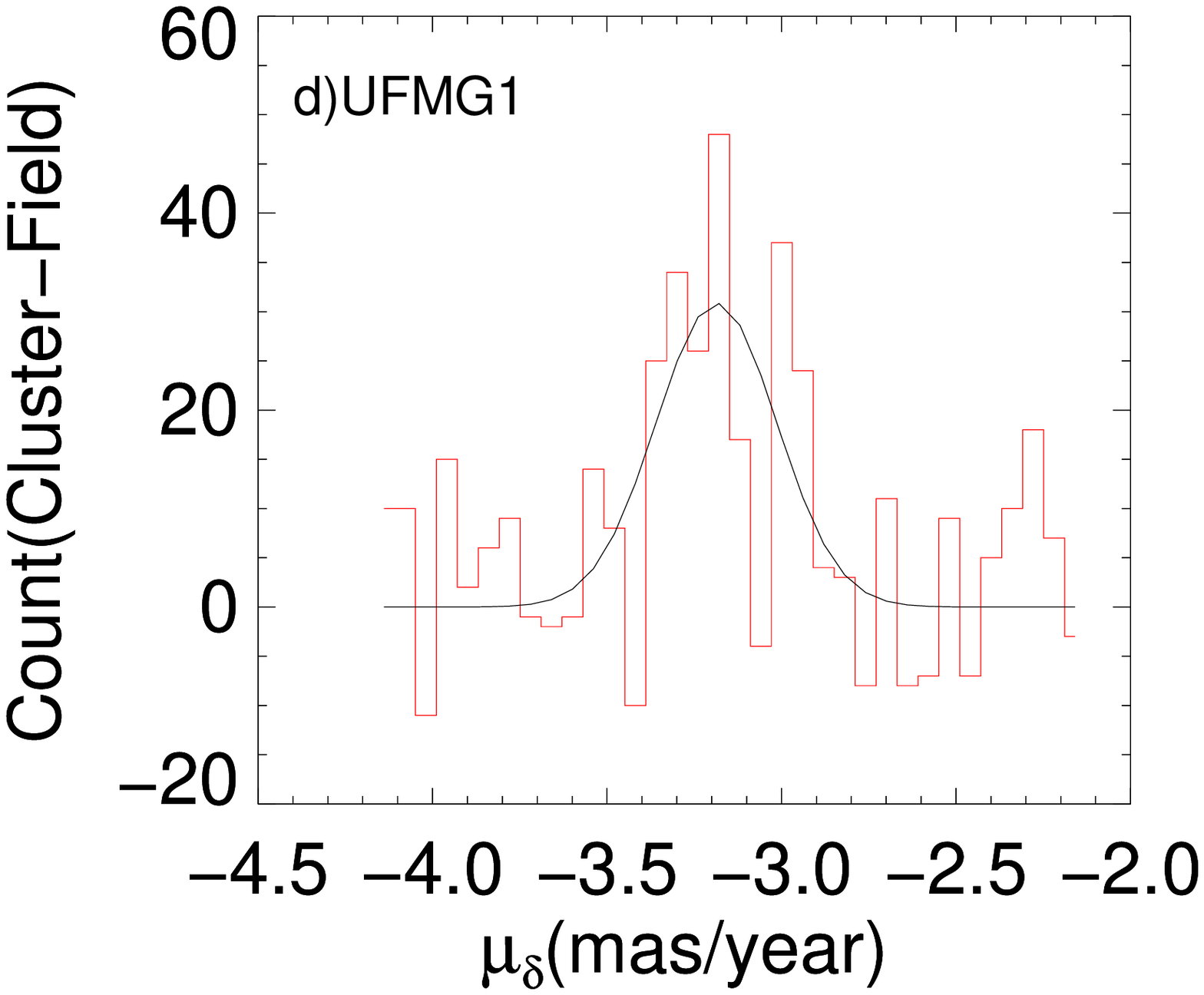} \\ \vspace{0.75cm}

\includegraphics[width=0.22\linewidth]{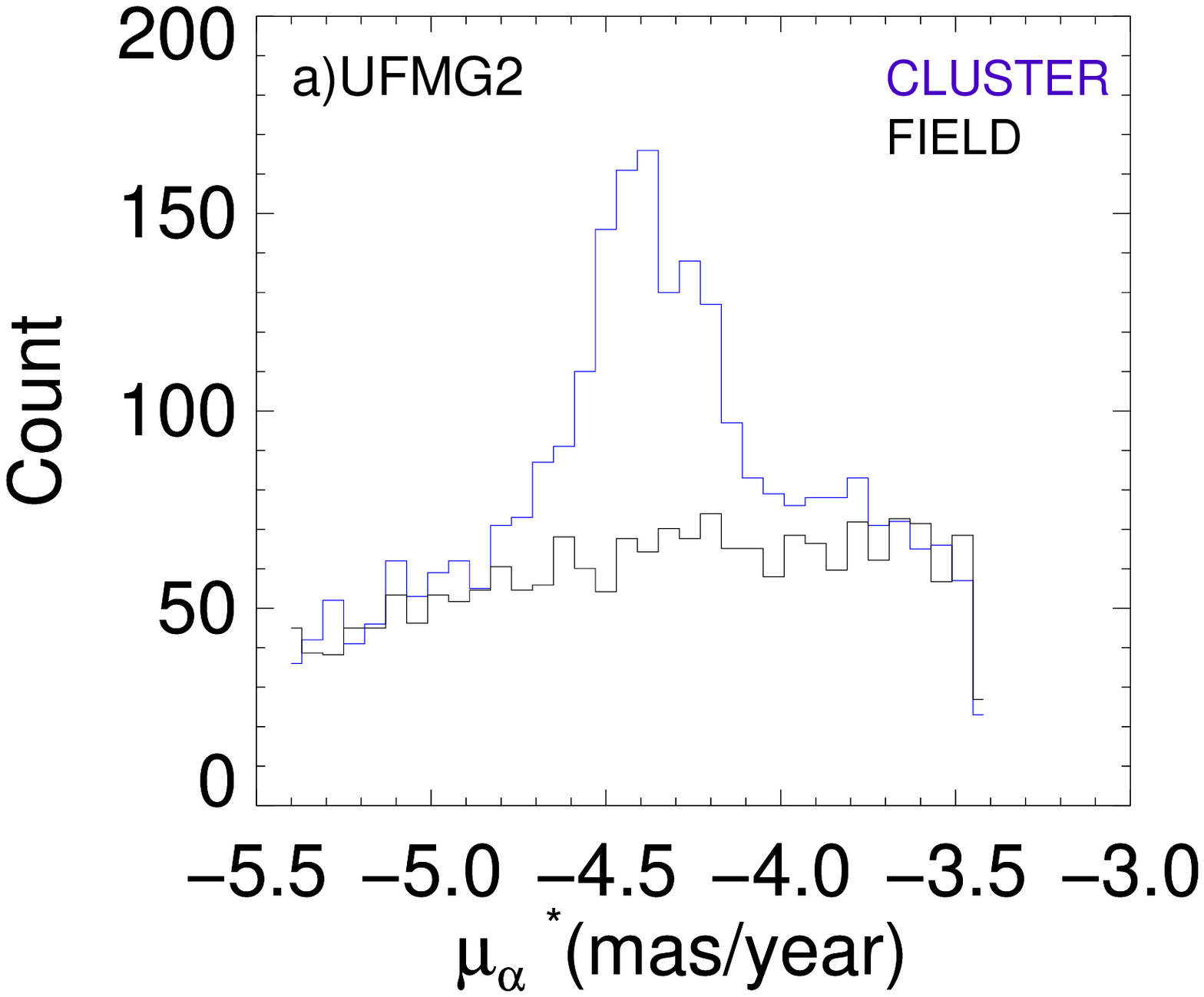}
\includegraphics[width=0.22\linewidth]{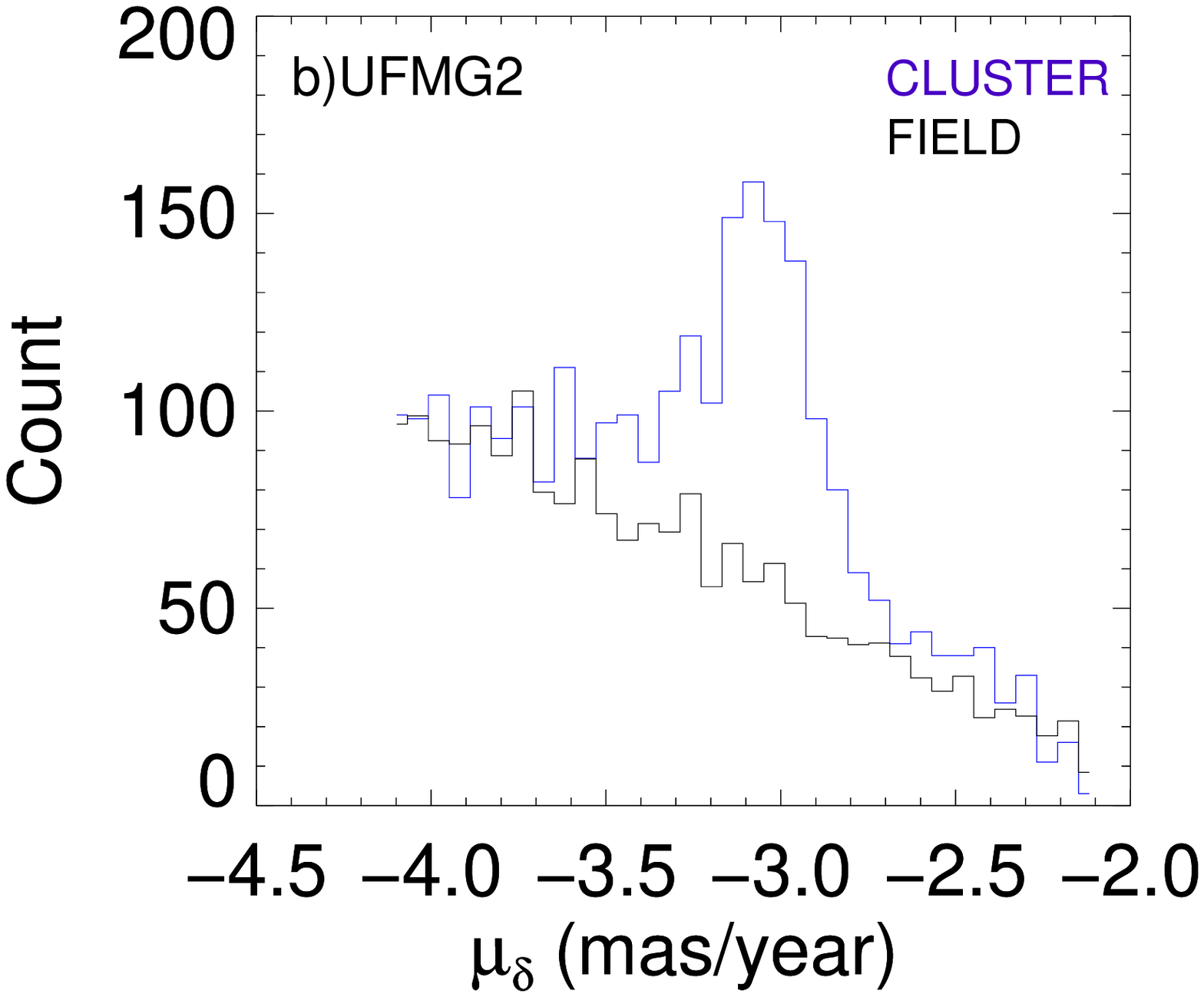} \hspace{0.9cm}
\includegraphics[width=0.21\linewidth]{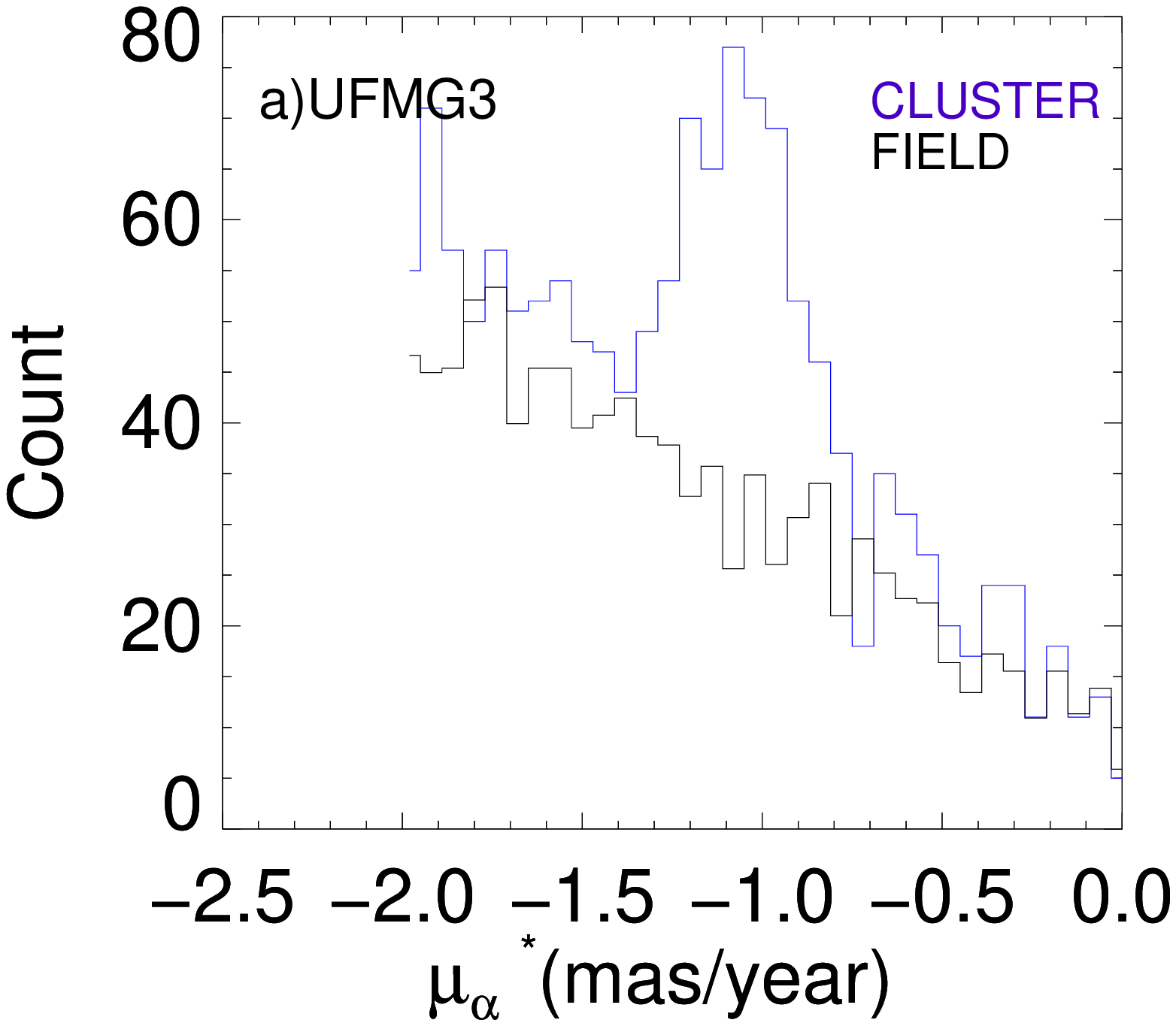}
\includegraphics[width=0.22\linewidth]{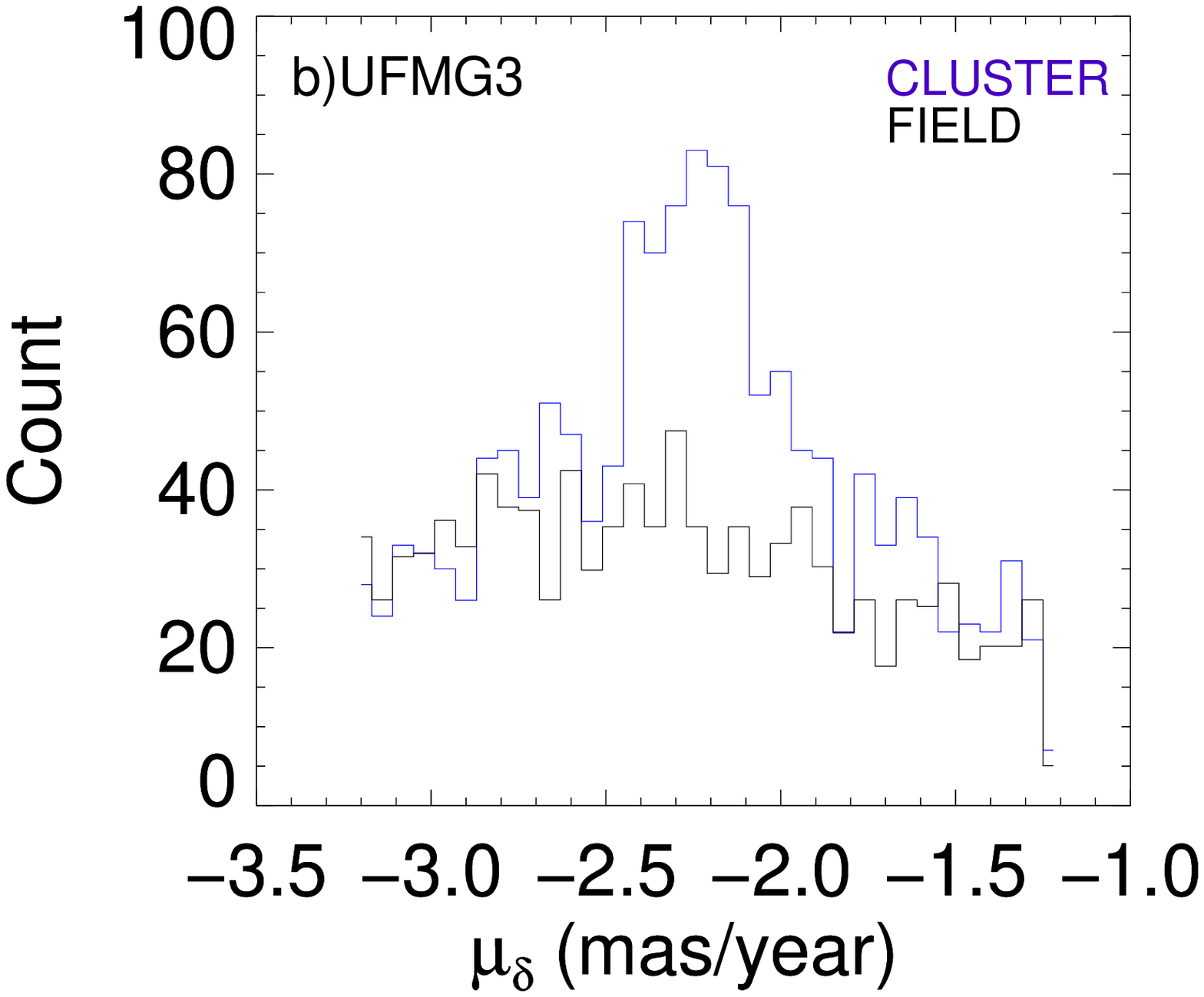} \\

\includegraphics[width=0.22\linewidth]{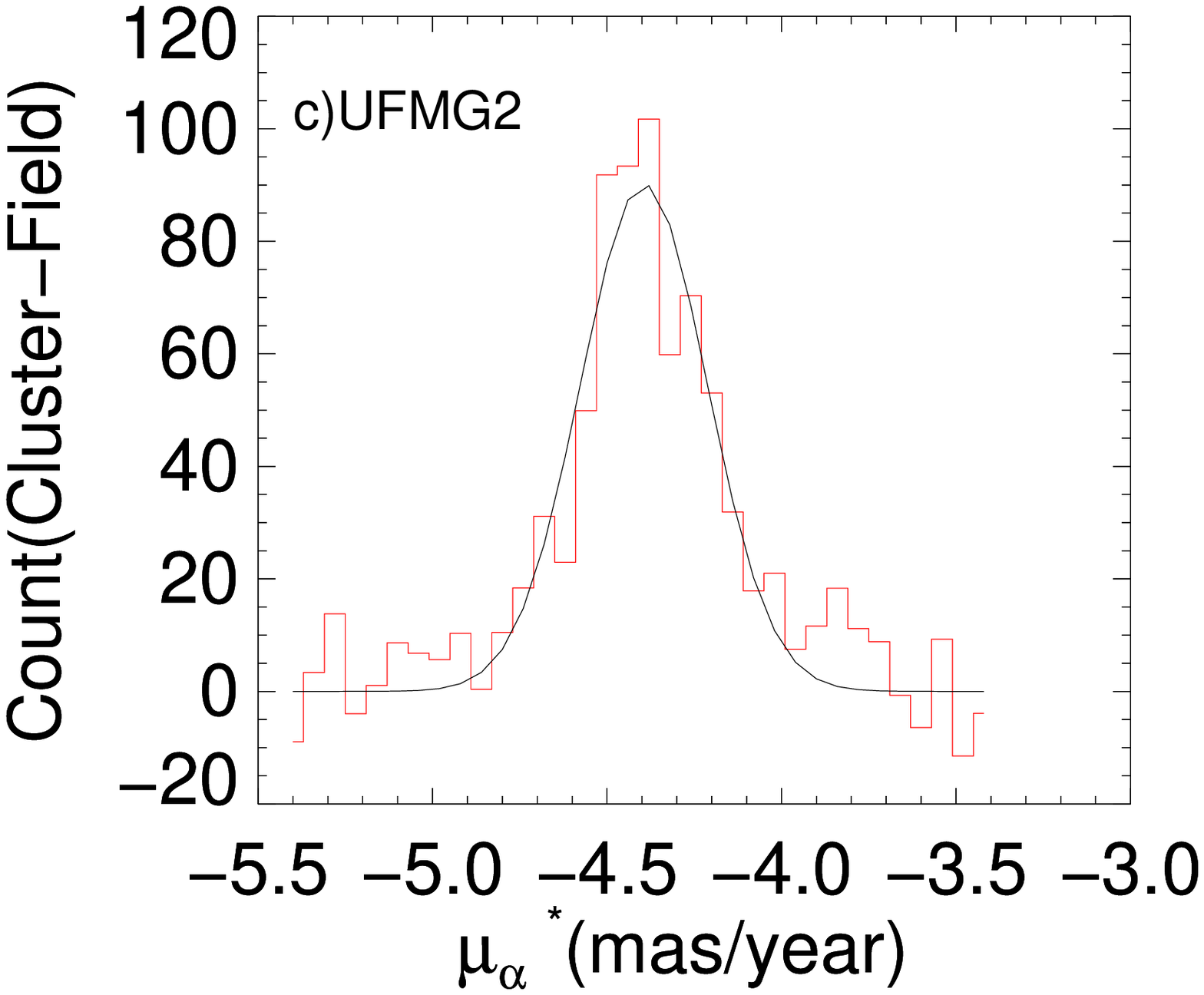} 
\includegraphics[width=0.22\linewidth]{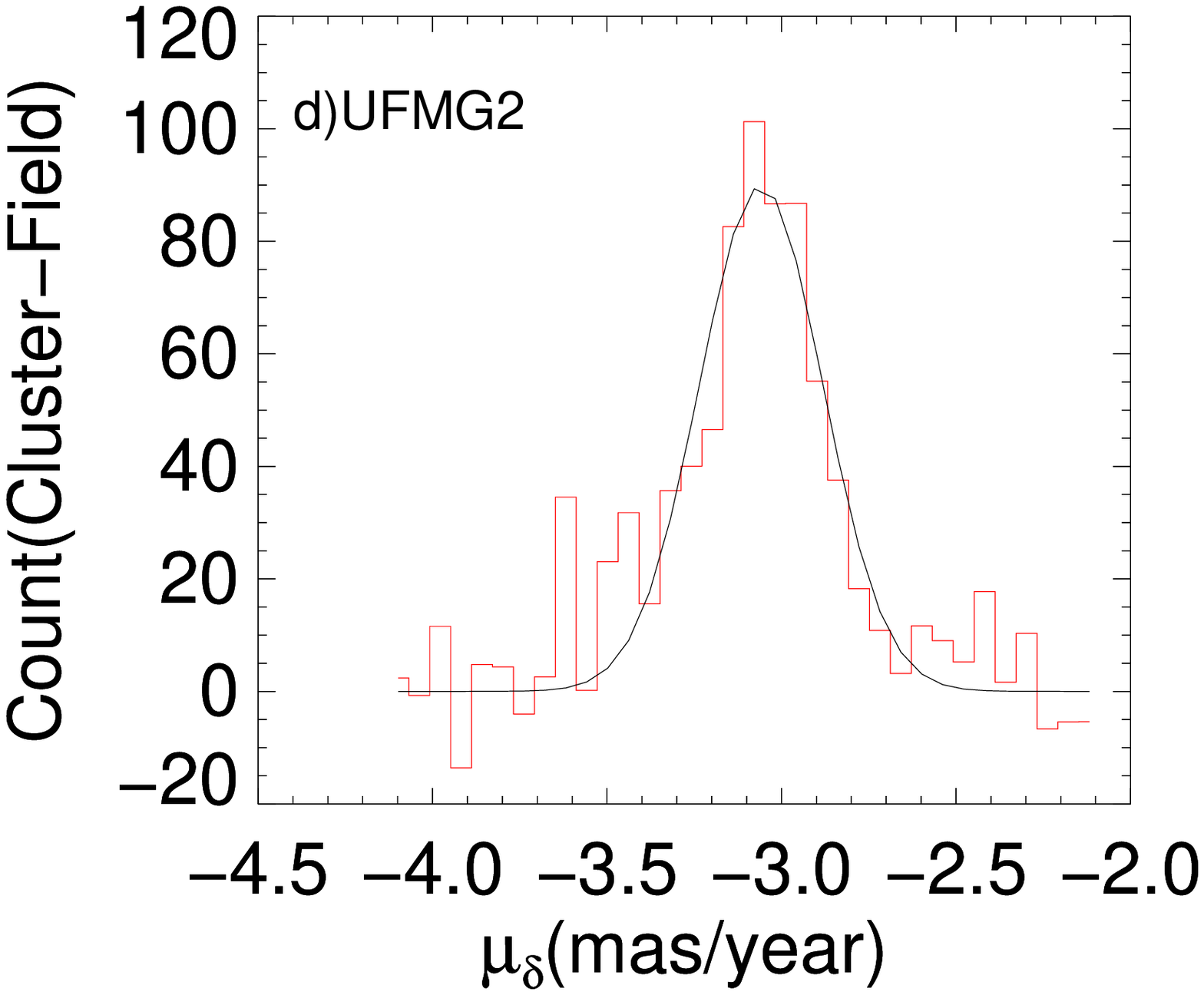} \hspace{0.75cm}
\includegraphics[width=0.22\linewidth]{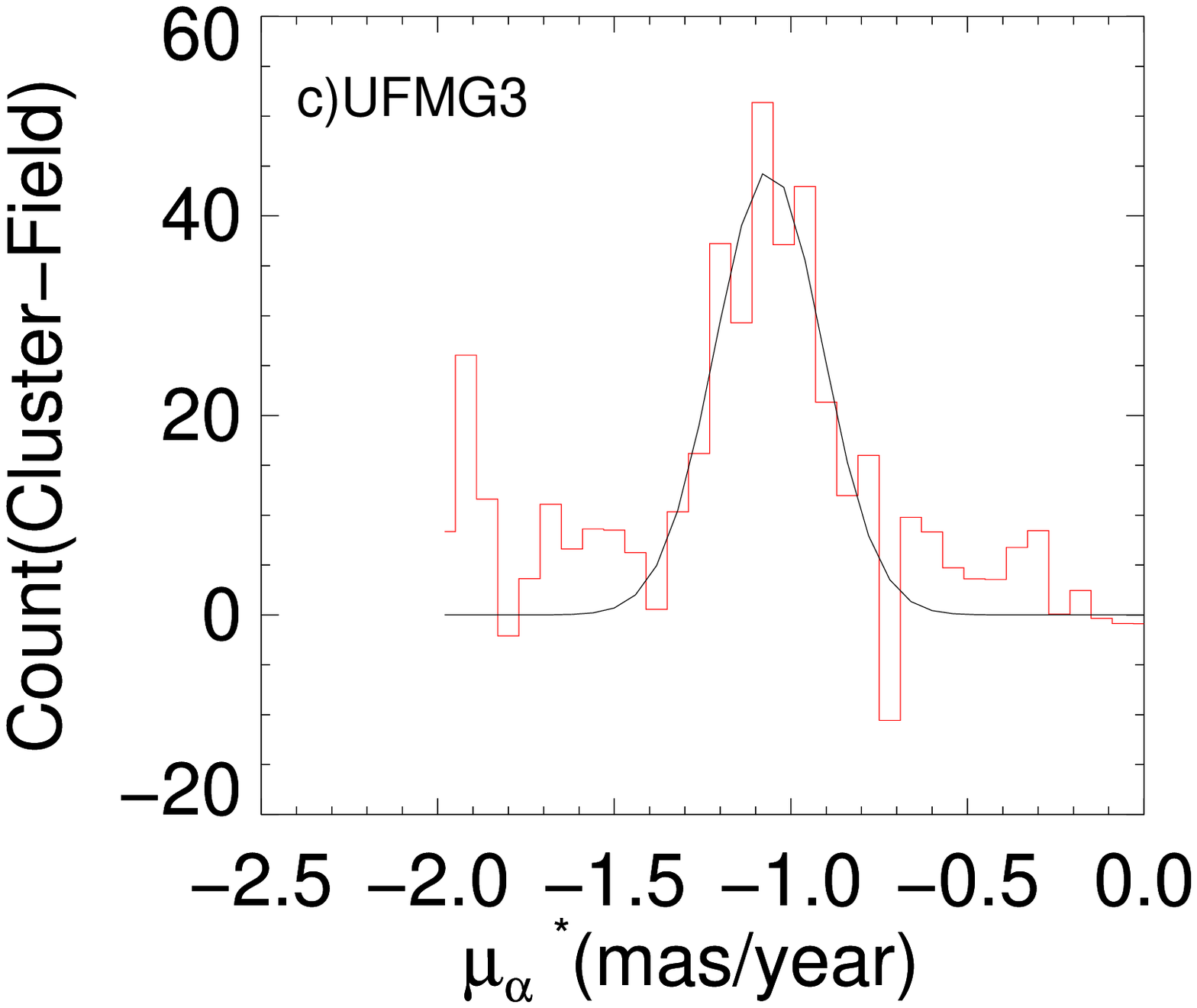}
\includegraphics[width=0.22\linewidth]{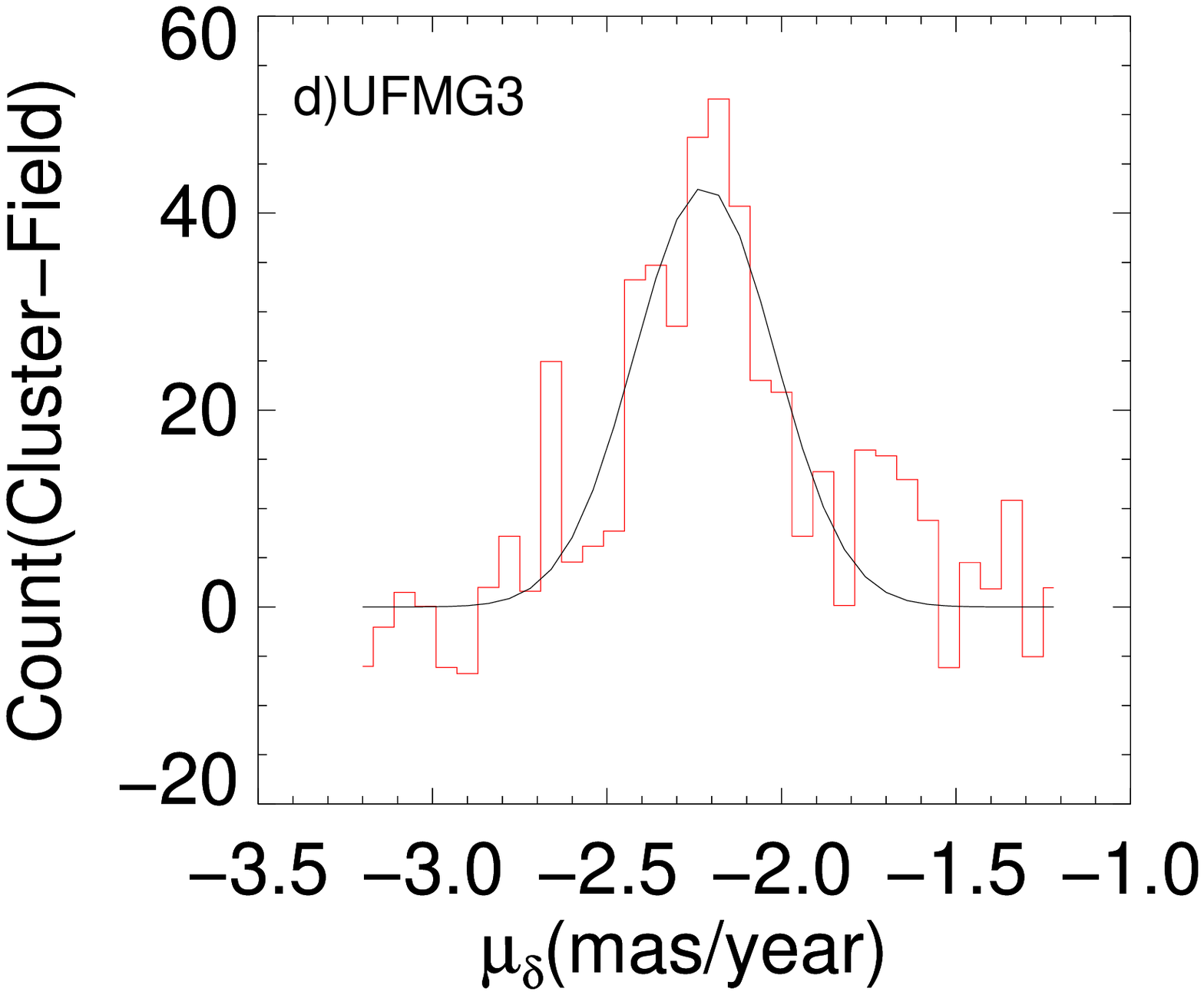}

\caption{(Panels (a) and (b)) Histograms of the proper motion components $\mu_{\alpha}^{*}$ and $\mu_{\delta}$ inside the limiting radii and an adjacent field with the same respective area. (Panels (c) and (d)) Gaussian fits over the resulting proper motion components for the clusters NGC\,5999 (top left panels), UFMG\,1 (top right panels), UFMG\,2 (bottom left panels) and UFMG\,3 (bottom right panels).}
\label{fig:gauss_pm}
\end{figure*}

 The  most probable values of the proper motion distribution centres were obtained from  Gaussian fittings on the resulting histograms (Fig.~\ref{fig:gauss_pm}). In this procedure, proper motion outliers were removed from the sample, i.e., those stars with proper motion components outside 2-$\sigma$ of the mean values obtained in the Gaussian fittings.  The mean proper motion components and the 1-$\sigma$ deviation found for each cluster are given in Table~\ref{tab:parameters}.

%\begin{figure}
%\centering
%\caption{(Panels (a) and (b)) Histograms of the proper motion components $\mu_{\alpha}^{*}$ and $\mu_{\delta}$ inside the limiting radii and an adjacent field with the same respective area. (Panels (c) and (d)) Gaussian fits over the resulting proper motion components for the cluster UFMG\,1.}
%\label{fig:gauss_pm1}
%\end{figure}

%\begin{figure}
%\centering
%\caption{(Panels (a) and (b)) Histograms of the proper motion components $\mu_{\alpha}^{*}$ and $\mu_{\delta}$ inside the limiting radii and an adjacent field with the same respective area. (Panels (c) and (d)) Gaussian fits over the resulting proper motion components for the cluster UFMG\,2.}
%\label{fig:gauss_pm2}
%\end{figure}

%\begin{figure}
%\centering
%\caption{(Panels (a) and (b)) Histograms of the proper motion components $\mu_{\alpha}^{*}$ and $\mu_{\delta}$ inside the limiting radii and an adjacent field with the same respective area. (Panels (c) and (d)) Gaussian fits over the resulting proper motion components for the cluster UFMG\,3.}
%\label{fig:gauss_pm3}
%\end{figure}

%The resulting samples of the clusters limited by the gaussian parameters can be seen in the Fig.~\ref{fig:amostra_gauss_memb_pm}

\subsubsection{Colour-magnitude  diagram decontamination}
\label{sect:cmdd}

\begin{figure*}
\centering
\includegraphics[width=0.36\linewidth]{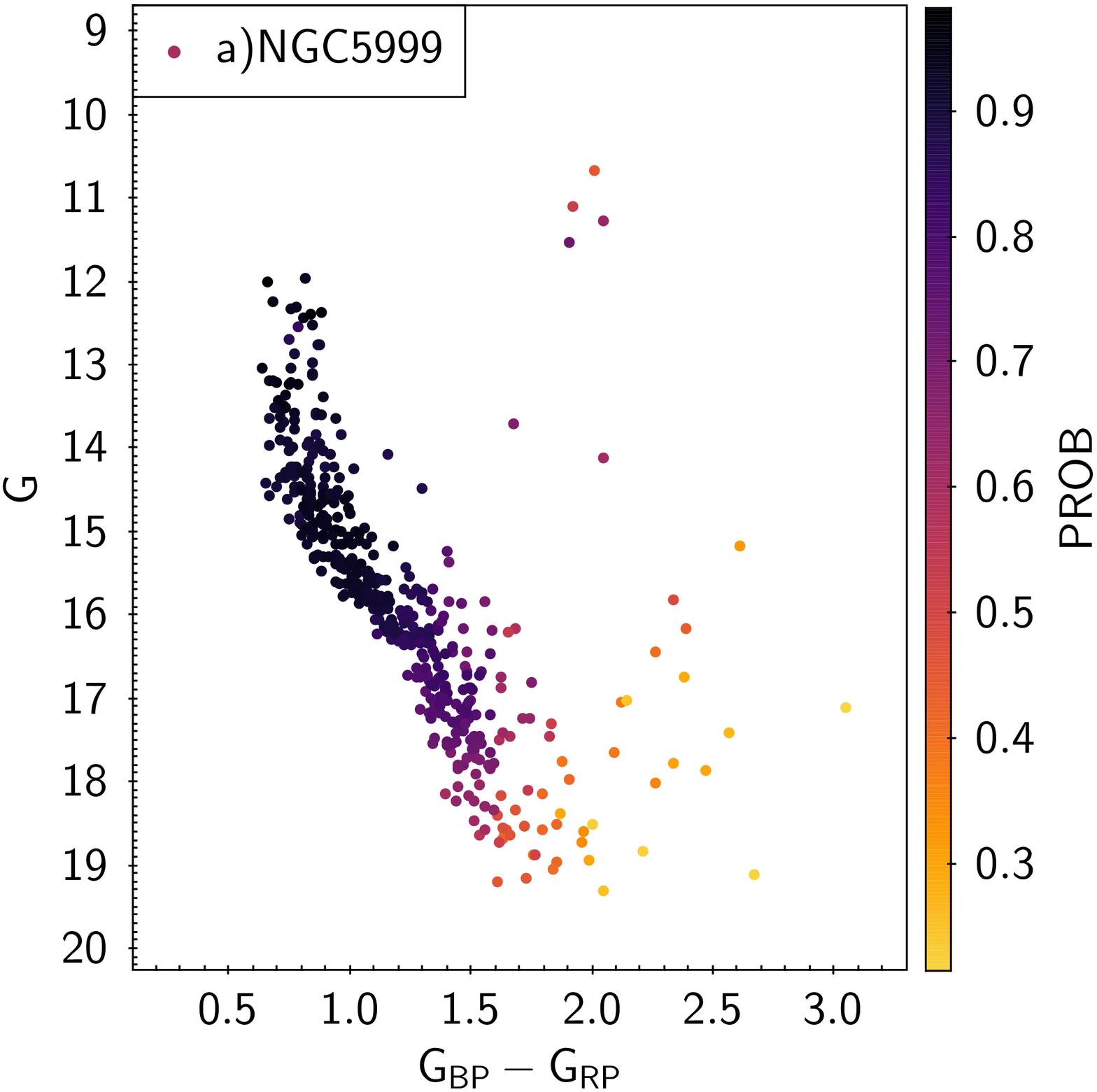}
\includegraphics[width=0.36\linewidth]{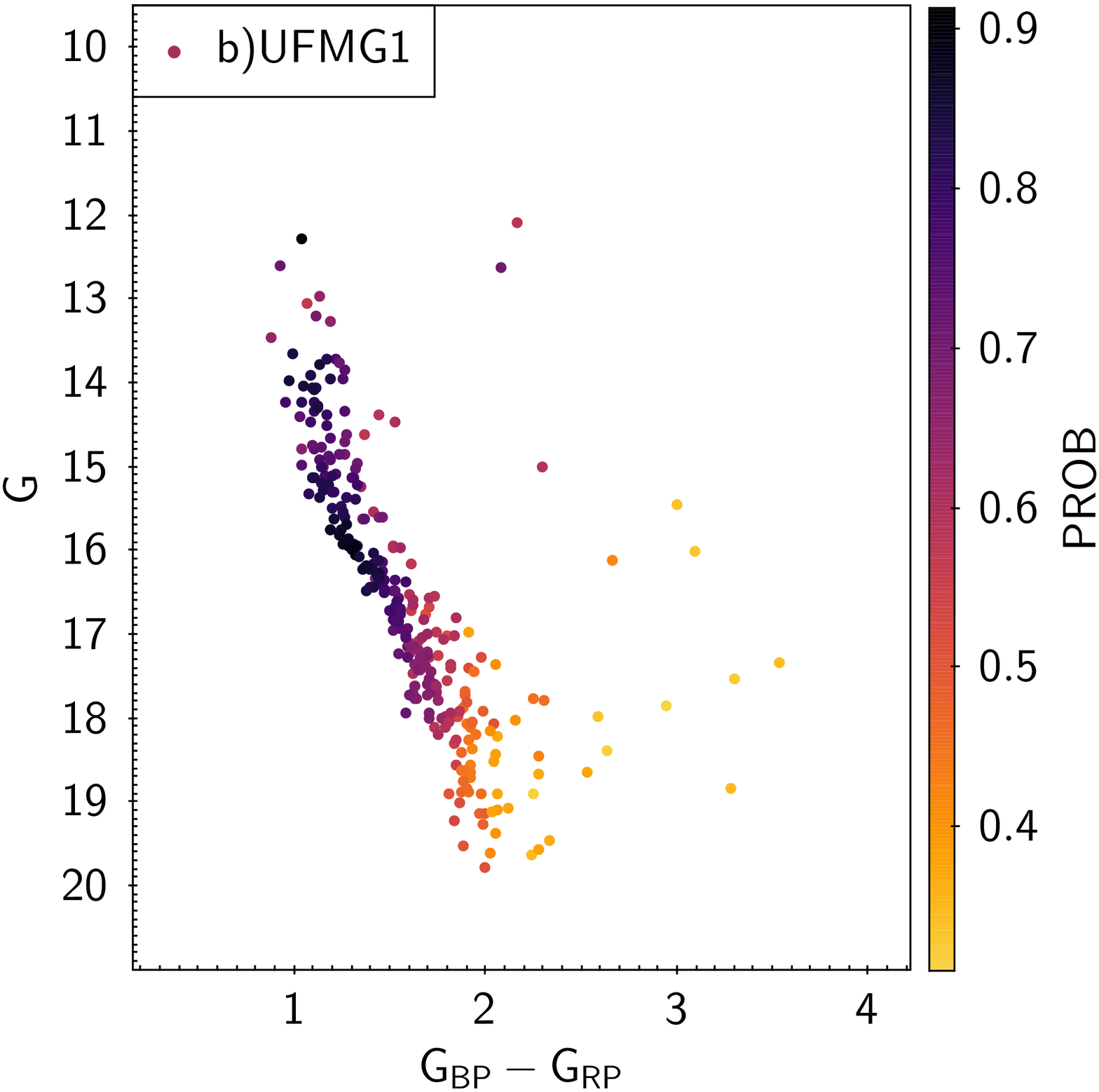} \\
\includegraphics[width=0.36\linewidth]{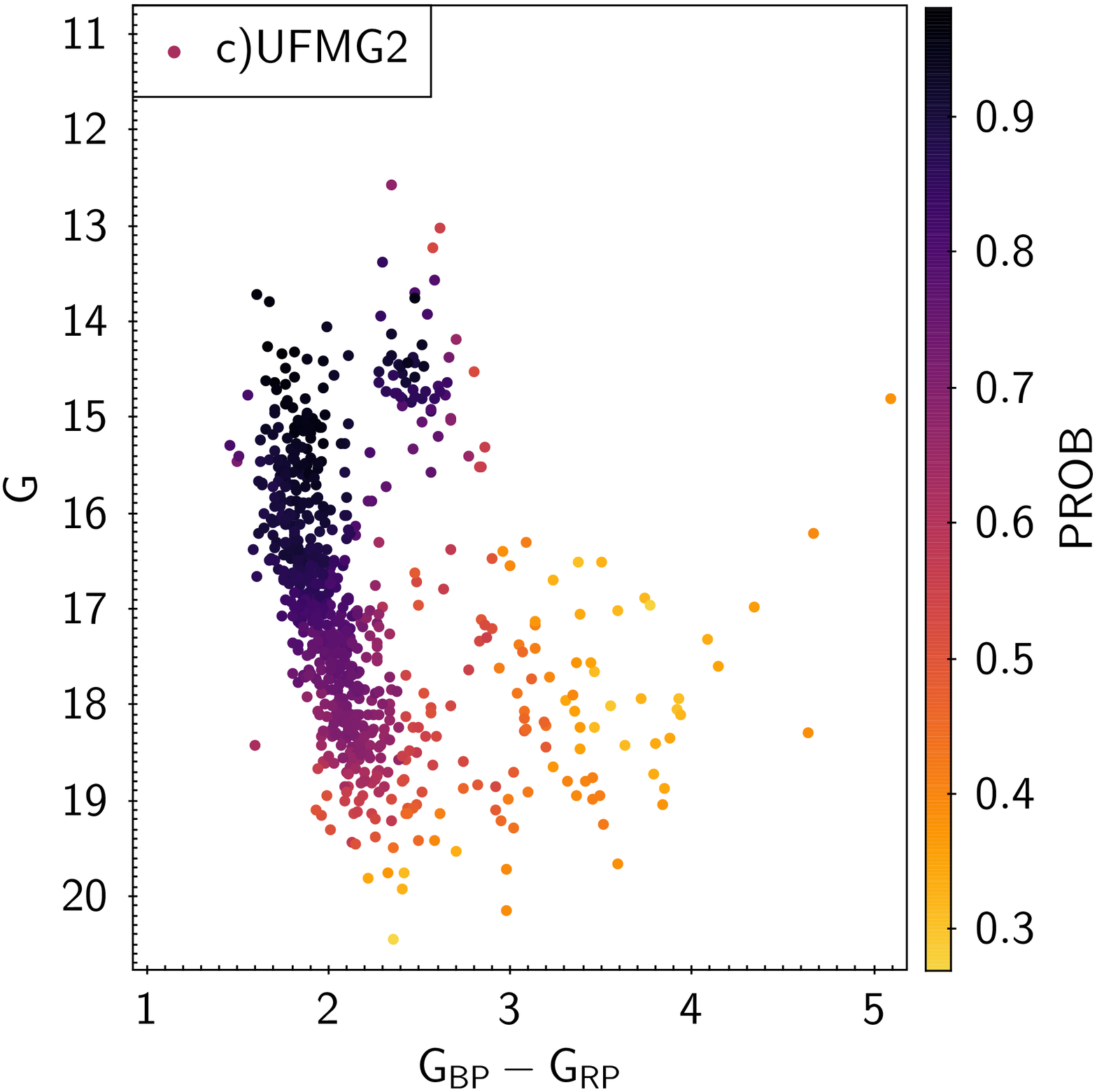}
\includegraphics[width=0.36\linewidth]{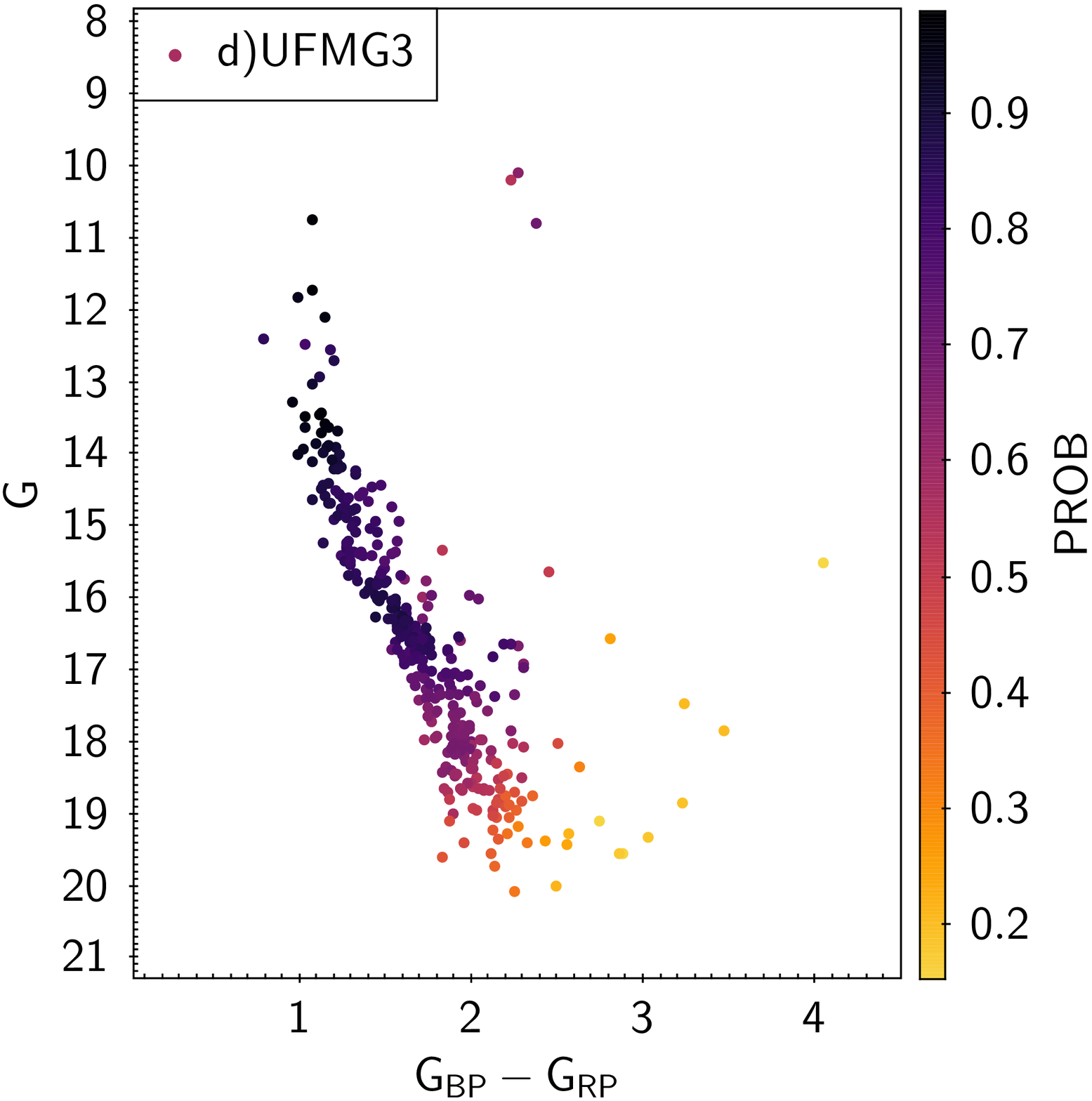}
\caption{CMD for NGC\,5999, UFMG\,1, UFMG\,2 and UFMG\,3 after applying the cleaning tool. The side bar shows the mean membership probability}
\label{fig:decont}
\end{figure*}

After the proper motion selection of members, CMDs were built for each cluster and the
CMD cleaning tool devised by \cite{Maia:2010} was applied
to statistically evaluate and remove the field population from
the clusters CMDs. The method has been recently updated
in \cite{Angelo:2018} to employ control fields of arbitrary
shapes, assigning photometric membership probabilities to
cluster stars based on the local overdensity in the cluster's
CMD relative to the field CMD and on their distance to the
cluster centre, according to the relation:

\begin{equation}
P \propto e^{-\rho_{fld}/\rho_{clu}} e^{-r/r_{lim}}
\label{eq:pert}
\end{equation}

\noindent
where $\rho_{clu}$ and $\rho_{fld}$ are the local density in the cluster and
field CMDs, respectively, and $r$ measures a star distance to
the cluster centre.   The membership probabilities associated to the stars in 
each cluster CMD are shown in Fig.~\ref{fig:decont}. The definitive member list 
was defined by selecting stars with photometric membership greater than 60\%. 
The resulting number of selected member stars ($N$) is shown in Table~\ref{tab:parameters}
for each cluster.

%CMD and parallax distributions of the most probable members of each cluster are shown in  Fig. \ref{fig:cmd}.}

%\subsubsection{Parallax distribution}

%To assure that the sample selected in proper motion space contains only good member candidates, we further analyzed the parallax distribution of the stars in each cluster. This was done in a similar way as for the proper motion analysis, by performing Gaussian fittings over the parallax distribution of the star sample already selected in proper motion space. 

%Also similarly to the proper motion analysis, the parallax distributions were filtered by excluding stars with parallax value and parallax errors beyond $3-\sigma$ of the mean values as defined by the Gaussian fittings. 
%The parallax distribution for each cluster region and their respective Gaussian fittings are shown in Fig.~\ref{fig:gauss_plx_sig}.

%\begin{figure}
%\centering
%\includegraphics[width=0.45\linewidth]{figs_novas_eps/dados_cortes_pm_gauss_1_ngc5999.eps}
%\includegraphics[width=0.45\linewidth]{figs_novas_eps/dados_cortes_pm_gauss_1_ufmg1.eps} \\
%\includegraphics[width=0.45\linewidth]{figs_novas_eps/dados_cortes_pm_gauss_1_ufmg2.eps}
%\includegraphics[width=0.45\linewidth]{figs_novas_eps/dados_cortes_pm_gauss_1_ufmg3.eps}\\
%\caption{Gaussian fittings over the samples filtred by the proper motion analysis for NGC\,5999 (a), UFMG\,1 (b), UFMG\,2 (c) and UFMG\,3 (d).}
%\label{fig:gauss_plx_sig}
%\end{figure}

\subsection{Astrophysical and structural parameters}

%{\textcolor{red} {As will be discussed in Sect.~\ref{sect:distinf} the parallax histograms of our targets show that they all present tight distributions with a FWHM of $\lesssim$\,0.1 mas, thus confirming the validity of our selection criteria.}}

%{\textcolor{red}{ 
%However, although the parallaxes distributions indicate that all clusters present similar distances, with UFMG\,1 being a little closer than the other ones, this was not the scenario found when distance is estimated from the distance modulus obtained from the isochrone fitting (see next section). Furthermore, the spread in the parallaxes of any particular cluster corresponds to a much larger depth ($\sim$\,200 pc) than one would expect considering only the cluster size along the line of sight (a few pc).
%%}}

 In the sequence, we performed isochrone fittings on the proper motion filtered (\ref{sect:pms}) 
and decontaminated CMD (\ref{sect:cmdd}) samples to determine the clusters' astrophysical parameters. 
Structural parameters were obtained by \cite{King:1962} model fittings to the proper motion filtered 
samples. Since the CMD cleaning tool operates on the star distances from the clusters' centre 
(Eq.~\ref{eq:pert}), we refrained to use the photometrically cleaned sample so as to not bias the 
model fittings. So, we only used the proper motion selected samples to get structural 
parameters. 

 Given the reported offsets between distances as obtained from parallaxes and CMD distance 
moduli \citep[e.g][]{Cantat:2018} and the rather different methodology employed in each case, we have 
opted to keep the photometric analysis completely independent from the parallaxes. We will address the 
parallax inversion problem in Sect. \ref{sect:distinf}, making full use of the distance moduli derived 
with the photometry. In addition, since we have not applied any parallax constraint to our data, it can 
be used as \emph{a posteriori} check to see how clustered the cleaned samples are in parallax space.

%\begin{figure}
%\centering
%\includegraphics[width=0.49\linewidth]{fig4a.pdf}
%\includegraphics[width=0.49\linewidth]{fig4b.pdf}
%\caption{Colour-magnitude diagram with the most probable members of each cluster (left). Histograms showing the parallax distribution of the most probable members of each cluster.}
%\label{fig:cmd_plx_memb}
%\end{figure}

\subsubsection{Isochrone fittings}
\label{sect:cmd}

A set of PARSEC-COLIBRI isochrones \citep{Marigo:2017} 
was employed to perform fittings on the decontaminated samples to determine age, metallicity, distance and reddening. A reddening law \citep{Cardelli:1989,Odonnell:1994} was used to convert $E(G_{BP}-G_{RP})$ to $E(B-V)$. 
We did not use the extinction values, as quoted in the Gaia catalogue, because they are only satisfactory when applied, in a statistical sense, to large samples \citep{Andrae:2018} and therefore are not useful to correct the photometry of smaller groups. Instead, we derived extinction towards the clusters from the isochrone fitting. 

 We have found the best-fitting isochrone by carefully inspecting the matching of key evolutionary regions such as the lower main sequence, the turnoff, and the giant clump across several isochrones covering a range of ages and metallicities. Then, to evaluate the uncertainty in the parameters, since reddening and distance modulus $(m-M)$ produce only a shift of the isochrone, we changed them simultaneously to get a maximum deviation from the central solution that still encompass the data towards both bluer/brighter and redder/dimmer extremes. 

\begin{figure}
\includegraphics[width=0.95\linewidth]{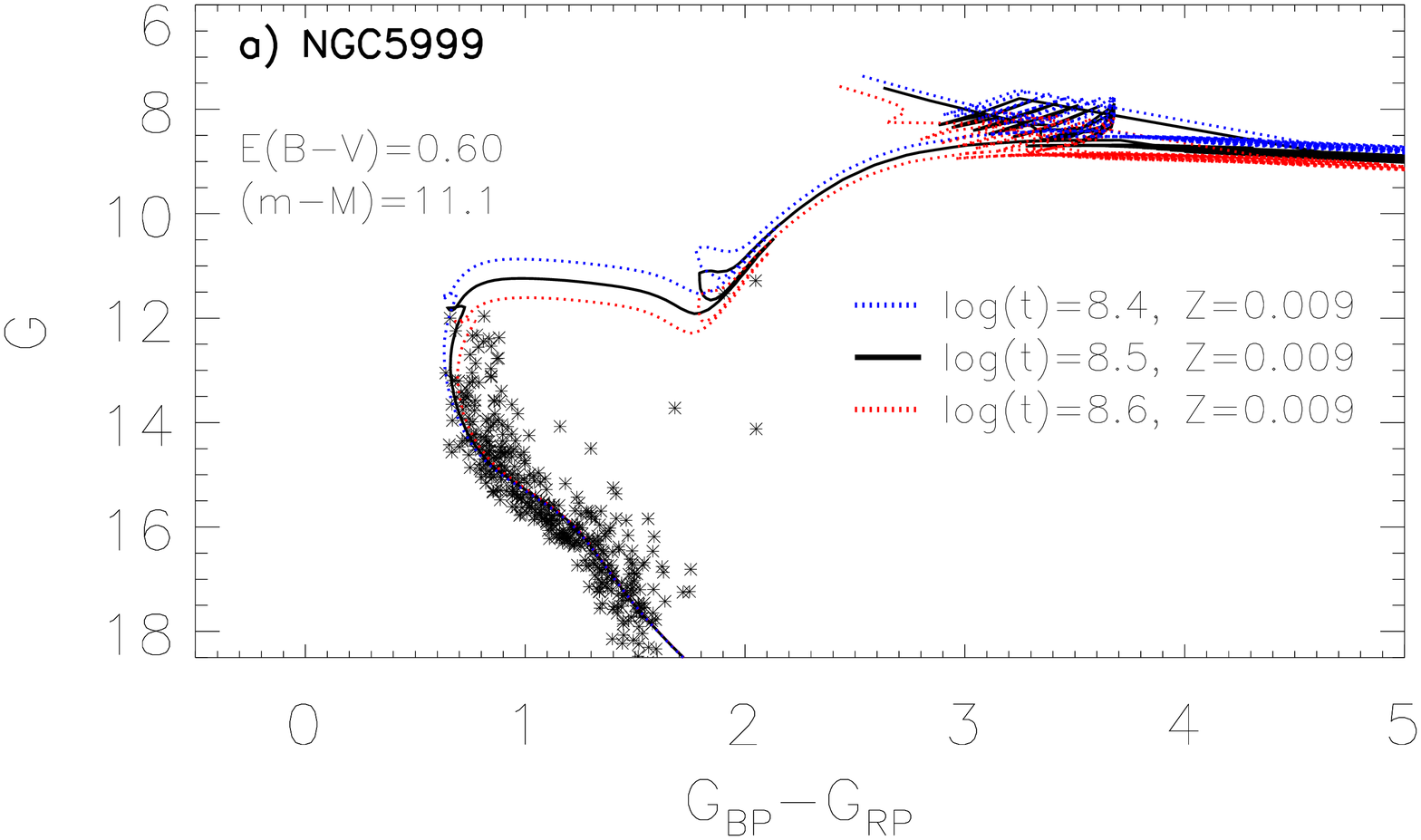} \\
\includegraphics[width=0.95\linewidth]{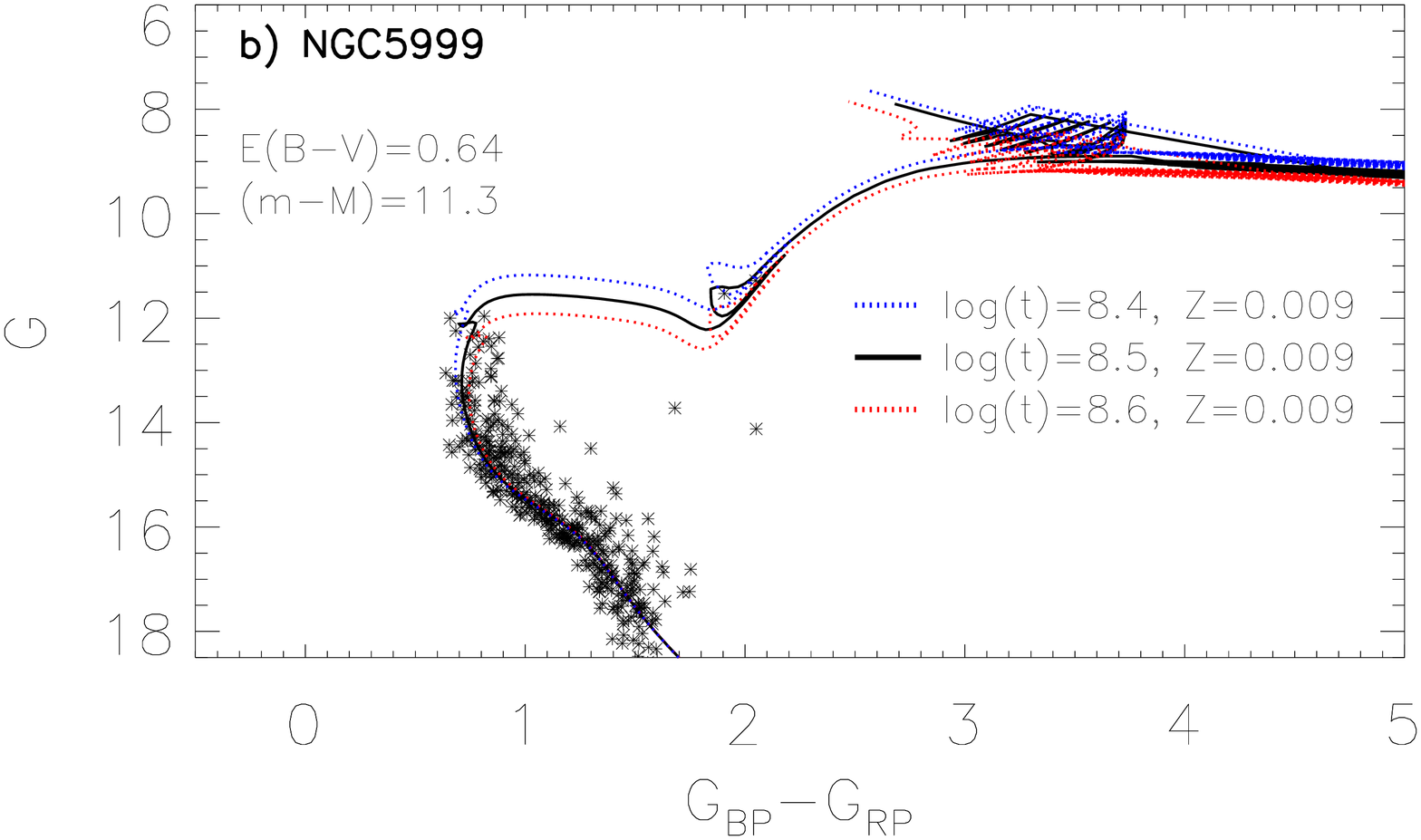} \\
\includegraphics[width=0.95\linewidth]{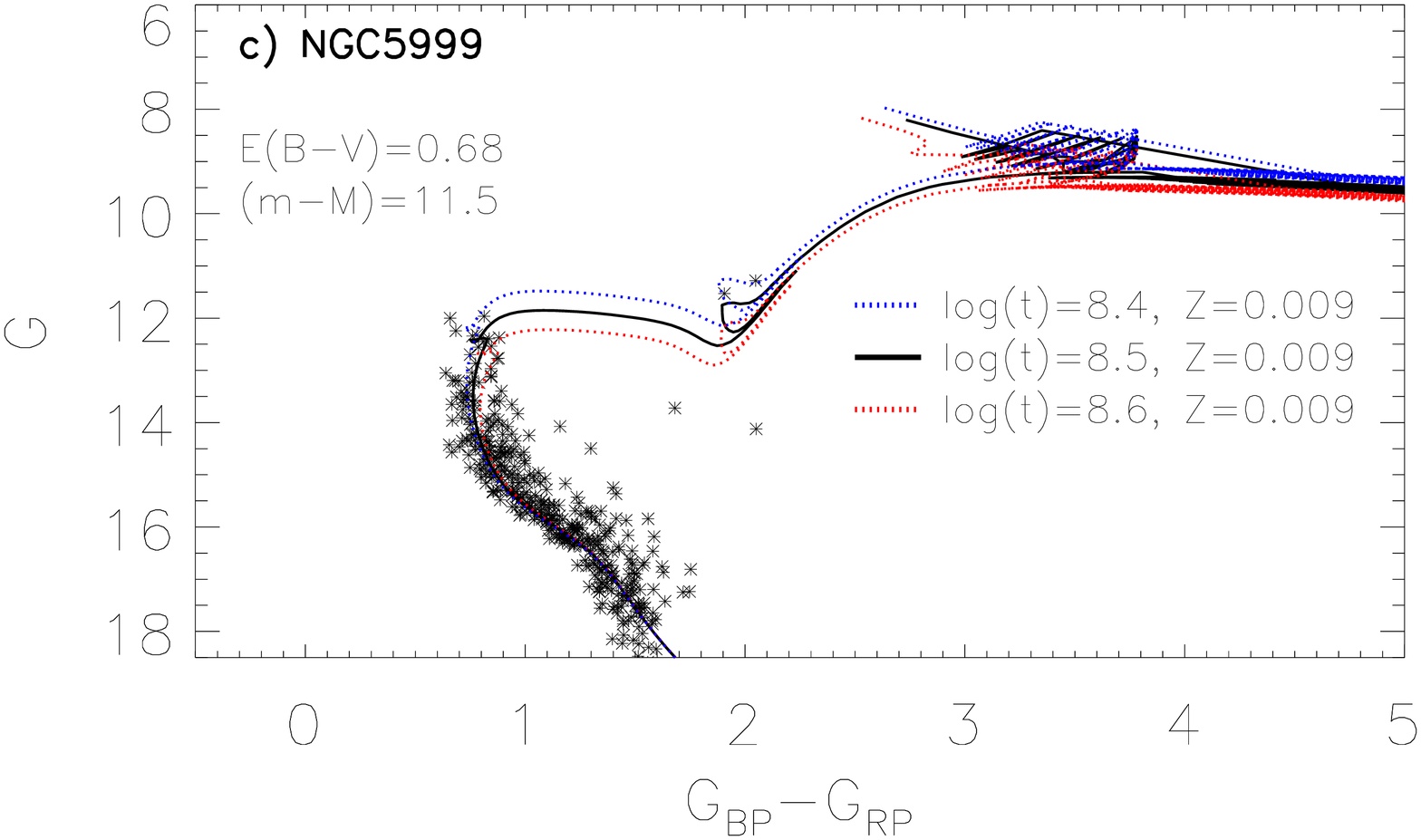}
\caption{PARSEC-COLIBRI isochrone fitting (solid line) over the cleaned CMD for NGC\,5999 and  the corresponding age uncertainties (dashed lines). The three isochrones have the same metallicity. From panels (a) to (c) one can see the effect of varying the colour excess and the distance modulus.}
\label{fig:isoc1}
\end{figure}

\begin{figure}
% \centering
\includegraphics[width=0.490\linewidth]{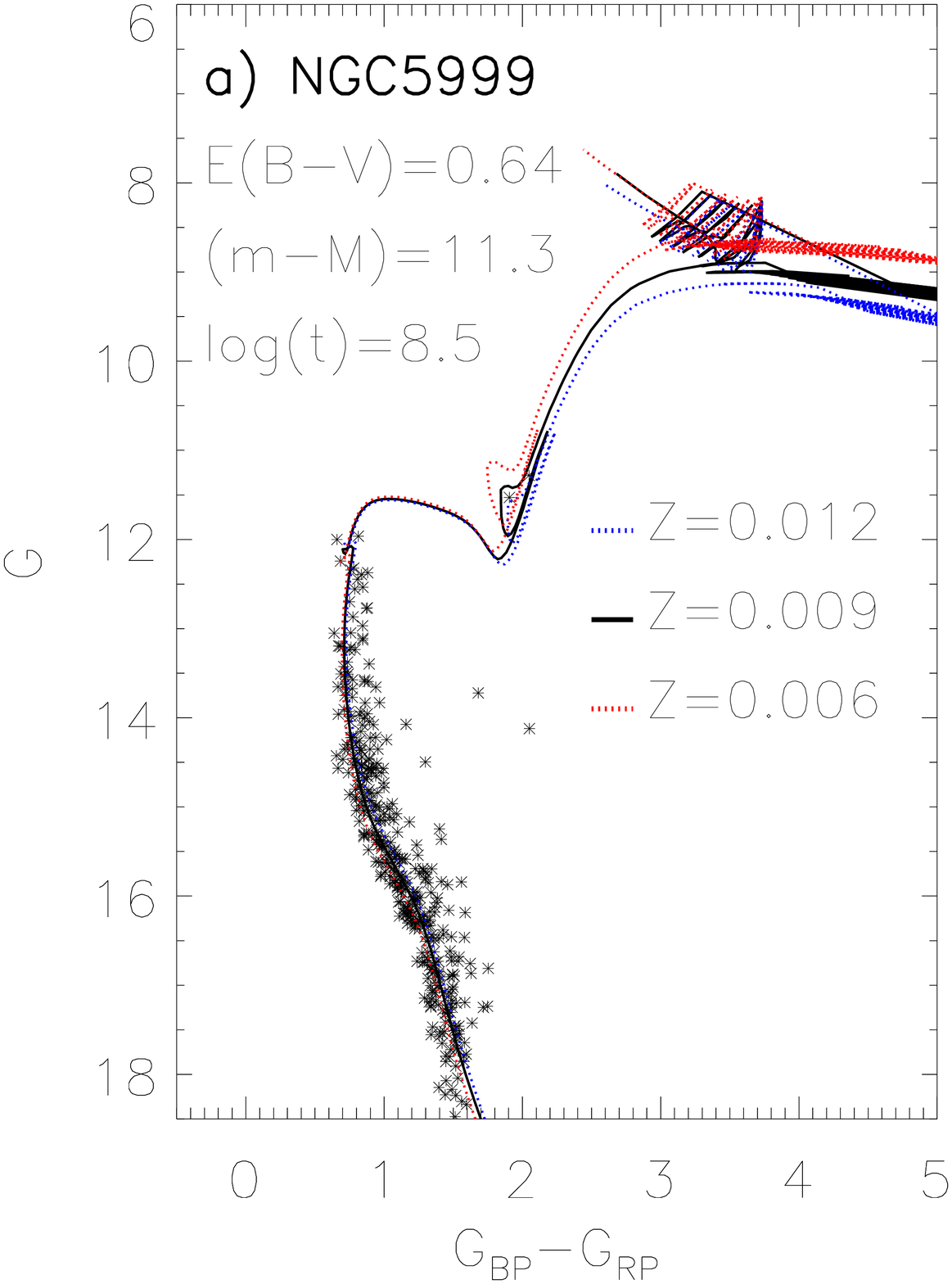}
\includegraphics[width=0.490\linewidth]{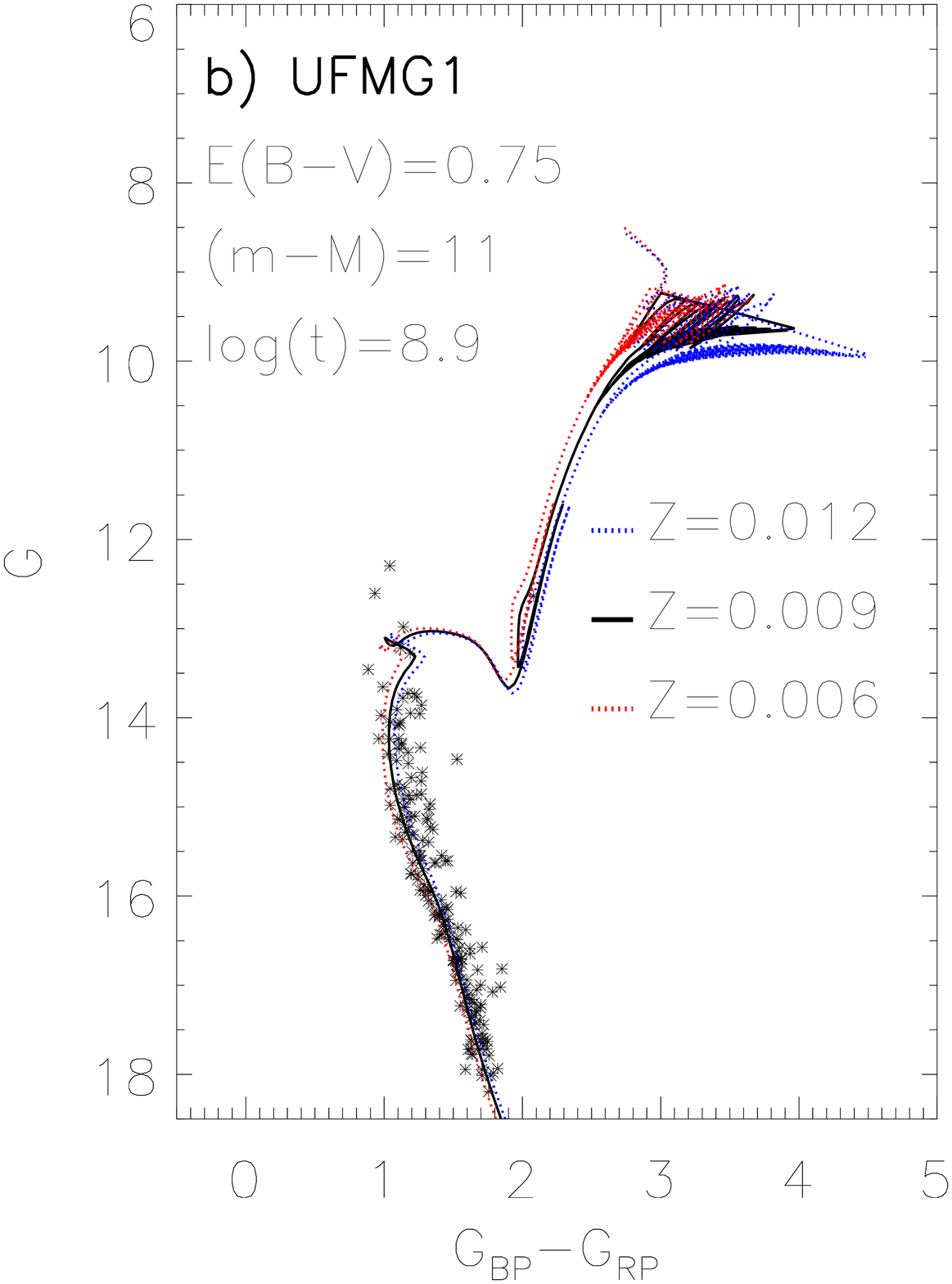} \\ \vspace{0.15cm}
\includegraphics[width=0.492\linewidth]{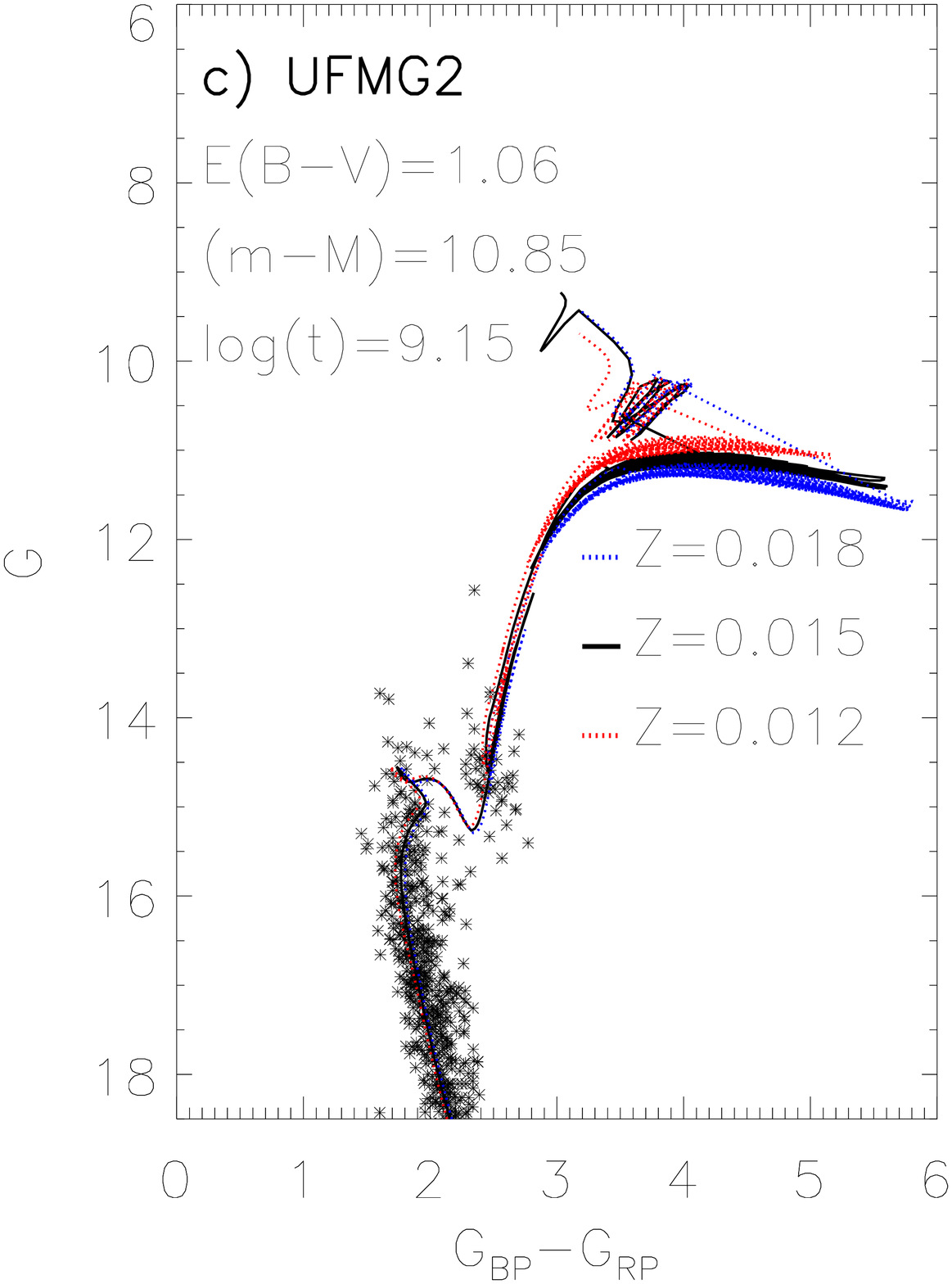}
\includegraphics[width=0.491\linewidth]{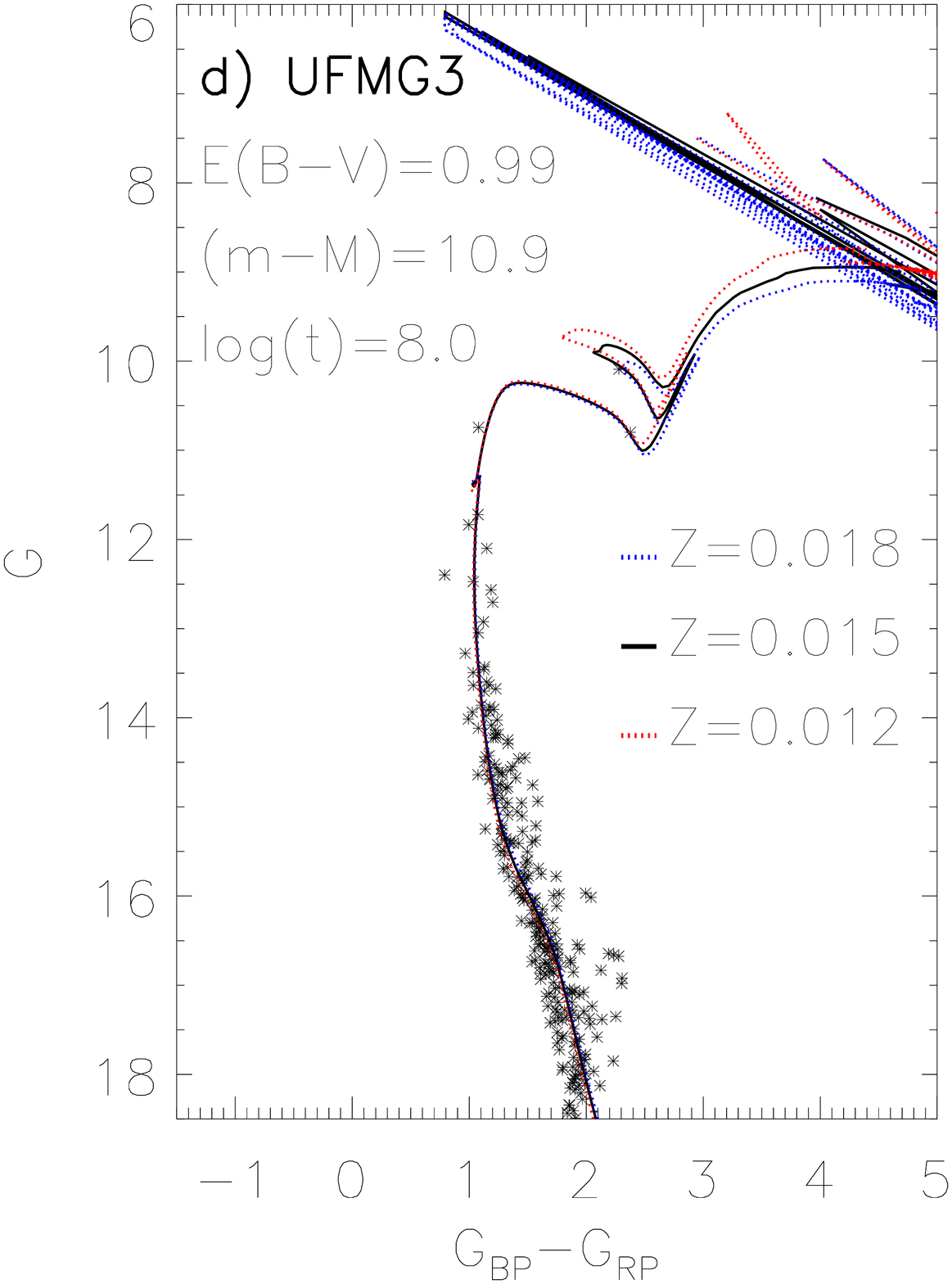}
\caption{PARSEC-COLIBRI isochrone fitting (solid line) over the cleaned CMD for NGC\,5999 (a), UFMG\,1 (b), UFMG\,2 (c) and UFMG\,3 (d) and  the corresponding age uncertainties in metalicity (dashed lines). The adopted values of the colour excess, distance module and age are indicated in the panels.}
\label{fig:isoc2}
\end{figure}

As an example, Fig.~\ref{fig:isoc1} presents this procedure for NGC\,5999, where three isochrones of different ages and same metallicity are overplotted on the cluster CMD. Note how the giant clump provides a strong constraint to the match. The same procedure was applied to the other clusters. The continuous line in the central panel of Fig. ~\ref{fig:isoc1} gives the best isochrone match, which  was then refined by exploring adjacent metallicities from the central value. Fig.~\ref{fig:isoc2} shows  the final match (continuous black line) for all clusters, making evident the influence of metallicity on the evolved stellar population.

%For each clusters we first assume a solar metalicity and searched for an isochrone that was capable to countain the stars from the main sequence until the apparent turn-off point, without looking to the giants. During this procedure we vary the color excess and the distance modulus. With his first procedure, we cold assume an approximate value for the distance modulus and color excess. Them we fixed the distance modulus and color excess and change the age to until fit the giants, during this procedure we could refine the distance modulus and color excess, we can see the effect of the changes in $log(t)$, distance modulus and color excess for the cluster NGC\_5999 in the Fig.~\ref{fig:isoc1}. Then we searched for the best metalicity by keeping constant the age, distance modulus and color excess to find the best metalicity. The best isochrones fitted can be seen in the Fig.~\ref{fig:isoc2}.

%Colour filters have been used to reject stars too distant from the fitted isochrone. 
In summary, the best-fitting isochrones along with conservative estimates that encompass reasonable matches to the stars loci, provided the uncertainties in the parameters. All parameters derived from the isochrone matches are presented in Table~\ref{tab:parameters}.

\subsubsection{King model fittings}

\begin{figure*}
\centering
\includegraphics[width=0.48\linewidth]{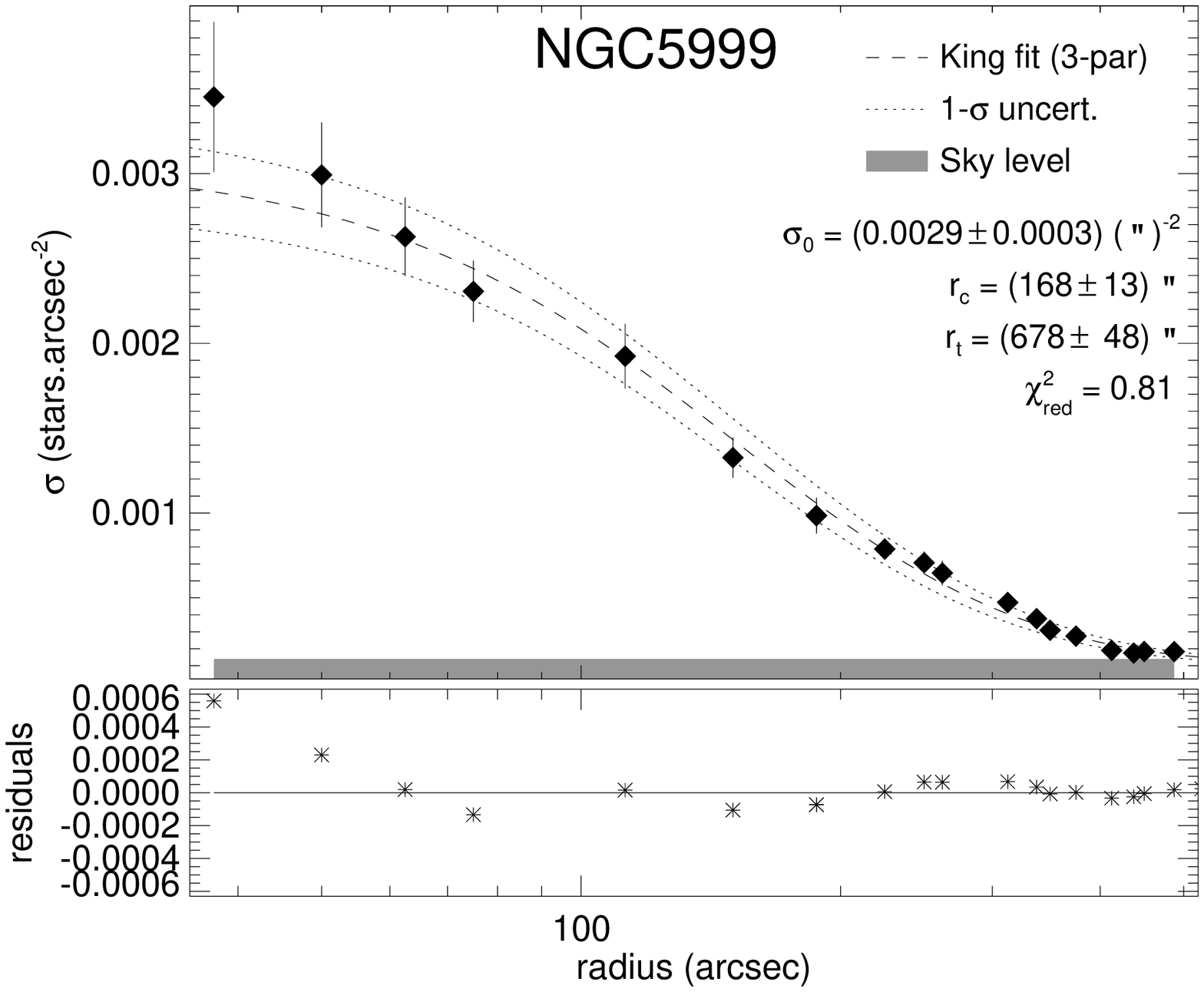} \hspace{0.25cm}
\includegraphics[width=0.48\linewidth]{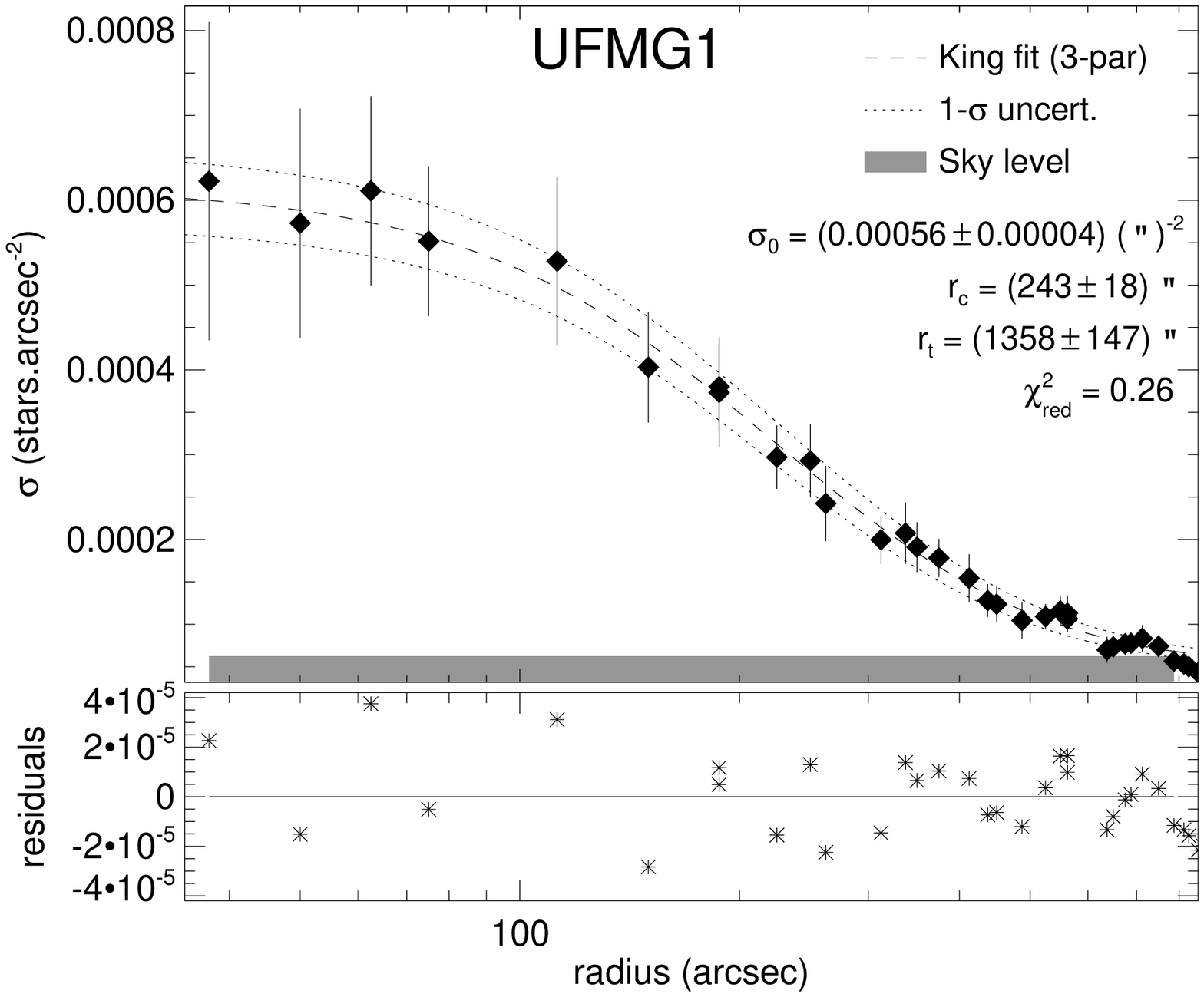} \\ \vspace{0.25cm}
\includegraphics[width=0.48\linewidth]{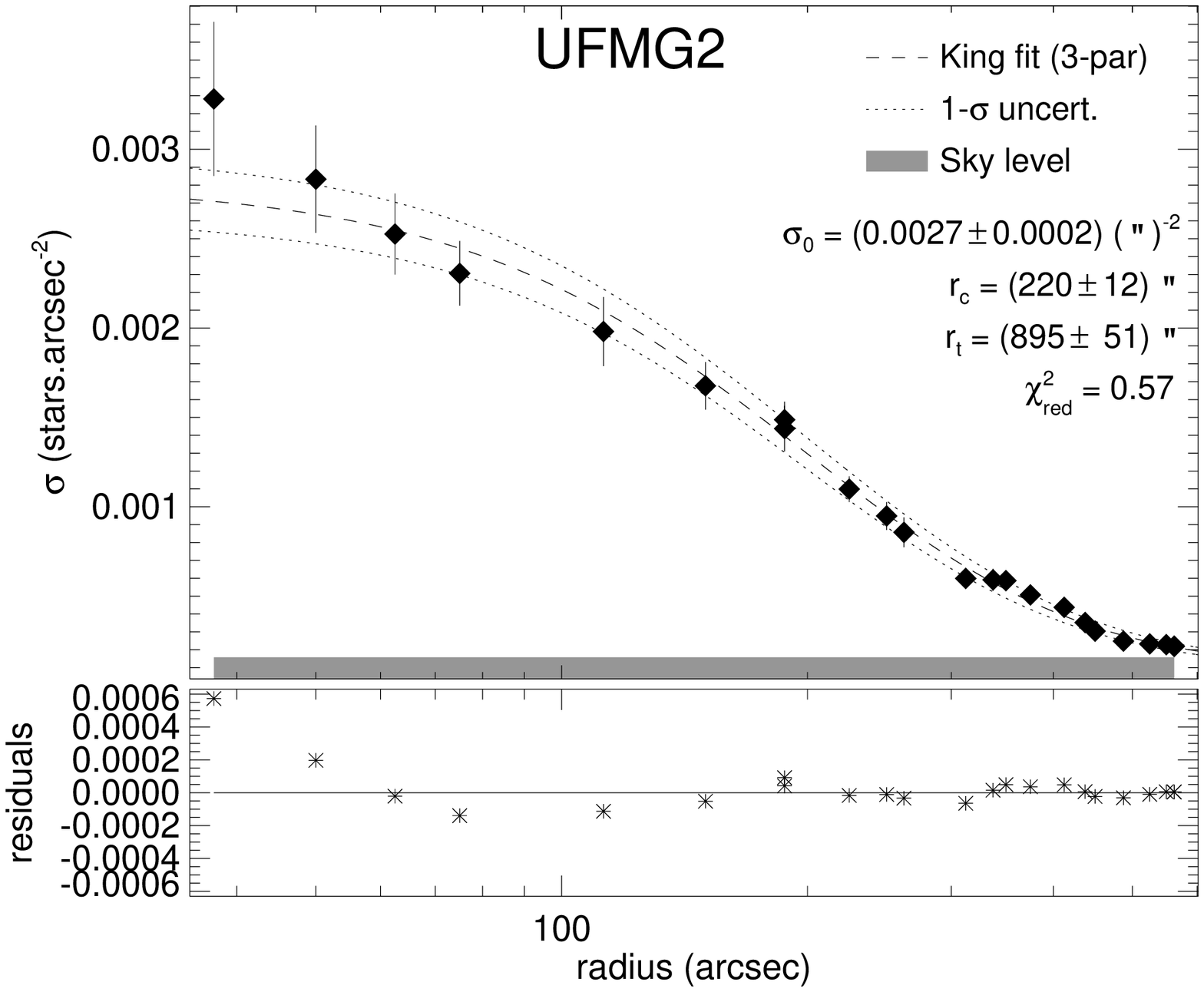} \hspace{0.25cm}
\includegraphics[width=0.48\linewidth]{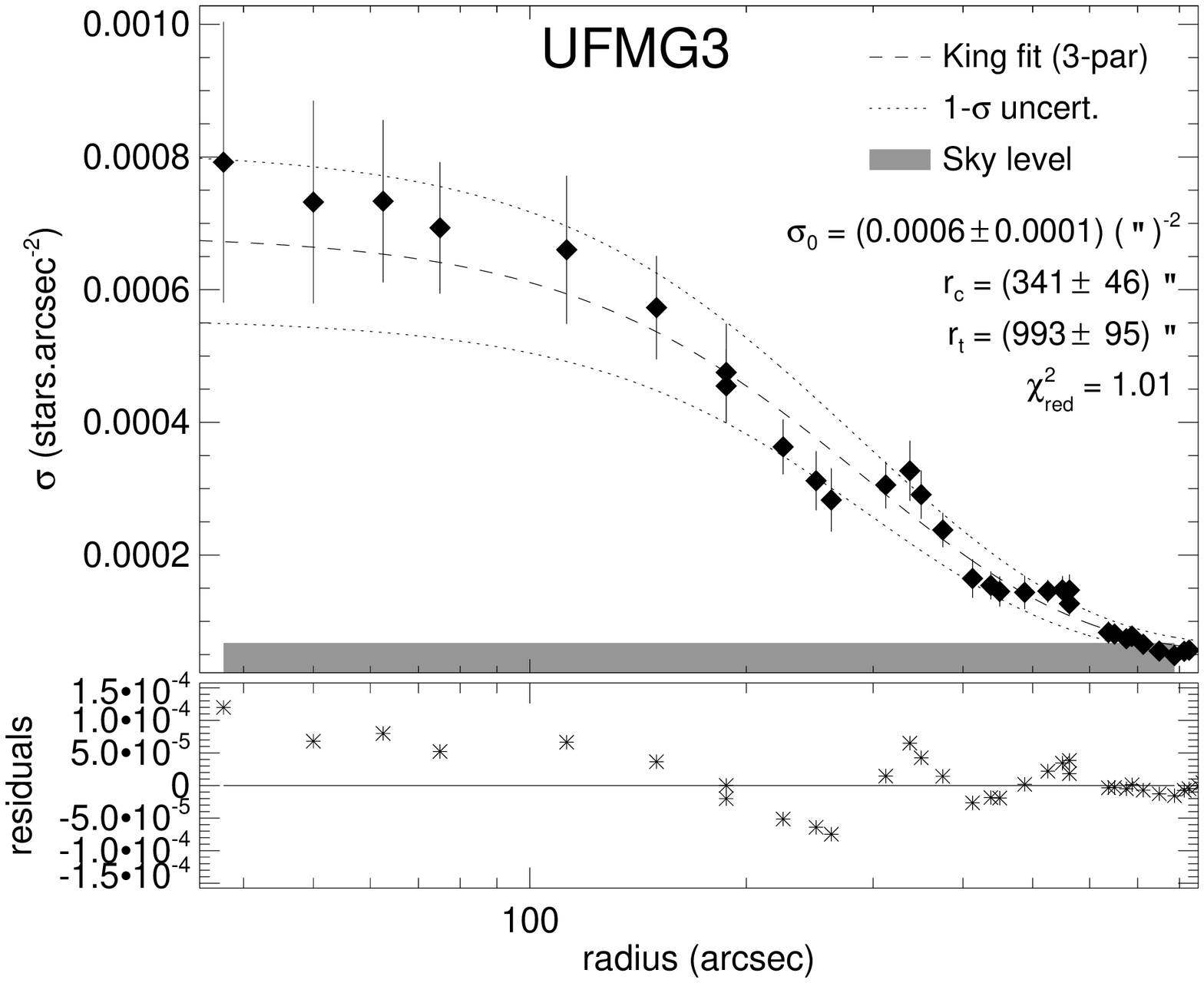}
\caption{Best-fitting of three-parameter King models to the radial density profile  for the studied clusters 
(dashed line) with envelope of 1-$\sigma$ uncertainties (dotted lines). Error bars correspond to poissonian noise. The sky level and its fluctuation is indicated by the grey bar. The fitting residuals are also presented in the lower panels.}
\label{fig:rdp_king}
\end{figure*}

The structural parameters central density ($\sigma_\circ$), core ($r_c$) 
and tidal ($r_t$) radii were obtained by fitting \cite{King:1962} models to
the RDP of each cluster member stars. Four sets of radial 
bins were used to represent the radial density profile of the clusters. Fig.~\ref{fig:rdp_king} shows the best King models (dashed lines) fitted to each cluster radial density profile and the respective 1-$\sigma$ uncertainties  (dotted lines). Error bars correspond to Poisson uncertainties.  Table~\ref{tab:parameters} presents the obtained parameters.

%\begin{figure}
%\centering
%\includegraphics[width=0.9\linewidth]{figs_novas_eps/rprof_ufmg2_pag3_king3.eps}\\
%\includegraphics[width=0.9\linewidth]{figs_novas_eps/rprof_ufmg3_pag3_king3.eps}
%\caption{Best fitting three-parameter King profiles for the clusters UFMG\,2 and UFMG\,3 for the determination of their core and tidal radii.}
%\label{fig:rdp_king2}
%\end{figure}

%\begin{figure}
%\centering
%\includegraphics[width=0.6\linewidth]{fig6.pdf}
%\caption{Difference between the mean parallax (plx[GAIA]) and the %expected value given by our distance modulus (plx[mod]). The error %bars represent the quadratic sum of our uncertainty (from distance %modulus) and that of the Gaia parallax.}
%\label{fig:plx_plx}
%\end{figure}

\subsection{Distance inference}
\label{sect:distinf}

Distances for the individual stars in each cluster were calculated using Markov Chain Monte-Carlo (MCMC) simulations in a bayesian framework, according to the prescription by \citet[][hereafter BJ15]{Bailer-Jones:2015}. The exponentially decreasing space density (EDSD) prior defined in BJ15 was employed using the typically adopted scale length of 1.35 kpc \citep[e.g. see][]{Astraatmadja:2016}. 

\begin{figure}
\includegraphics[width=0.475\linewidth]{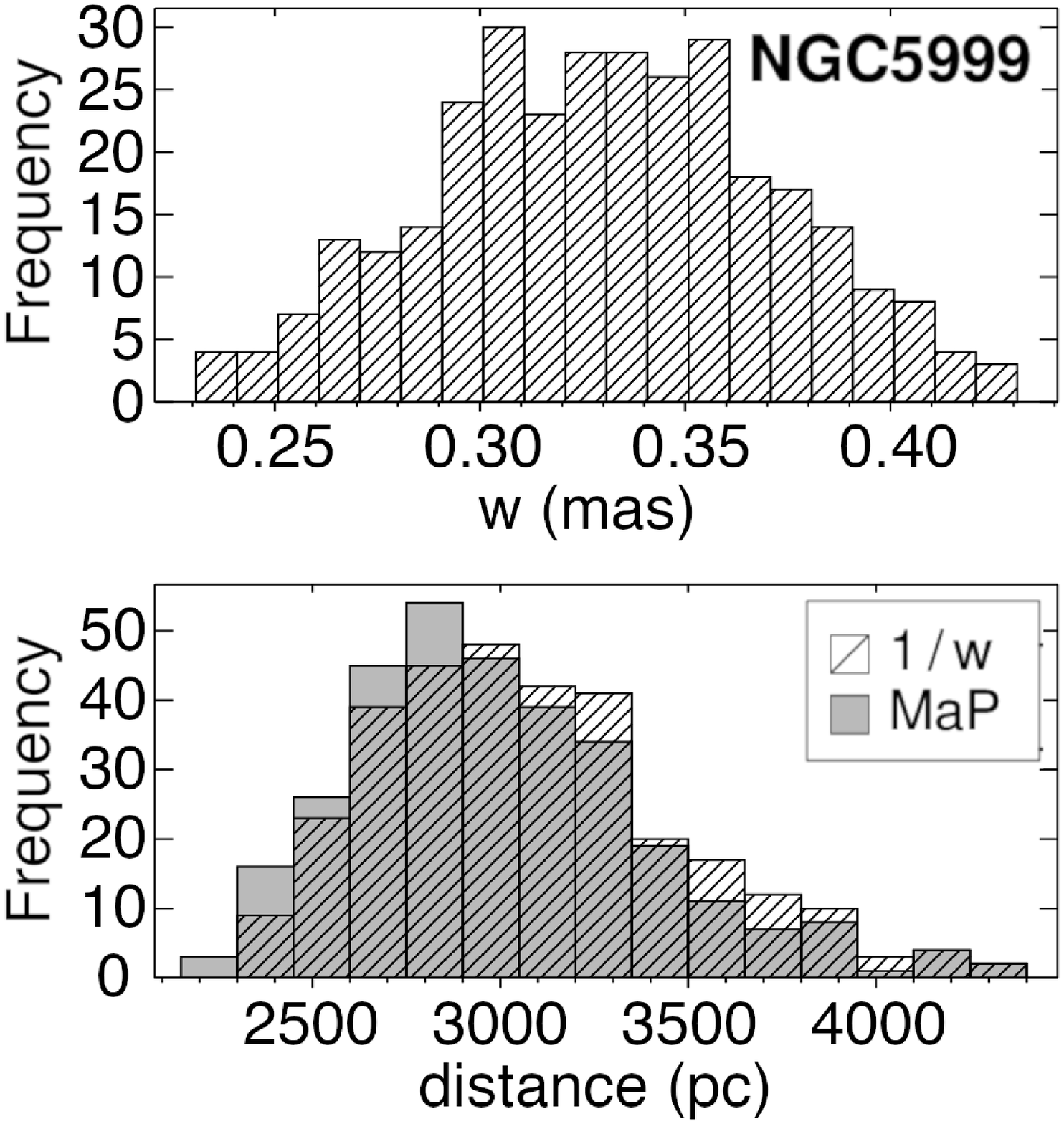}
\hspace{0.2cm}
\includegraphics[width=0.475\linewidth]{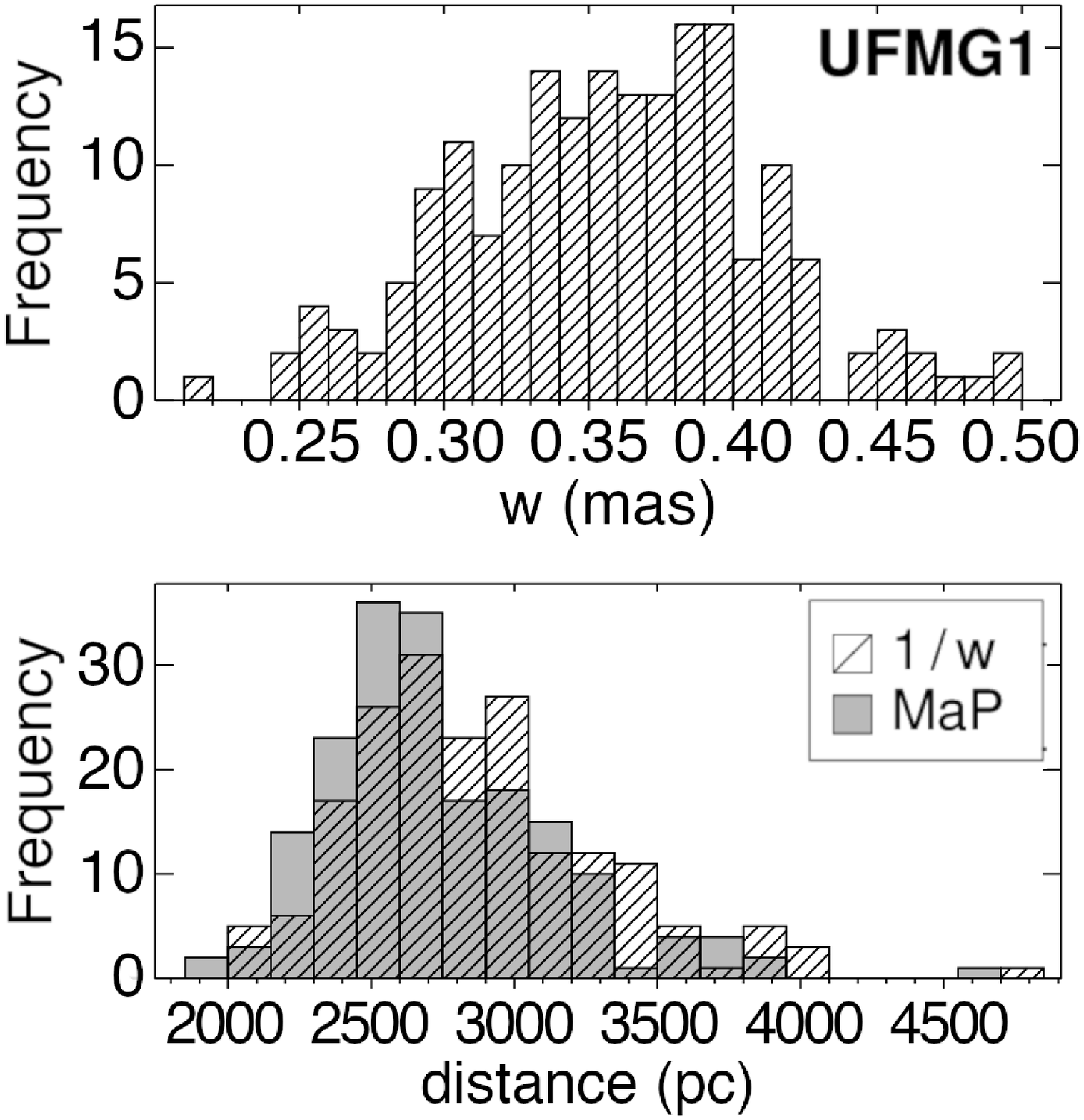} \\\vspace{0.1cm}

\includegraphics[width=0.475\linewidth]{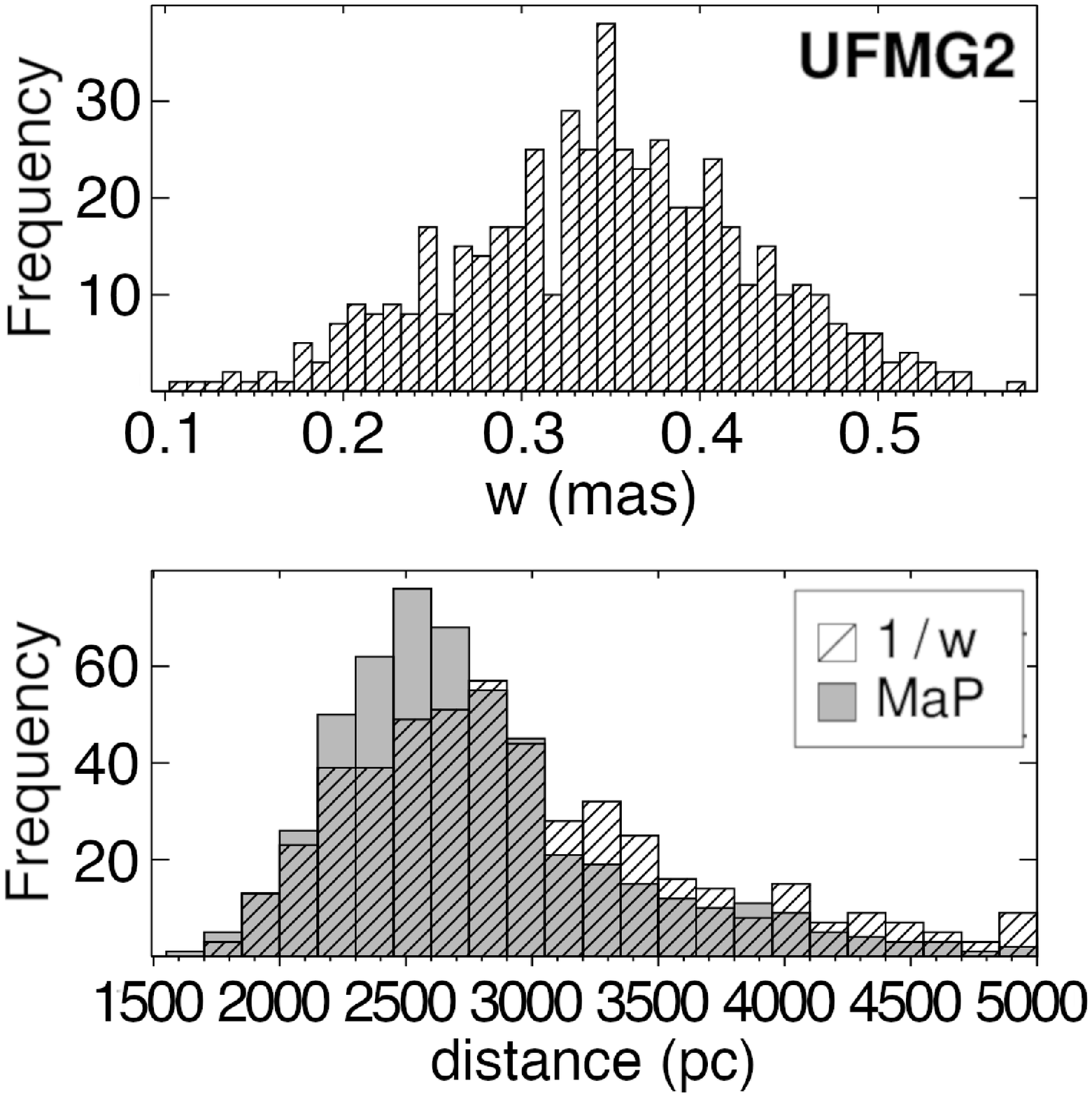} 
\hspace{0.2cm}
\includegraphics[width=0.475\linewidth]{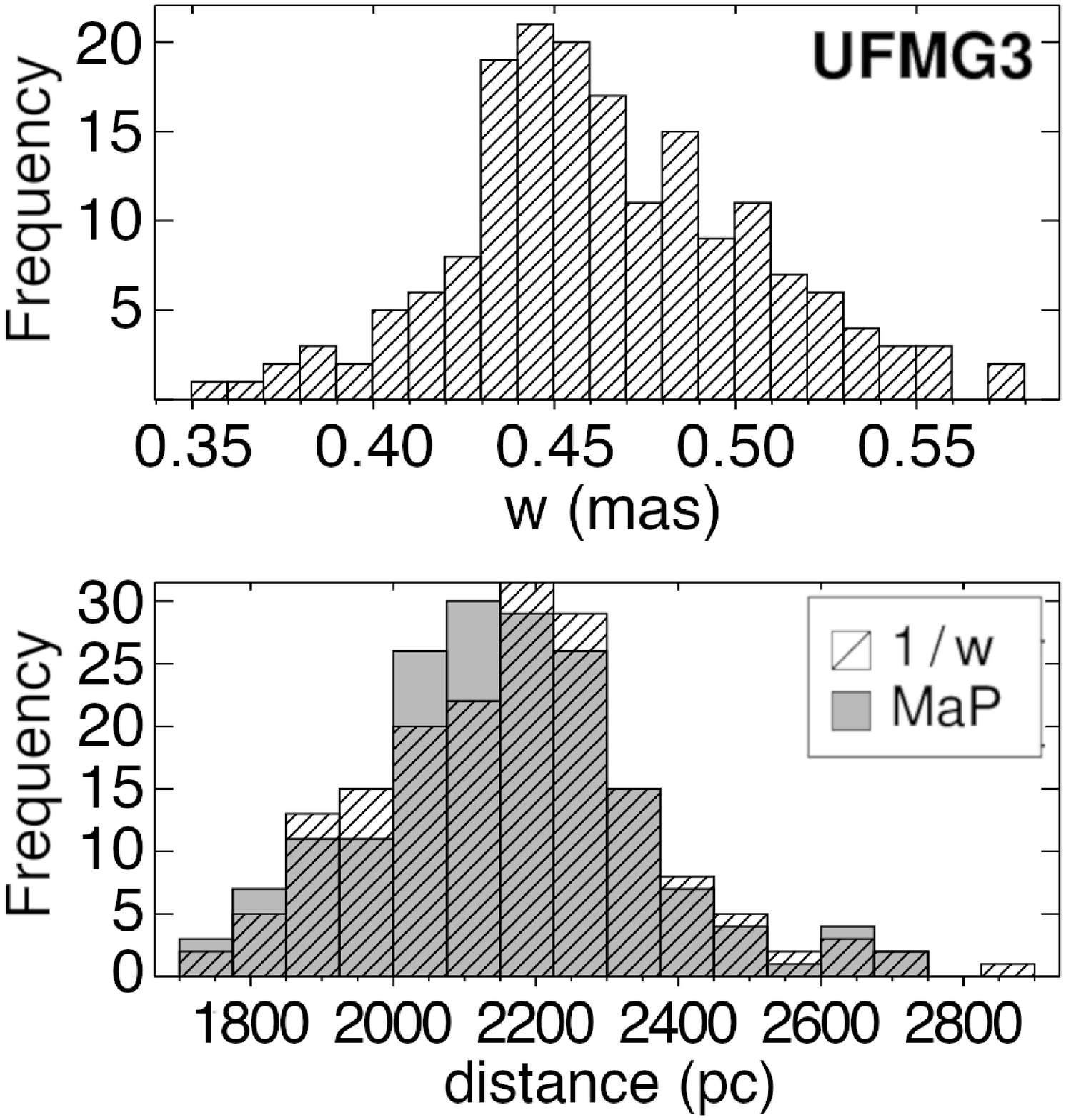}
\caption{Parallaxes ($w$) and the inferred distance (MaP) distributions of 
member stars for the clusters under investigation obtained from MCMC 
simulations.}
\label{fig:plx}
\end{figure}

The distance for each star was calculated as the mode of its posterior distribution (also called maximum a posteriori - MaP) and the corresponding uncertainty using the 68.27\% ($\sim$ 1-sigma) confidence interval ($c.i.$) around this value. The resulting distance (MaP) distribution of our targets is compared with the parallaxes ($w$) distribution in Fig.~\ref{fig:plx}. It can be seen that the MaP distribution does not deviate severely from the distances obtained by simply inverting the parallaxes ($1/w$), but it does show significant differences. 

Concerning parallaxes, we have added in quadrature the random errors resulting from the bayesian model (1-sigma of the stellar parallax distribution shown in Fig.~\ref{fig:plx}) to the systematic uncertainty of 0.1\,mas affecting the astrometric solution in Gaia DR2. Furthermore, the clusters' parallaxes were obtained by adding 0.03\,mas to those derived from the bayesian model, corresponding to the zero-point offset of Gaia DR2 parallaxes \citep{Luri:2018}. The final distances and uncertainties are then determined by inverting the parallaxes and by propagating their errors as calculated above. Table~\ref{tab:plx} presents these results compared to those obtained from the isochrone fittings.

%Table~\ref{tab:plx} shows the median clusters parallaxes, after the systematic correction reported for the Gaia DR2 solution (i.e. $-0.03$ mas). Uncertainties were also corrected to account for the 0.1 intrinsic parallax dispersion of the data, due to internal correlations \citep{Luri:2018}. Cluster distances were calculated from the mode of the stellar distances distribution inferred using the MCMC code and were also subject to the corrections mentioned. Table~\ref{tab:plx} compares these results with those obtained from isochrone fitting. 

Although the errors are large, it can be seen that distances from both methods appears to agree on the limit of their uncertainties for NGC\,5999 and UFMG\,3 and nearly so for UFMG\,1 and UFMG\,2. This behaviour was also found in previous works \citep[e.g.][]{Cantat:2018}, where isochronal distances agreed with parallaxes ones for some clusters and did not for others. Several factors could contribute to this difference such as the calibration of GAIA photometry, issues with the model isochrones and/or their fitting process and/or systematics in the astrometric solution of GAIA DR2.

\begin{table}
\centering
\caption{Distances comparison of studied clusters}
\def\arraystretch{1.3}
\begin{tabular}{l c c c} \hline
& $w$ & {\sc mcmc} & (m-M) \\
& [mas] & [kpc] & [kpc] \\ \hline
NGC\,5999 & $0.383\pm0.108$ & $2.61^{+1.25}_{-0.66}$ & $1.82\pm0.19$ \\
UFMG\,1 & $0.425\pm0.113$  & $2.35^{+1.01}_{-0.56}$ & $1.58\pm0.16$ \\
UFMG\,2 & $0.426\pm0.130$ & $2.35^{+1.23}_{-0.62}$ & $1.48\pm0.15$ \\
UFMG\,3 & $0.502\pm0.108$ & $1.99^{+0.63}_{-0.39}$ & $1.51\pm0.14$ \\
%$\Delta w$ (mas) & -0.20 & -0.24 & -0.28 & -0.19 \\ \hline
%$\tilde{d}$ (pc) & $1949$ & $1751$ & $1620$ & $1585$ \\ 
%$\sigma_{\tilde{d}}$ (pc) & 107 & 121 & 118 & 72  \\
%$\widetilde{c.i.}$ (pc) & $^{+149}_{-110}$ & $^{+124}_{-98}$ & $^{+123}_{-102}$ & $^{+108}_{-83}$ \\
\hline
\end{tabular}
\def\arraystretch{1.0}
\label{tab:plx}
\end{table}

Recently, \citet{Bailer-Jones:2018} used a smooth Galaxy model to improve the EDSD prior by calculating a length scale as a function of the Galactic longitude and latitude of each source in Gaia DR2, thus determining distances for the entire catalogue. When compared to the distances reported by their work, our results show very similar distances distributions for our targets, with nearly identical mode values. This is a expected result given that the EDSD is a weak prior and that the length scale reported by their galactic model towards the studied clusters ($1480-1580$ pc) is close to the length scale adopted.

\section{Clusters in the sky region}

\begin{figure}
\centering
\includegraphics[width=0.999\linewidth]{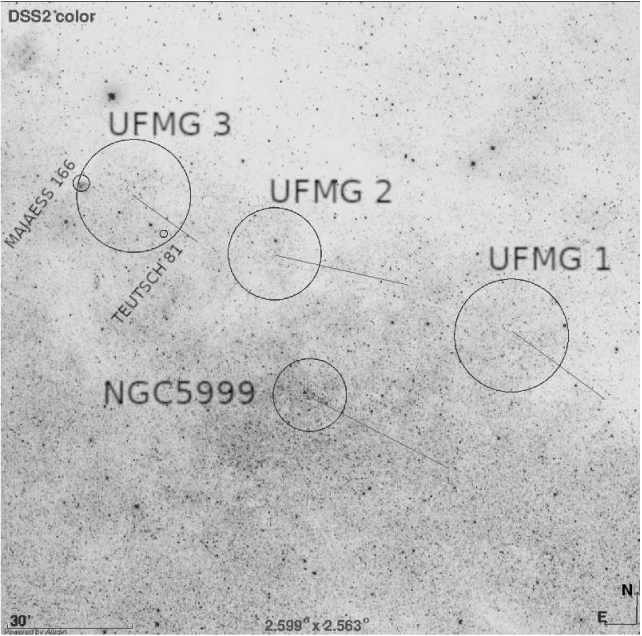}
\caption{2.5$\times$ 2.5 square degrees DSS2 image (Galactic coordinates) of the sky region containing the discovered clusters and NGC\,5999 together with their proper motions in 0.5\,Myr (lines). The circles represent the limiting radius of the clusters. The catalogued clusters Majaess\,166 and Teutsch\,81 are also shown. Although they are nearby, these two clusters do not confuse with UFMG\,3, as discussed in the text. }
\label{fig:clusters_dss} 
\end{figure}

Fig.~\ref{fig:clusters_dss} shows the field panoramic in Galactic coordinates built from a DSS2 colour image covering 2.5$\times$ 2.5 degrees. The cluster sizes are 
indicated by circles and their proper motion in $0.5$\,Myr represented by lines. 
%Catalogued smaller clusters Majaess\,166 and Teutsch\,81 are also shown.

One of the clusters (UFMG\,3) is located near two other known clusters: Majaess\,166 \citep{Majaess:2013} and Teutsch\,81 \citep{Kronberger:2006}, which appears to be different objects located $\sim$ 13\,arcmin from UFMG\,3 central coordinates. Teutsch\,81 was catalogued as a suspected open cluster candidate on the basis of visual inspection of DSS and 2MASS images. Given the small diameter reported by \cite{Kronberger:2006}, i.e., 1.8\,arcmin, which is much smaller than that found for UFMG\,3 (see Sect. \ref{sect:4.2}), we dismiss as a coincidence in this case. Majaess\,166 was identified as a very young cluster with a size of 4\,arcmin \citep{Majaess:2013} and it is easily distinguished from UFMG\,3 in DSS images for its nebulous nature, the cluster being still embedded in the progenitor gas cloud.  

\begin{figure}
\centering
\includegraphics[width=0.999\linewidth]{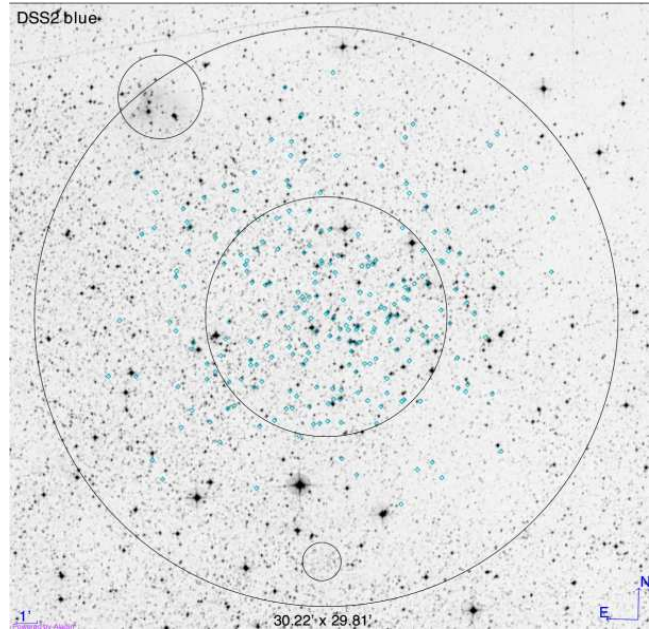}  \vspace{0.3cm}
\caption{DSS2 image in the vicinity of UFMG\,3, showing its core and tidal radius (large circles), and its selected members (tiny coloured diamonds). The limiting radius of clusters Majaess\,166 and Teutsch\,81 are represented by smaller circles in the top and bottom of the image, respectively.}

%\caption{ DSS2 images of the clusters Majaess\,166 (top-left),  Teutsch\,81 (top-right) and UFMG\,3 (bottom). Circles around Majaess\,166 and Teutsch\,81 correspond to their limiting radii, while circles around UFMG\,3 indicate its limiting and core radius.} 
%It can be seen that UFMG\,1 UFMG\,3 are both sparser than UFMG\,2.}
\label{fig:ra_de_maja_teut_ufmg3}
\end{figure}

 Since there is some superposition of the area covered by the three clusters, we further searched for additional evidences that would clarify their natures as distinct stellar systems.  
Fig.~ \ref{fig:ra_de_maja_teut_ufmg3} shows a zoomed DSS2 image of the three clusters area with circles indicating their limiting radii and, in the case of UFMG\,3, also its core radius.
%Note the presence of the relatively bright, main sequence member stars within UFMG\,3 core radius. 
The UFMG\,3 members are also shown, confirming that most of them lie within its core radius, particularly the relatively bright stars belonging to the red clump and turnoff stars. Given the size differences and the disposition of the stellar content of UFMG\,3, it is clear that these are distinct objects.
%The spatial distribution of all stars within the literature radius for the known clusters and UFMG\,3 members according to our analysis is presented in Fig.~\ref{fig:ra_dec_ufmg3_t81_m166}. A region of 15\,arcmin radius around UFMG\,3 central coordinates, therefore encompassing Teutsch\,81 and Majaess\,166, was explored in the astrometric space. Fig.~\ref{fig:teutsch81_pm} shows the VPD with colour-coded parallaxes for this region depicting Teutsch\,81 stars. The same data was used to build a similar VPD for Majaess\,166 (Fig.~\ref{fig:majaess166_pm}). In both diagrams, UFMG\,3 members stand out as a clear overdensity clump around  $\mu_{\alpha}^{*}\sim -1$ and $\mu_{\delta}\sim-2$, distinct from the astrometric parameters of the stars sampled towards the other clusters. 
 The proper motion of Majaess\,166 based on UCAC4 data, $\mu_{\alpha}^{*}=-2.17 \pm 0.53$ and $\mu_{\delta}=-2.70 \pm0.70$ \citep{Dias:2014}, is the only additional information for this cluster and it differs from the proper motion of UFMG\,3 (Table~\ref{tab:parameters}). 
 
 %The few stars in Majaess\,166 region with Gaia proper motion similar to those of UFMG\,3 stars are out of UFMG\,3 limiting radii. In addition, 
 
 Being probably distant clusters projected in the direction of UFMG\,3, Teutsch\,81 and Majaess\,166 could be better investigated using near-infrared bands, possibly allowing their stellar content to be distinguished from field stars.

%The only data available for Teutsch\,81 are its centre coordinates and its apparent size. The top-right pannel of the Fig.~\ref{fig:ra_de_maja_teut_ufmg3} shows that in fact there exist a stellar overdensity in its centre coordinate. When analysing the proper motions of stars within Teutsch 81 perimeter,  we note that UFMG3 stars have different proper motions. The Fig~\ref{fig:teutsch81_pm} shows the proper motion for a field of 15 arcmin radius from UFMG3 central coordinate, we overplotted the stars located inside Teutsch 81 perimeter. It is possible to note an overdensity around  $\mu_{\alpha}^{*}\sim -1$ and $\mu_{\delta}\sim-2$, they are UFMG3 probable members and we note no relation between the clusters populations.

\section{Concluding remarks}

We report the serendipitous discovery of three star clusters projected towards 
the Norma constellation around the open cluster NGC\,5999, in the Sagittarius 
spiral arm. The discovery was possible due to filtering procedures on Gaia DR2 
data. A colour-magnitude filter applied to select young to intermediate-age 
main-sequence stars and red clump giants turned evident the presence of NGC\,5999 
and the three new stellar groups in proper motion space. We then determined their 
centres and limiting radii by analysing their projected stellar distributions. 
Gaia photometry and proper motions were employed to characterize the new clusters 
at tandem with isochrone fitting of their stellar populations, from which a 
consistent set of astrophysical parameters was obtained.

All clusters have well-defined main sequences and contain at least two giants, 
facilitating the isochrone fitting procedure. The oldest cluster (UFMG\,2) has a 
developed red giant clump and presents a broader main sequence, possibly 
indicating differential reddening and/or stellar populations with slow rotators 
(bluer stars) and fast rotators (redder stars), as exhibited by the open cluster 
M\,11 \citep{Marino:2018}. 

King-profile fittings over the clusters RDP have shown that our targets present
tidal boundaries between 5-10 pc, typical of relaxed open clusters. Their 
concentration parameter ($c = \log\,{r_t/r_c}$) corroborate our derived ages, 
indicating a more dense core ($c \sim 0.65$) for evolved targets NGC\,5999, 
UFMG\,1 and UFMG\,2, possible as result of internal stellar collisions and energy 
equipartition during their long term evolution and a sparser central structure 
($c=0.46$) for the much younger UFMG\,3.

When compared to previous studies of NGC\,5999 \citep{Santos:1993, 
Piatti:1999,Dias:2002,Roeser:2010,Kharchenko:2013,Moni:2014} we have found consistent 
values of colour excess, age, distance and tidal radius. In addition, owing to the 
high quality of the photometric and astrometric data from Gaia DR2 we were able to 
achieve a considerably smaller uncertainty (at least 50\%) for its distance estimate. 
For UFMG\,1, UFMG\,2 and UFMG\,3 we are providing the structural and astrophysical 
parameters for the first time.

By employing the bayesian inference method described in BJ15 to estimate individual 
stellar distances of the studied clusters members and by properly applying the systematic
corrections to GAIA DR2 data, we have found that their probable distances are marginally
consistent with those found by isochrone fitting. Although the distances inferred from the
parallax are systematically higher, their relatively large uncertainties puts them within
the 1-sigma range of the isochronal distances. This systematic difference cannot be attributed 
to the adopted methodology as it is also present in the distances calculated by 
\citet{Bailer-Jones:2018} for our targets. While this effect has been seen 
before \citep[e.g.][]{Cantat:2018}, we are not sure of its exact origin. 
It could be attributed to uncertainties linked to the model isochrones, the 
photometry and/or the astrometric solution of Gaia DR2.
%{\bf The choice of an appropriate prior in our likelihood model for all clusters stars delivered distances consistent with those found by isochrone fittings considering a systematic offset of 0.2-0.3\,mas.}

%It appears that this methodology is needed if one is to found realistic star cluster distances in the small parallaxes domain, where relative uncertainties are large. In the absence of such corrections (i.e. adopting the EDSD prior with a null shift) resulted in cluster distances up to 2.3 times larger in relation to the determined values with uncertainties up to 10 times bigger. Unfortunately, since all our clusters present similar parallax ranges ($0.20-0.55$ mas) and distances ($1.5-2.0$ kpc), we cannot say if this correction is constant or distance dependent. 

The membership (Sect.~\ref{sect:membership}) and distance 
(Sect.~\ref{sect:distinf}) derived for the clusters stars 
are available as electronic tables through 
Vizier\footnote{http://cdsarc.u-strasbg.fr/vizier/cat/J/MNRAS/{\bf vol/page}}.

%negrito

%As warned by \cite{Cantat:2018}, \cite{Babusiaux:2018} and \cite{Lindegren:2018}, we found discrepancies between distances estimated from Gaia parallax and from CMD distance modulus. Our analysis indicates that Gaia parallaxes are smaller by a factor of at least 0.2 mas in relation to the parallaxes converted from distance modulus. Furthermore, the clusters stars showed a spread in parallaxes that is about 50$-$100 times larger than their expected size along the line of sight. 

\section*{Acknowledgements}

The authors wish to thank the Brazilian financial agencies FAPEMIG, CNPq and CAPES (finance code 001). F.~Maia acknowledge FAPESP funding through the fellowship n$^\mathrm{o}$ 2018/05535-3.
This research has made use of the VizieR catalogue access tool, CDS, Strasbourg, France. This work has made use of data from the European Space Agency (ESA) mission
{\it Gaia} (\url{https://www.cosmos.esa.int/gaia}), processed by the {\it Gaia}
Data Processing and Analysis Consortium (DPAC,
\url{https://www.cosmos.esa.int/web/gaia/dpac/consortium}). Funding for the DPAC
has been provided by national institutions, in particular the institutions
participating in the {\it Gaia} Multilateral Agreement. This  research  has  made  use  of  TOPCAT \citep{Taylor:2005}.   

%%%%%%%%%%%%%%%%%%%%%%%%%%%%%%%%%%%%%%%%%%%%%%%%%%



\bibliographystyle{mnras}
\bibliography{references} % if your bibtex file is called example.bib

%\begin{thebibliography}{99}
%\end{thebibliography}

%%%%%%%%%%%%%%%%%%%%%%%%%%%%%%%%%%%%%%%%%%%%%%%%%%

%%%%%%%%%%%%%%%%% APPENDICES %%%%%%%%%%%%%%%%%%%%%

%\appendix

%%%%%%%%%%%%%%%%%%%%%%%%%%%%%%%%%%%%%%%%%%%%%%%%%%

% Don't change these lines
\bsp	% typesetting comment
\label{lastpage}
\end{document}